\documentclass[3p,11pt]{elsarticle}

\usepackage{lipsum}
\makeatletter
\def\ps@pprintTitle{%
 \let\@oddhead\@empty
 \let\@evenhead\@empty
 \def\@oddfoot{}%
 \let\@evenfoot\@oddfoot}
\makeatother

\usepackage{latexsym}
\usepackage{graphicx}  

\usepackage[T1]{fontenc}
\usepackage[latin1]{inputenc}
\usepackage{dsfont}
\usepackage{amssymb}
\usepackage{amsmath}
\usepackage{amsfonts}
\usepackage{latexsym}
\usepackage{color}

\usepackage{moreverb}
\usepackage{epstopdf}

\newcommand\BibTeX{{\rmfamily B\kern-.05em \textsc{i\kern-.025em b}\kern-.08em
T\kern-.1667em\lower.7ex\hbox{E}\kern-.125emX}}

\usepackage{graphicx, caption, subcaption}

\begin{document}

\begin{frontmatter}

\title{{\bf A QFT approach to robust dual-rate control systems}\tnoteref{dpi}}

\author[1]{Alfonso Ba\~nos\corref{cor1}}

\author[2]{Juli\'an Salt}

\author[2]{Vicente Casanova}

 \tnotetext[dpi]{The work of A. Ba\~nos has been supported by {\em FEDER-EU} and {\em Ministerio de Ciencia e Innovaci\'on (Gobierno de Espa\~na)} under project DPI2016-79278-C2-1-R; the work of J. Salt and V. Casanova under Grant RTI2018-096590-B-I00.}

\cortext[cor1]{Corresponding author}

\address[1]{{Dept. Inform\'atica y Sistemas}, {Universidad de Murcia}, {Spain}}

\address[2]{{Dept. Ingenier\'ia de Sistemas y Autom\'atica}, {Universidad Polit\'ecnica de Valencia}, {Spain}}

%\address[3]{\orgdiv{Org Division}, \orgname{Org Name}, \orgaddress{\state{State name}, \country{Country name}}}

%\corres{*Facultad de Inform\'atica, Universidad de Murcia, 30100 Murcia, Spain. \email{abanos@um.es}}

\begin{abstract}
A dual-rate control system is a hybrid system composed of continuous-time and discrete-time elements with two sampling frequencies. In this work, a new frequency domain analysis and design approach, based on the Quantitative Feedback Theory (QFT) is developed, to cope with robust stability and tracking specifications. Tracking specifications are considered not only in the discrete-time but also in continuous-time, that allow a precise description of the intersample behavior (ripples), and characterization of frequencies below and beyond the Nyquist frequency.   Several illustrative examples and a case study has been developed.
\end{abstract}


%\jnlcitation{\cname{%
%\author{Williams K.}, 
%\author{B. Hoskins}, 
%\author{R. Lee}, 
%\author{G. Masato}, and 
%\author{T. Woollings}} (\cyear{2016}), 
%\ctitle{A regime analysis of Atlantic winter jet variability applied to evaluate HadGEM3-GC2}, \cjournal{Q.J.R. Meteorol. Soc.}, %\cvol{2017;00:1--6}.}

%\maketitle

\end{frontmatter}

%\footnotetext{\textbf{Abbreviations:} ANA, anti-nuclear antibodies; APC, antigen-presenting cells; IRF, interferon regulatory factor}

{\em Notation}: 
$\mathbb{R}_{\geq 0}$ is the set of non-negative real numbers, $\text{L}_{pe}(\mathbb{R}_{\geq 0})$, or simply  $\text{L}_{pe}$,  is the extended space of $\text{L}_{p}(\mathbb{R}_{\geq 0})$ (or simply  $\text{L}_{p}$), which is the normed space of Lebesgue measurable functions $f:\mathbb{R}_{\geq 0} \rightarrow \mathbb{R}$ with $\|f\| = (\int_0^\infty |f(t)|^p dt)^{1/p} < \infty$ for $1 \leq p < \infty$, and $\|f\| = \text{ess sup}_{t\in \mathbb{R}_{\geq 0}} |f(t)|$ for $p = \infty$. 
For a function $f$ of bounded variation on $(a,b)$, $f(t^+) = \lim_{\varepsilon \rightarrow 0, \varepsilon >0} f(t+\varepsilon)$, and  $f(t^-) = \lim_{\varepsilon \rightarrow 0, \varepsilon >0} f(t-\varepsilon)$, for any $t \in (a,b)$ . 
For a dual-rate system with sampling periods $T_f , T_s \in \mathbb{R}$, and $T_f = T_s/N$, with a integer $N >0$, the complex z-variables are $z_f = z$ and $z_s = z^N$. ${\mathbb C}$ is the set of complex numbers, ${\mathbb D} \subset {\mathbb C}$ is the open unit disk, and ${\mathbb D}^c ={\mathbb C} \setminus {\mathbb D}$, where $\setminus$ stands for set difference. 

%$\mathbb{R}^n$ is the $n$-dimensional Euclidean space, and $\mathbf{x} = (x_1,\cdots,x_n) \in \mathbb{R}^n$ is a column vector; $\|\mathbf{x}\|$ is the euclidean norm.  $\mathbb{B}$ is the closed unit ball in $\mathbb{R}^n$ centered at the origin. For a set $K \subset \mathbb{R}^n$, $\text{con}(K)$ denotes the convex hull of $K$, $\overline{K}$ is its closure, and $\text{int }K$ is the interior of $K$.  $\mathcal{S}_{\mathcal H}(\xi)$ is the set of maximal solutions $\phi$ to the hybrid system ${\mathcal H}$ with $\phi(0,0) =\xi$. $\text{dom}$ stands for domain, and $\setminus $ denotes sets difference.

%\vspace{2cm}
%\textcolor{red}{A sampled-data system is periodically time-varying. Therefore, the robustness of a sampled-data system to uncertainties with a bounded L2-induced norm depends on whether the uncertainty is assumed linear and time-invariant (LTI) or time-varying and/or nonlinear. For LTI uncertainties, the necessary and sufficient conditions for robustness are quite difficult to use. In this project a class of systems has been considered, for which a significant simplification is possible, and for which the robustness condition reduces to an H-infinity norm bound of a finite-dimensional discrete transfer function [TS99]. The simplification requires some assumptions on the system, which, however, it is believed are not too restrictive.}
%\vspace{0.2cm}

\section{Introduction}
A multirate (MR) control system is defined as a hybrid system composed of continuous-time and discrete-time elements (plant, controllers and filters), where two or more variables are sampled or updated at different frequencies \cite{kalman1959general,kranc1957input,jury1977sampled}. Since many years ago these systems have been considered in industrial environments where chemical analyzers are needed \cite{morant1986model, li2003application}, or in visual feedback applications in robotics \cite{hutchinson1994multi, sim2002multirate}; in all these cases post-processing requirements need a time interval that for a real-time process control request could be long. With these restrictions is not viable to keep and ideal single frequency in the control loop.  In the last years, remote trajectory control of autonomous vehicles \cite{lozano2012autonomous,cuenca2019remote} and efficient energy saving in networked based control systems \cite{cuenca2011delay,zhang2017analysis,alcaina2019energy} also required the use of MR systems. In every these cases, the control problem is that with the mentioned restricted frequency of measurement, far away from the ideal one, is not possible to assure the correct performance of the system. MR control systems allow to achieve a performance close to the projected one with no frequency restrictions. A dual-rate (DR) system is a MR system where there are only two sampling frequencies. The case with slow output and fast input called MRIC (multirate input control) is especially important. In a DR system it is usual to consider an integer relation between the sampling periods and without jitter between both sequences. Different control design methods have been introduced for these kind of systems \cite{ araki1986pole, cimino2010design,salt2014multiratesen}. A big number of these contributions were inspired in classical time-domain or state-space approach single rate methods. It was also introduced the optimal $H^{\infty}$ design in frequency domain \cite{chen1994h, saagfors1998h} for MR systems but an iterative problem was the ripple of the system response. Some authors faced the robust control problem for MR systems \cite{lee1992robust,lall2001lsr}. Nevertheless there was not a frequency-based analysis or design method inspired in classical techniques and, even more, assuming robust control. There was an inherent difficulty due to the complexity of the MR frequency response. In recent years, some contributions allow to make easier those purposes \cite{araki1986multivariable, araki1993frequency, thompson1986gain,salt2014new}. 

In this work, the Quantitative Feedback Theory (QFT) \cite{Horowitz93} is postulated as an efficient technique for analysis and design of DR control systems, including system with potentially large uncertainty. Being a sound and well-developed frequency domain technique, it is believed that QFT will be a unique framework for understanding how slow and fast sampling from the DR controller interact with the plant continuous dynamics, being a goal the efficient characterization of ripples and their removal with a proper controller design. QFT dates back to the seminal works of Isaac Horowitz \cite{Horowitz63} in the late fifties of the past century, that pioneered the analysis and design of linear and time-invariant systems with large uncertainty \cite{HorowitzSidi72}. Although somehow aside of the mainstream robust control research, %and sometimes in clear controversy, 
over the years QFT has been extended to cope with uncertainty in linear and time-varying systems \cite{Horowitz75} , nonlinear systems \cite{Horowitz76, HorowitzBanos01,Banos07}, systems with multiple-input multiple-outputs \cite{YanivHorowitz86,ElsoGilGarcia13}, multiloop \cite{Horowitz93,BanosHorowitz00}, etc., and has been also successfully applied in practice \cite{GarciaSanzBook}. Specifically regarding (single-rate) sampled-data control systems, several QFT approaches have been developed. The classical approach is based on the application of continuous-time QFT through the use of the $w$-domain with the bilinear transformation\cite{HorowitzLiao86}.  A much more solid approach \cite{YanivChait93} includes continuous-time tracking and gain and phase margin problems, %has been developed by Yaniv and Chait. 
in line with many others \cite{francis1988stability,ChenFrancis91} that focus on the continuous-time response of a continuous plant under sampled data control, and has been a clear inspiration for the present work.

The main contribution of this work is the development of a QFT approach for DR control systems having plants with potentially large uncertainty. It is mainly focused on the problem of robust stability and continuous-time tracking, and is specially focused on the slow-rate controller as design element. Several others performance specifications like disturbance rejection may be considered by using the developed framework. Some specific contributions are:
\begin{itemize}
\item The quantification of the continuous-time response in the frequency domain under DR control, that will allow the efficient characterization of ripples.
\item A Nyquist-like theorem for the robust stability of DR control systems, and the formulation of worst-case gain and phase margins.
\item Continuous-time tracking restrictions over the slow-rate controller, for a given fast-rate controller and prefilter, with performance specifications below and beyond the slow Nyquist frequency. 
\end{itemize}

As a result of the proposed approach, a number of new boundaries are developed that guaranty robust stability and continuous-time tracking. The next design steps are standard in QFT and will not be developed here in detail. Templates and boundaries computation are well-developed (note that only boundaries will be shown in the different examples along this work).  %(they has been computed with the algorithm by Moreno et al.). Finally, loop shaping of the nominal open-loop gain function (with the slow-rate controller as a key part).
The nominal open-loop gain shaping may be manually performed in simple cases, eventually with the aid of some computer toolbox \cite{BorghesaniChaitYaniv00,Gutman96,GarciaSanz08,RubinGutman19}. Additionally, automatic loop-shaping techniques are also available \cite{ChaitChenHollot99,GarciaSanzGuillen00,NatarajKubal06,CerveraBanos08}.  

 It is worthwhile to mention that the proposed QFT approach can be also applied to design single-rate controllers with continuous-time specifications beyond the Nyquist frequency, extending previous work \cite{YanivChait93} that suffered from that limitation. Also, it is useful for analyzing and designing DR controllers for plants with small or no uncertainty, although its full potential is clearly obtained for the case of large uncertainty.

In Section 2, besides some basic preliminary results the DR control problem is formulated. Section 3 is about analysis of DR control systems in the frequency domain; firstly, a motivational example is investigated by using several analysis tools available in the literature, then new frequency-domain tools are proposed. As a result, a Nyquist-like theorem for exponential and $L_p$-stability of the DR control system is developed. Also, properties of the continuous-time signal spectra are derived that will be the basis for QFT approach to be developed in Section 4. Here, with the focus on robust stability and tracking (including continuous-time tracking), a detailed QFT-based method is formulated to solve the DR control problem, for systems with potentially large uncertainty. Finally, Section 5 is devoted to a case study for a reaction wheel inverted pendulum.

\section{Preliminaries and problem statement}
For a continuous-time signal $x:\mathbb{R}_{\geq 0} \rightarrow \mathbb{R}$ and  a sampling time $T$, a sampler $S_T$ is a system that produces a  sampled-data signal $x^T = S_T x:\mathbb{N}_{\geq 0} \rightarrow \mathbb{R}$, given by $x^T(n) = x(nT)$ for $n \in \mathbb{N}_{\geq 0}$. As it is well known \cite{BraslavskyTesis}, if $x$ is a function of bounded variation in every finite interval of $\mathbb{R}_{\geq 0}$, then the spectra of the signals $x$ and $x^T$ are related by {(strictly speaking, $x$  must also have a Laplace transform with abcissa of convergence  $\sigma <0$)}:
%Note that $x$ is not necessarily continuous, but has at most finite jump discontinuities) 

\begin{equation}
X^T(e^{j\omega T}) = \frac{x(0^+)}{2}+ \sum_{k=1}^{\infty}\frac{x(kT^+)-x(kT^-)}{2} e^{-j\omega kT}+ \frac{1}
{T} \sum_{n=-\infty}^{\infty} X(j(\omega + n\frac{2\pi}{T})) \label{eq:sampsignal}
\end{equation}

\noindent where $\omega \in (-\infty,\infty)$. Obviously, if $x$ is a continuous function over $\mathbb{R}_{\geq 0}$ then the usual expression $X^T(j\omega) = \frac{1}
{T} \sum_{n=-\infty}^{\infty} X(j(\omega + n\frac{2\pi}{T}))$ is recovered. 

On the other hand, a zero-order hold $H_T$, with sampling time $T$, is a system that acts over a discrete-time signal $x^T$ and produces a continuous-time signal $x = H_T x^T$ given by $x(t) = x^T(n)$ for $nT \leq t <(n+1)T$ and $n \in \mathbb{N}_{\geq 0}$. In addition, and with some abuse of notation, the zero-order hold can be characterized by the function 

\begin{equation}
H_T(j\omega) = \frac{1-e^{-j\omega T}}{j\omega }
\label{eq:holdfunction}
\end{equation}

\noindent and the spectra of the signals $x$ and $x^T$ are related simply by $X(j\omega) =  H_T(j\omega)X^T(e^{j\omega T})$.

\begin{figure}[h] 
   \centering
    \includegraphics[width=7.5cm]{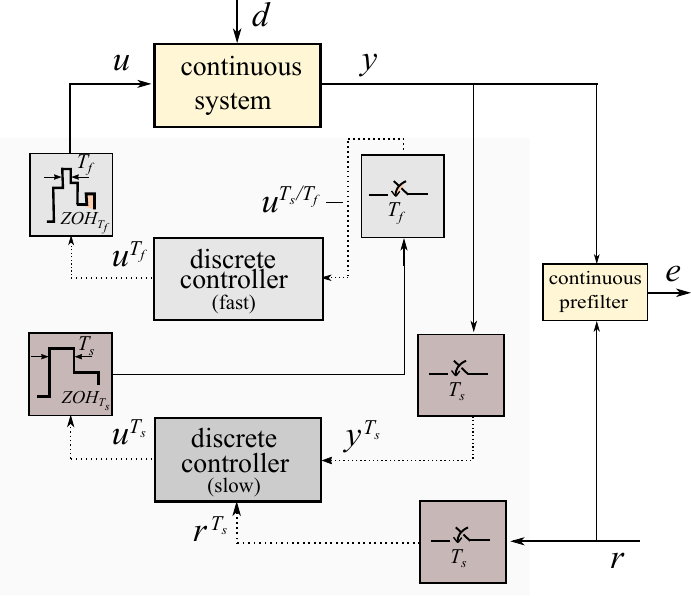}
    \caption{A dual-rate control system with discrete controllers working at a fast/slow sampling time. }
    \label{fig:controlsetup}
\end{figure}

%\subsection{Multirate control systems}
Consider the DR control system of Fig. \ref{fig:controlsetup} that will be the control setup to be investigated in this work, where all the signals are scalar.
A continuous-time system, with transfer function $P(s)$, is controlled by a multirate controller working with two sampling periods $T_s$ and $T_f$. It is assumed that $T_s \geq T_f$ and $T_s$  will be referred to as the {slow} sampling time, and $T_f$ as the {fast} sampling time. More specifically, the controller consists of two discrete-time controllers:  a {slow} controller with two-degrees of freedom, with transfer functions $F_L(z_s)$ and $G_L(z_s)$, acting over signals sampled every $T_s$ time units, and a {fast} controller $G_R(z_f)$, acting over signals sampled every $T_f$ time units (note that the z-transform uses different values $z_s$ or $z_f$ to emphasize dependence on the sampling period $T_s$ or $T_f$, respectively).

This work is focused on robust stability and tracking problems considering continuous-time responses. A previous QFT approach to sampled-data control \cite{YanivChait93} will be used as reference for approaching the QFT dual-rate control problem. More specifically the following closed-loop objectives are considered: robust stability with worst-case gain and phase margins, robust discrete-time tracking, and robust continuous-time tracking. A generic control design problem is: given a set of system transfer functions $\cal{P}$, and the prefilters $F(s)$ and $F_L(z_s)$, find discrete controllers $G_R(z_f)$ and $G_L(z_s)$ to meet the above objectives. Here, this  problem is approached starting with a previously designed fast controller $G_R(z_f)$, thus the focus is on how to design the slow controller $G_L(z_s)$ to satisfy the closed-loop specifications.
The z-transform of the slow controller output $u^{T_s}$ is given by:
\begin{equation}
U^{T_s}(z_s) = G_L(z_s)(F_L(z_s)R^{T_s}(z_s) - Y^{T_s}(z_s))
\label{eq:slowcont}
\end{equation}
where, in addition, the continuous-time error signal $e$ has the s-transform 
\begin{equation}
E(s) = F(s)R(s) - Y(s)
\label{eq:error}
\end{equation}
and thus $U^{T_s}(z_s) = G_L(z_S)(E^{T_s}(z_s))$, once that $F_L(z_s) = {\cal Z}\{F(s)H_{T_s}(s)\}$. 
The output of the slow controller $u^{T_s}$ is resampled with the fast sampling time obtaining the fast controller input $u^{T_s/T_f}$. This operation is modeled by using a combination of a zero-order hold $H_{T_s}$ and a sampler with sampling time $T_f$. Finally, the output of the fast controller $u^{T_f}$ is processed by a zero-order hold $H_{T_f}$ producing the system input $u$. 
Note that in contrast to other QFT approaches based on tracking error specifications \cite{Eitelberg00,Boje03,ElsoGilGarcia13}, here the tracking specification is based on (4) and the continuous prefilter $F$ and its discretization $F_L$ are firstly designed \cite{YanivChait93}, and then the emphasis will be on the design of the discrete feedback controller $G_L$ for the continuous-time responses closely follow the reponse of the prefilter $F$. 

It is also assumed that the exogenous signals and the system and controllers transfer functions satisfy the following standing assumption. 
%and the disturbance $d$

\vspace{0.125cm}
{\bf Assumption 1}:
\begin{itemize}
\item The reference signal $r$ and disturbance $d$ are functions in $\text{L}_{1e}(\mathbb{R}_{\geq 0})$ (signals in $\text{L}_{p}(\mathbb{R}_{\geq 0})$, $1\leq p \leq \infty$, such as steps, ramps, sinusoids, etc., are included; impulses are excluded). 
\item The system transfer funcion $P(s)$ is rational and strictly proper.
\item \textcolor{black}{The prefilter $F(s)$ is rational, strictly proper, and minimum-phase; and, in addition, the discrete prefilter $F_L(z_s)$ is the discretization of $F(s)$ as given by $F_L(z_s) = {\cal Z}\{F(s)H_{T_s}(s)\}$.}
\item The controllers $G_L(z_s)$ and $G_R(z_f)$ are rational and proper, and in addition $T_f = T_s/N$, that is $z_s = z_f^N$ ($N$ is a positive integer). By notational simplicity $z = z_f$ and $z_s = z^N$ may be used.
%\item \textcolor{black}{The controllers and prefilters are initially at rest}.
\end{itemize}

%%%%%%%%%%%%%%%%%%%%%%%%%%%%%%%
\section{Frequency domain analysis of dual-rate control systems}
Analysis of DR and in general multirate (MR) sampled systems in the frequency domain has been developed since early contributions\citep{araki1986multivariable, thompson1986gain} to the field of digital control, trying to overcome the basic difficulty that multirate sampled systems are time-varying.  In particular, several seminal works introduced switch decomposition\cite{sklansky1955analysis,kranc1957input} and frequency decomposition\cite{coffey1966stability} techniques, that has been the basis for future developments.  More recently, a relevant approach has been the lifting technique \cite{khargonekar1985robust,bamieh1991ltl}, that transforms the periodic system into a linear time invariant one considering every signal referred to the least common multiple
of all the periods of the MR system. The frequency domain analysis of multirate systems may be performed by using singular value decomposition(SVD) of the lifted MIMO system. Also, a number of works have extended the switch decomposition method of Kranc to very general cases obtaining which has been called a generalized Bode diagram (GBD) \cite{whitbeck1980multirate,salt2019dual}. By using a GBD, it is possible to analyze the several harmonic components of a DR sampled system as interleaved fragments of the frequency response of a particular single-rate system. 

In the following, the lifted system SVD technique and, with some more detail, the GBD technique are applied to a DR control system to analyze its intersample behavior and motivate the QFT analysis and design technique to be developed in this work.  

\subsection{A motivational example}
%{\bf Example}: The above result is illustrated with an example (adapted from $\cdots$), that will be work out in different directions in the rest of this work.
Consider \cite{SaltSala14,salt2005mbm} the system with transfer function 
\begin{equation}
P(s) = \frac{1.5}{(s+0.5)(s+1.5)}
\label{eq:Pex}
\end{equation} 
and a DR controller with sampling times $T_s = 0.4 \text{ s}$ and $T_f = \frac{0.4}{3} \text{ s}$ and thus $N = 3$, given by the slow and fast controllers (by simplicity a case without prefilter is analyzed, that is $F(s) = F_L(z^3) =1$) 
\begin{equation}
G_L(z^3) = \frac{z^9 - 1.296z^6 + 0.5636 z^3 - 0.1721}{z^9 - 2.131z^6 + 1.365 z^3 -0.2344}
\label{eq:GL}
\end{equation} 
and
\begin{equation}
G_R(z) = \frac{26.31z^4 - 85.24 z^3 + 102.1z^2 -53.32 z +10.21}
{z^4 - 1.469z^3 - 0.2344 z^2 + 1.225 z - 0.5089}
\label{eq:GR}
\end{equation} 
respectively. The goal of this DR controller is to emulate the design specifications obtained by the continuous-time PID controller $G_c(s) = 7.5(1 + 0.2 s + \frac{1}{3s})$, that will be used for comparison. It is desired that the DR controller achieves similar closed-loop performance but satisfying design implementation constraints such a a slow output sampling and fast input sampling of the system \eqref{eq:Pex}.   

\begin{figure}[t]
\begin{center}
{\includegraphics[width=\textwidth]{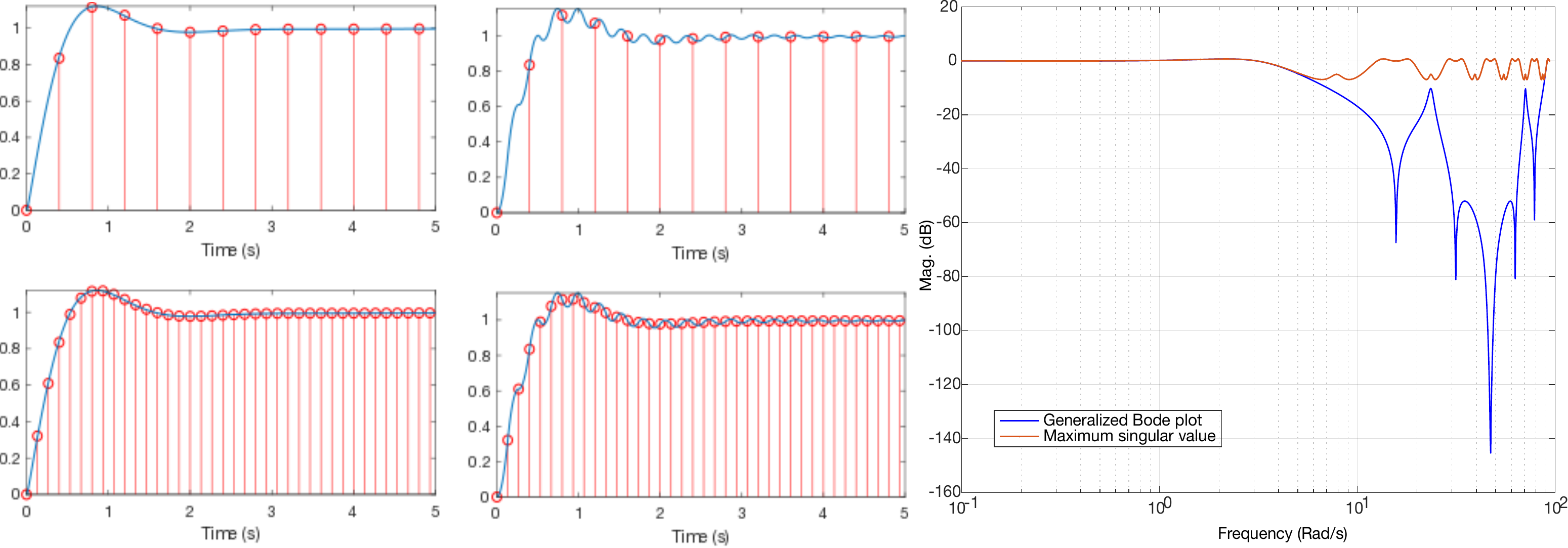}}
\caption{DR control system: (left) Closed-loop step response of the PID controller and the DR controller at slow sampling -up- and at fast sampling -down-; (center) Closed-loop step response of the DR controller and their slow sampling -up- and fast sampling -down-; (right) Generalized Bode plot and maximum singular value vs frequency.}
\label{fig:yrippleBodeSV}
\end{center}
\end{figure}

A simulation of this DR control system has been performed, and results are shown in Figure \ref{fig:yrippleBodeSV}. Although its performance in terms of unit step tracking seems to be correct in comparison with the PID controller, both at the slow and fast sampling periods % $T_s = 0.4 s$ and at the fast sampling period $T_f = \frac{0.4}{3} s$ 
(see Figure \ref{fig:yrippleBodeSV}-left), the step response of the DR controller (see Figure \ref{fig:yrippleBodeSV}-center) exhibits a ripple that degrades the intersample behavior and is clearly unacceptable in control practice.  This ripple is is obtained at a frequency $\omega_{ripple} = \frac{3 \pi}{0.4} \approx 23.6 \text{ Rad/s} $ wich is exactly the fast Nyquist frequency, that is  $\omega_{ripple} = \frac{\pi}{T_f}$.

Now, the question is if some of the previously developed methods for frequency analysis is able to detect this intersample behavior in a efficient way. Figure \ref{fig:yrippleBodeSV}-right shows both the (magnitude) GBD and SVD plots. Note that the oscillating intersampling behavior or ripple is due to the folding of high frequencies, and this alias at $\omega_{ripple}  \approx 23.6 \text{ Rad/s} $  is barely distinguishable in the SVD diagram, that makes very difficult if not imposible to estimate the frequency and amplitude of the ripple using SVD. However, in the GBD both the ripple amplitude and frequency are clearly depicted. This is explained in detail in the following, discussing some limitations of the technique that has been a main motivation for this work.

\begin{figure}[htbp]
\begin{center}
{\includegraphics[width=\textwidth]{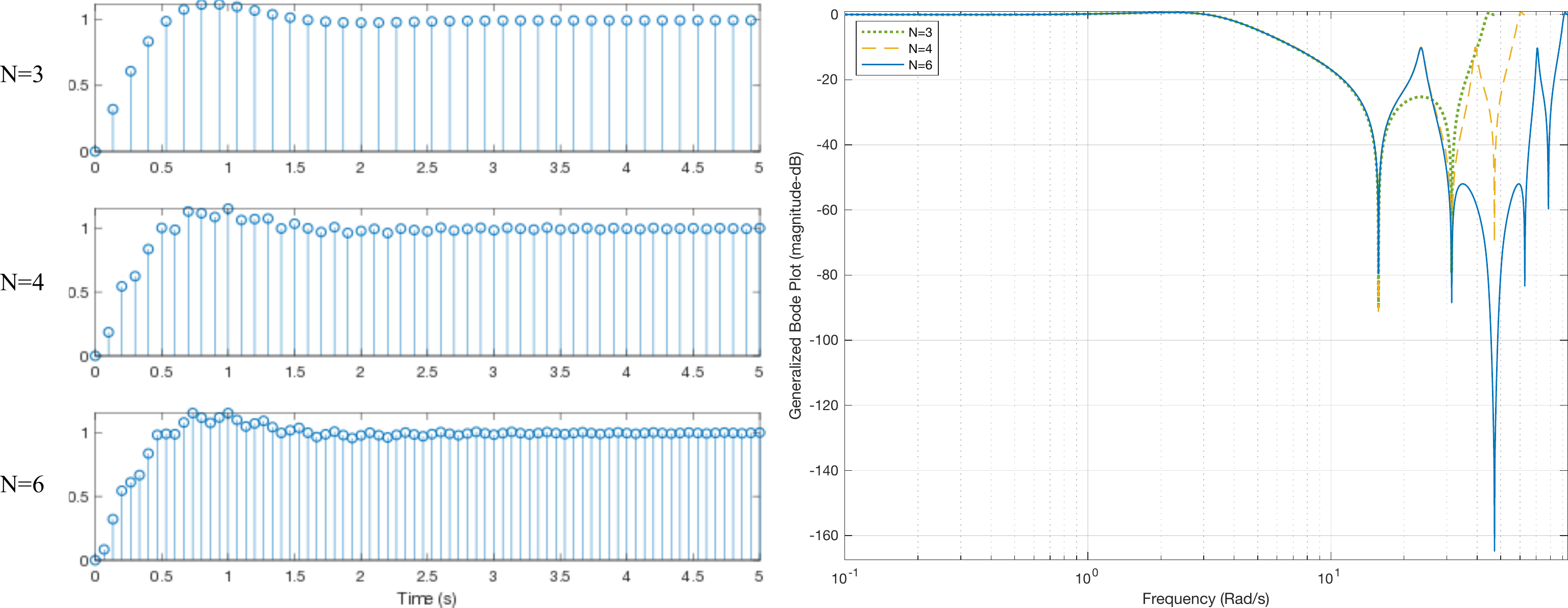}}
\caption{DR control system: (left) Closed-loop step response of the DR controller at several sampling periods $T_f = T_s/N$, for $N = 3$, $4$, and $6$; (right) Generalized Bode plot for $N = 3$, $4$, and $6$ (note that  the GBD plot of Figure \ref{fig:yrippleBodeSV}-right corresponds to $N=6$.}
\label{fig:GBP346}
\end{center}
\end{figure}

The GBD technique\cite{salt2014new} allows the computation of the frequency response from $r^{T_s}$ to $y^{Tf}$ (see Fig. \ref{fig:controlsetup}) by using only one Bode plot, and even for a more general case in which $N_f$ and $N_s$ are coprime integers (being $N_sT_s = N_fT_f$). It is understood that this "frequency response" does not give a single sinusoidal output for a sinusoidal input, in fact for a input $r^{T_s}(k) = e^{j\omega T_s k}$ the output is a sum of components $y_r(k) = \hat{y}_r e^{j\omega_r T_f k}$ with frequencies $\omega_r = \omega + r\frac{2\pi}{N_fT_f}$, $r = 0, 1, \cdots, N_f-1$. And the GBD plot is used to compute $\hat{y}_r$  at the $N_f$ frequency points $\omega_r$, for $r = 0, 1, \cdots, N_f-1$. Note that this technique only allows to analyze the frequencies appearing in the sampling of the signal of interest (in this case the output $y^{Tf}$). For a finer intersample behavior analysis, the usual practice\cite{keller1990new,er1991practical} is to sample the output at a faster sampling period and then to obtain the corresponding frequency components from the GBD. 

For this example, and also in this work (see Assumption 1), $N_s = 1$ and $N_f = N$ which are clearly coprime, and thus the GBD can be applied. 
Figure \ref{fig:GBP346} shows step responses and GBD plots of the DR control system example sampled at several sampling periods with $N = 3, 4$, and $6$. Note that for $N=3$, with Nyquist frequency $\omega_{Nyquist (N=3)} = \frac{3\pi}{0.4} \approx 23.56$ Rad/s, the GBP does have a relatively small value in magnitude, under $-20$ dB, for frequencies between approximately $1$ Rad/s and $23.56$ Rad/s,  and thus a ripple is not expected in the sampled signal as it is observed in the corresponding time response (Figure \ref{fig:GBP346}). 
However, for the case $N = 4$ with a frequency range $[0,\omega_{Nyquist (N=4)}] \approx [0, 31.42]$ Rad/s, the GBD clearly shows one (only) peak at the ripple frequency $\omega_{ripple} \approx 23.56$ Rad/s, that is also observed at the time response. Finally, for  $N = 6$ with a frequency range $[0,\omega_{Nyquist (N=6)}] \approx [0, 47.12]$ Rad/s, the GBD only shows a significant peak at the frequency $\omega_{ripple} $, clearly observed also in the time response.

Although the GBD technique allows the analysis of ripples occurrence, and in general frequency domain analysis of DR systems, a drawback is that is only allows the frequency analysis of the continuous-time signals of interest (like the control signal and the closed-loop output)   in a somehow indirect way through their samples. Obviously, this can be partly alleviated by using a large value of $N$, but this is always an approximated analysis. Another more important issue that hampered the application of GBD in control practice, is that it is not obvious how to use GBD to design DR controllers, specially for systems with large uncertainty.  

In the rest of this work, after developing a  basic extension of the GBD technique to directly obtain continuous-time signal spectra, this result will be used as basis to develop a QFT-based methodology of  robust DR controller design, that will be specially useful for systems with large uncertainty.

\subsection{A Nyquist-like theorem for nominal closed-loop stability}

Before analyzing the frequency response of the DR control system, it is necessary to substantiate a stability result. By simplicity, the case of no uncertainty in $P$ is considered here (robust stability is developed en Section 4.1). Also note that, by Assumption 1, $T_f = T$ and $T_s = NT$. 
% Since our approach is based on the frequency domain, a Nyquist-like theorem will be developed, adapting previous results to our control setup. In our setup, $1 + G_L(e^{j\omega T_s})H_L(e^{j\omega T_s})$ denotes the characteristic polynomial of the slow-sampled closed-loop system, and $G_L(e^{j\omega T_s})H_L(e^{j\omega T_s})$ its open-loop gain function. By simplicity, the case of no uncertainty in $P$ is considered here (robust stability is developed en Section 4.1). Also, note that $T_f = T$ and $T_s = NT$. 
%Consider a general feedback interconnection (with no exogenous signals) $(D_1,D_2)$ of two finite-dimensional, causal and discrete-time systems $D_1$ and $D_2$. It is said that $(D_1,D_2)$ is {internally stable} if for 
The DR control system of Fig. \ref{fig:controlsetup} is now modeled at different signal levels, from the continuous-time signals to the discrete-time signals given by the sampling with the fast and slow sampling periods. With some abuse of notation, the plant and the different controllers are now represented as time domain operators (see Fig. 4): $P$ is the continuous-time LTI plant, $P_R = S_TPH_T$ is its "fast" discretization (which is the zero-order hold equivalent of $P$ at the fast sampling), and $K = H_TK_RS_T$ is a continuous-time controller, %where $H_T$ is a zero-order hold with sampling time $T$, and $S_T$ is a sampler of a continuous-time signal with sampling period $T$.
where $K_R$ is a discrete-time controller with input and output $y^{T_f} $ and $u^{T_f}$, respectively, given by $K_R = G_R Q_N^* G_L S_N$. Here, $G_R$ and $G_L$ are the fast and slow controllers, respectively, and $S_N$ and $Q_N^*$ will be defined in the following. Finally, $P_L = S_NP_RG_RQ_N^*$. 

Here $S_N$ is a sampler of a discrete-time signal $x$ that gives the discrete-time signal $(S_N x)(k) = x(kN)$, for $k\geq0$, and in the setup of Fig. \ref{fig:controlsetup} $y^{T_f} = S_Ty$ and $y^{T_s} = S_N y^{T_f} = S_N S_T y$ are obtained. Moreover $Q_N^*$ represents the operation in Fig. \ref{fig:controlsetup} corresponding to the zero-holding and resampling of $u^{T_s}$ for obtaining $u^{T_s/T_f}$. $Q_N^*$ has been referred to as a $Q$-upsampler\cite{Dasgupta99} (with $Q =[1,1,\cdots,1]$) in contrast to the zero padding upsampler $S^*_N$ (corresponding to $Q =[1,0,\cdots,0]$) used in previous seminal works\cite{francis1988stability}, which is the adjoint of  $S_N$.

%A discrete-time linear system $G$ is time-invariant if $U^*GU = G$ 
% The backward shift $U$ and the forward shift $U^*$ 
As a result, the DR control system of Fig. \ref{fig:controlsetup} corresponds to $(P,K,F)$ in Fig. \ref{fig:Nyquist1}-a. If exogenous inputs are not considered, that is $r=d=0$, the autonomous DR control systems is denoted by $(P,K)$. Moreover, $(P_R,K_R)$ and $(P_L,K_L)$ correspond to discrete-time models of $(P,K)$ with fast and slow sampling, respectively (Fig. \ref{fig:Nyquist1}-b,c,d).

For the DR system $(P,K)$, the state ${\bf x}(t)=({\bf x}_P,{\bf x}_K)(t)$ is sufficient information for the computation of all future values of all signals\cite{francis1988stability} in the absence of exogenous inputs. By definition, $(P,K)$ is exponentially stable if there exist positive constants $\alpha$ and $\beta$ such that for every initial time $t_0$ and every initial state $ {\bf x}(t_0)$, $\|{\bf x}(t)\| \leq \|{\bf x}(t_0)\|\beta e^{\alpha(t-t_0)}$. A similar definition can be stated in discrete-time for the feedback systems $(P_R,K_R)$ and $(P_L,K_L)$ in Fig. \ref{fig:Nyquist1}.2-4. On the other hand, the DR system $(P,K,F)$ is $\text{L}_p$-stable if the operators from $d$, $r$ to $e$, $u$ are bounded from $\text{L}_p$ to $\text{L}_p$, for $1\leq p \leq\infty$.

Consider the following standing assumption:

\vspace{0.5cm}

{\bf Assumption 2}:
{\begin{enumerate}
\item (Non-pathological fast sampling of the continuous-time plant) None of the points $jk\frac{2\pi}{T_f}$, $k\neq0$ is a pole of $P$, and if $s_p$ is a pole of $P$ in CRHP then $s_p + jk\frac{2\pi}{T_f}$ is not a pole of $P$, for $k\neq0$. 
\item (Non-pathological slow sampling of the fast discrete-time plant) If $z_p$ is a pole of $P_R$ in ${\mathbb D}^c$ then $z_pe^{jk\frac{2\pi}{N}}$ is not a pole of $P_R$, for $k = 1, 2, \cdots,N-1$.
\item ( No unstable hidden modes with fast and slow sampling) There is not pole-zero cancellations of the products $P_RG_R$ and $P_LG_L$ in ${\mathbb D}^c$.
%\item (No unstable hidden modes with slow sampling) There is not pole-zero cancellations of the product $P_LG_L$ in ${\mathbb D}^c$.
\item (stability of the fast controller) The poles of $G_R$ are in ${\mathbb D}$. 
\end{enumerate}

\begin{figure}[htbp]
\begin{center}
{\includegraphics[width=0.75\textwidth]{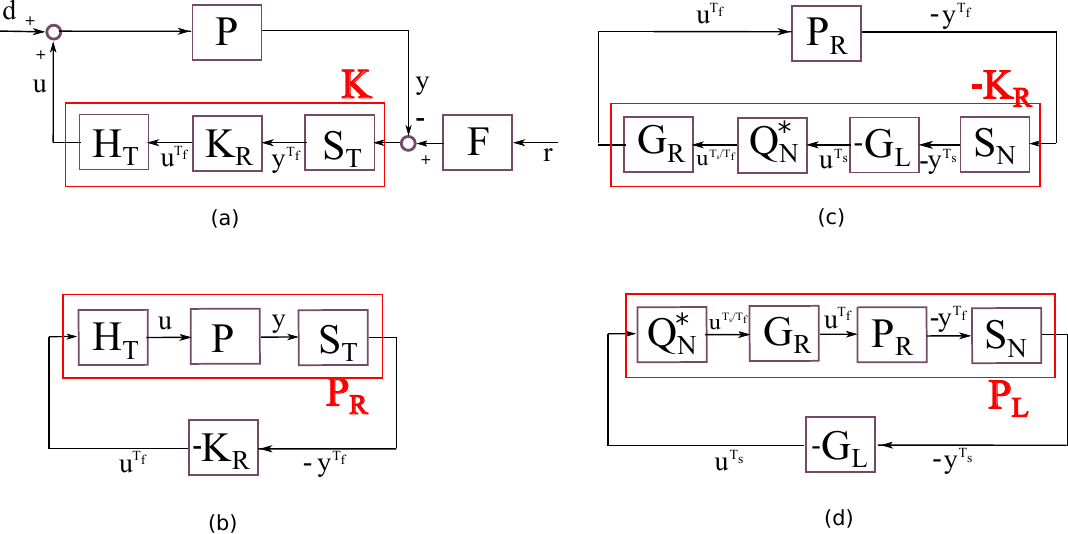}}
\caption{Representation of the DR control system as a feedback system interchanging different signals models: a) Dual-rate system $(P,K,F)$, with continuous-time signals $u$ and $y$ and exogenous signals $r$ and $d$; the system $(P,K)$ is obtained by doing $d = r = 0$;  b) Discrete-time system $(P_R,K_R)$, with fast-sampled signals $u^{T_f}$ and $y^{T_f}$, and internal signals $u$ and $y$; c) Discrete-time system $(P_R,K_R)$ with several internal signals: slow-sampled signals $u^{T_s}$ and $y^{T_s}$ and the upsampled signal $u^{T_s/T_f}$; d) Discrete-time system $(P_L,G_L)$ with slow-sampled signals  $u^{T_s}$ and $y^{T_s}$, and internal signals $u^{T_r}$, $y^{T_r}$, and $u^{T_s/T_f}$.}
\label{fig:Nyquist1}
\end{center}
\end{figure}

%{\bf Proposition 1}: Assume that the discretized plant  $H_L = [P \cdot G_R \cdot H_{T_s}]^{T_s}$ is free of unstable hidden modes, and that the product $H_L \cdot G_L$ has no pole-zero cancellations in ${\mathbb D}^c$.  Then, the dual-rate control system of Fig. \ref{fig:controlsetup} is  stable if and only the Nichols plot of $(H_L \cdot G_L)(e^{j\omega T_s})$ does not intersect the  point $(-180^\circ,0 \text{ dB})$ and the net sum of crossings of the ray ${\bf R}_0 := \{(-180^\circ,r) : r > 0 \text{ dB} \}$ is equal to the number of poles of  $H_L \cdot G_L$ (including multiplicities) in  ${\mathbb D}^c$. 

Since our approach is based on the frequency domain, a Nyquist-like theorem will be developed, adapting previous results\cite{francis1988stability,ChenFrancis91,cohen1994stability,ChenBallance98} to our control setup. %In our setup, $1 + G_L(e^{j\omega T_s})P_L(e^{j\omega T_s})$ denotes the characteristic polynomial of the slow-sampled closed-loop system, and $G_L(e^{j\omega T_s})P_L(e^{j\omega T_s})$ its open-loop gain function. 
In the following, the full Nichols plot of $P_L \cdot G_L$ refers to the plot of $|P_L(e^{j\omega T_s}) \cdot G_L(e^{j\omega T_s})|$ in dB against $\angle{P_L(e^{j\omega T_s}) \cdot G_L(e^{j\omega T_s}})$ in the domain [-360,0] degrees, for $\omega \in [0,2\pi/T_s]$. Also the half Nichols plot corresponds to the segment of the full Nichols plot for $\omega \in [0,\pi/T_s]$. The next result is based on the number of crossings\cite{cohen1994stability,ChenBallance98} of the full Nichols plot. Afterwards, the result is adapted to the half Nichols plot in a remark. In this work, for simplicity the half Nichols plot will be also referred to as the Nichols plot. 

\vspace{0.25cm}
{\bf Proposition 1}: Under Assumptions 1 and 2, if in addition the full Nichols plot of $P_L \cdot G_L$ does not intersect the  point $(-180^\circ,0 \text{ dB})$ and the net sum of crossings of the ray ${\bf R}_0 := \{(-180^\circ,r) : r > 0 \text{ dB} \}$ is equal to the number of poles of  $P_L \cdot G_L$ (including multiplicities) in  ${\mathbb D}^c$ (the crossing condition), then the DR control system $(P,K)$ is exponentially stable and $(P,K,F)$ is input-output $\text{L}_p$-stable, for $1\leq p \leq\infty$.

\vspace{0.5cm}
{\bf Proof}: Firstly, consider the discrete-time feedback system $(P_L,G_L)$ (Fig. \ref{fig:Nyquist1}.d). The system $P_L = S_NP_RG_RQ_N^*$ will be shown to be time-invariant. Before that, some properties of the upsampler $Q_N^*$,  the backward shift $U$, and forward shift $U^*$  need to be elaborated. It easily follows that 
\begin{equation}
Q_N^*U = U^NQ_N^*
\end{equation}
\label{eq:propN1a}
and
\begin{equation}
U^{*N}Q_N^* = Q_N^*U^*
\label{eq:propN2}
\end{equation}
Some other well-known properties\cite{francis1988stability} in relation with $S_N$, besides $S_NS_N^* = I$ and $U^*U = I$, are
\begin{equation}
U S_N = S_N U^N
\label{eq:propN3}
\end{equation}
and
\begin{equation}
U^{*}S_N = S_NU^{*N}
\label{eq:propN4}
\end{equation}
Also, a discrete-time linear system $G$ is time-invariant if $U^*GU = G$, and is $N$-periodic if $U^{*N}GU^N = G$.

Now, from \eqref{eq:propN3}-\eqref{eq:propN4} it directly follows that
\begin{equation}
U^*P_L U = U^*S_NP_RG_RQ_N^*U = S_NU^{*N}P_RG_RU^NQ_N^*
\label{eq:propN5}
\end{equation}
In addition, by using the fact that $G_R$ and $P_R$ are time-invariant, and the identities (8)-\eqref{eq:propN2}, it results
\begin{equation}
U^*P_L U = S_NU^{*N}P_RG_RU^NQ_N^* = S_NP_RU^{*N}U^NG_RQ_N^* = P_L
\label{eq:propN6}
\end{equation}
that is, $P_L$ is time-invariant. Since the full Nichols plot of $P_L \cdot G_L$ satisfies the crossing condition, it is a standard result\cite{cohen1994stability,ChenBallance98} that $(P_L,G_L)$ is exponentially stable. 

Next, consider the stability of the discrete-time system $(P_R,K_R)$ (Fig. \ref{fig:Nyquist1}.b,c). In contrast with the above reasoning, now $K_R$ is not time-invariant. However, it will be shown that $K_R$ is $N$-periodic. This directly follows by using (8)-\eqref{eq:propN2} and the fact that $G_R$ and $G_L$ are time-invariant, that is
\begin{equation}
U^{*N}K_R U^N =  U^{*N} G_R Q_N^* G_L S_N U^N = G_RU^{*N}Q_N^* G_L S_N U^N  =  G_RQ_N^*U^*G_LU S_N = G_RQ_N^*G_LS_N = K_R
\label{eq:propN6}
\end{equation}
And thus, all the conditions of Theorem 1\cite{francis1988stability} are satisfied (note that in this Theorem the zero padding upsampler $S_N^*$ is used instead of the upsampler $Q_N^*$, and thus it is not directly applicable), and as a result the system $(P_R,K_R)$ is exponentially stable. Finally, the exponential stability of the DR system $(P,K)$ and the $\text{L}_p$-stability of the DR system $(P,K,F)$ follows by direct application of Theorem 4\cite{francis1988stability} and Theorem 7\cite{ChenFrancis91}, respectively.
$\Box$

\vspace{0.25cm}

{\bf Remark 2}. Note that the ray crossings have a positive sign when the full Nichols plot crosses from left to right, and a negative sign in the opposite direction\cite{ChenBallance98}. On the other hand, it is customary in QFT to work with the half Nichols plot (that will be referred to as Nichols plot in the rest of this work) of $P_L \cdot G_L$ as design element. Note that one crossing of the Nichols plot corresponds to two crossings of the full Nichols plot. In addition, some care with the crossings count is needed in the cases in which the Nichols plot starts or ends at the ray ${\bf R}_0$; in these cases, they should be counted as half crossings. Also, if there are poles of $P_L \cdot G_L$ in $z_s = 1$, there is a segment of the full Nichols plot from $\omega=0^-$ to $\omega=0^+$ (coming from the indentation of the Nyquist path at $z_s = 1$) that may produce crossings of ${\bf R}_0$: % The segment from $\omega = 0^-$ to $\omega = 0^+$ (at infinite gain) has to be taken into consideration for the crossings count: 
it counts as $-1$ crossing of the full Nichols plot and $-1/2$ crossing of the Nichols plot.

%These crossing are counted as half crossings of the Nichols plot.

%\vspace{0.25cm}

Moreover, as it is well-known\cite{yaniv1993direct}, stability margins are conveniently depicted in the Nichols plane (see Fig. \ref{fig:NicholsEx31}): if $P_L(e^{j\omega T_s})G_L(e^{j\omega T_s}) = le^\lambda$ then the gain margin is defined as GM =  $1/l$ at $\lambda = -180^\circ$, and the phase margin is PM =  $180^\circ + \lambda$ where $\lambda$ is the phase corresponding to $l = 1$.

\vspace{0.25cm}
{\bf Example 3}: Consider the Example of Section 3.1. To apply the stability Nyquist result of Prop. 1, firstly Assumption 2 must be checked (Assumption 1 easily follows):  
\begin{enumerate}
\item $P(s)$, given by \eqref{eq:Pex}, has real poles. Thus, fast sampling is non-pathological. 
\item Here $P_R(z) = \frac{1.197e-05 z + 1.194e-05}{z^2 - 1.992 z + 0.992}$, having zeros and poles in ${\mathbb D}$. No unstable cancellation is possible.
\item Poles of $P_R(z)$ are real, and thus slow sampling is non-pathological. 
%\item  $[P \cdot G_R \cdot H_{T_s}]^{T_s}$ is free of unstable hidden modes, $\cdots$ (this is related with non-pathological sampling)
\item Poles of $G_R(z)$, given by \eqref{eq:GR}, are in ${\mathbb D}$.
%\item There are no pole-zero cancellations of the product $H_L \cdot G_L$ in ${\mathbb D}^c$.  All the poles of $G_L(z)$ are in ${\mathbb D}$, however it has a zero in ${\mathbb D}^c$, in particular in $z=1$; thus, it only has to be checked that $H_L(z)$ does not have a pole in $z=1$. This is in fact the case because $H_L$ is the zero-order hold equivalent of $P(s)G_R(e^{sT_f})$, having all their poles in LHP, and thus these poles are mapped to ${\mathbb D}$.
\end{enumerate}

Once it is shown that the Nyquist test can be applied, it has to be checked that there are no crossing of the full Nichols plot of $P_L \cdot G_L$ with the ray $\bf{R}_0$, since $(P_L \cdot G_L)(z^N)$ has no poles in ${\mathbb D}^c$. Fig. \ref{fig:NicholsEx31} shows the full Nichols plot and the Nichols plot, and the fact that there are no crossings. As a result, exponential and $\text{L}_p$-stability of the DR control system directly follows.% is exponentially stable and $L_p$ stable. 

%%%%%%%%%%%%%%%%%%%%%%%%%%%%%%%%

\begin{figure}[htbp]
\begin{center}
{\includegraphics[width=\textwidth]{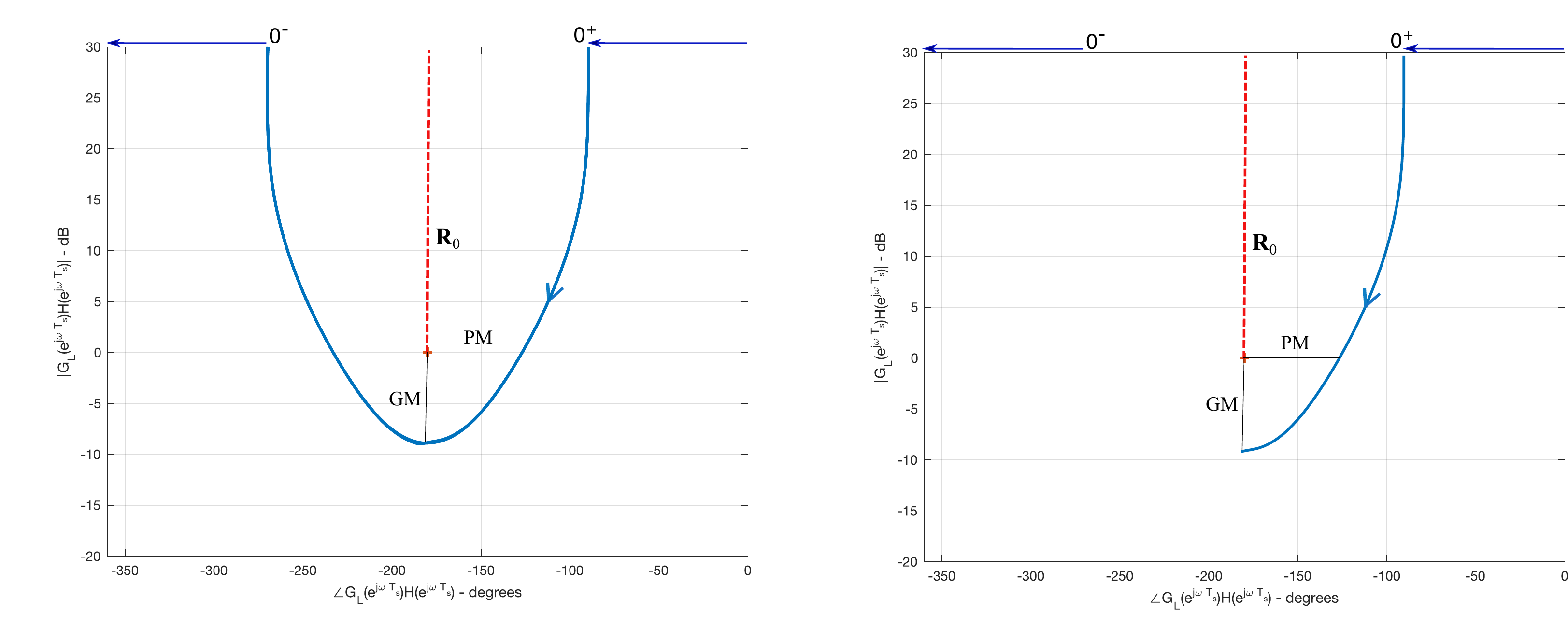}}
\caption{ DR control system Example of Section 3.1: Full Nichols plot (left) and Nichols plot (right) of $G_L \cdot P_L$, showing no crossings of the ray $\bf{R}_0$, and phase and gain margins.}
\label{fig:NicholsEx31}
\end{center}
\end{figure}

\vspace{0.25cm}

\subsection{Continuous-time signals spectra in dual-rate systems}
A basic goal of this work is to analyze, if possible, the frequency responses from the control system input $r$ to signals of interest like the control input $u$ and the output $y$.  Besides stability, this work is specifically devoted to tracking problems, and thus it is consider $d = 0$ in the rest of this work (the case of non zero disturbances can be approached by using a similar treatment), that is $Y(s) = P(s)U(s)$. Thus, the key question is if it is possible to establish a frequency response that relates the reference input $r$ to the output $y$.  It will be shown that indeed it is possible, although with some limitation. %as far as the slow and fast sampling times are related by $T_s = NT_f$ for some integer $N >0$.

Consider the discrete sensitivity frequency response $S_L(e^{j\omega T_s})$ defined as 
\begin{equation}
S_L(e^{j\omega T_s}) = \frac{1}{1+ G_L(e^{j\omega T_s})P_L(e^{j\omega T_s})} 
\label{eq:SL}
\end{equation}
where $P_L(e^{j\omega T_s})$ is the frequency response function corresponding to  $P_L = S_NP_RG_RQ_N^*$, 
%\begin{equation}
%P_L(e^{j\omega T_s}) = [P \cdot G_R \cdot H_{T_s}]^{T_s} (e^{j\omega T_s})
%\label{eq:HL}
%\end{equation}
and the complementary sensitivity frequency response $T({j\omega})$ defined as
\begin{equation}
T({j\omega}) = P(j\omega)G_R(e^{j\omega T_f}) H_{T_s}(j\omega) G_L(e^{j\omega T_s})S_L(e^{j\omega T_s}) 
\label{eq:T}
\end{equation}

\vspace{0.5cm}
The following result establishes the frequency responses from $r^{T_s}$ to $y^{T_s}$ and from $r^{T_s}$ to $y$. The existence of the first frequency response, establishing the frequency response at the slow sampling time, is more or less obvious once the fast sampling time is a multiple of the slow sampling time. However, the existence of a (exact) frequency relationship between the sampled signal $r^{T_s}$ and the continuous-time signal $y$ is less evident.
\vspace{0.5cm}

{\bf Proposition 4}: Consider the DR control system of Fig.1, and assume that the stability conditions of Prop. 1 are satisfied. %it is stable and that Asumption 1 holds. 
Then, for the case $Y(s) = P(s)U(s)$ ($d = 0$), the spectra of the system output $y$ and its slow sampling $y^{T_s}$ are given by
\begin{equation}
%Y(j\omega)=P(j\omega)G_R(e^{j\omega T_f})H_{T_s}(j\omega)G_L(e^{j\omega T_s})S_L(e^{j\omega T_s})F_L(e^{j\omega T_s}) R^{T_s}(e^{j\omega T_s}) 
Y(j\omega)=T(j \omega)F_L(e^{j\omega T_s}) R^{T_s}(e^{j\omega T_s}) 
\label{eq:spectrumY}
\end{equation}
and
\begin{equation}
Y^{T_s}(e^{j\omega T_s}) = \left ( 1- S_L(e^{j\omega T_s}) \right ) F_L(e^{j\omega T_s}) R^{T_s}(e^{j\omega T_s})
\label{eq:spectrumYTs}
\end{equation}
respectively.

\vspace{0.5cm}

{\bf Proof}: Since according to Assumption 1 $P$ is strictly proper, then the output $y$ is a continuous function and thus, using \eqref{eq:sampsignal}, 
\begin{equation}
Y^{T_s}(e^{j\omega T_s}) =  \frac{1}{T_s}\sum_{n=-\infty}^{\infty} Y(j(\omega + n\frac{2\pi}{T_s})) \label{eq:YTs}
\end{equation}
where the spectrum of $y$ is directly given by
\begin{equation}
Y(j\omega) = P(j\omega) H_{T_f}(j\omega)U^{T_f}(e^{j\omega T_f}) \label{eq:Y}
\end{equation}
and, in addition, 
\begin{equation}
U^{T_f}(e^{j\omega T_f})= G_R(e^{j\omega T_f}) U^{T_s/T_f}(j\omega) \label{eq:UTf}
\end{equation}
Moreover,  using \eqref{eq:sampsignal}, the fact that the signal $H_{T_s}(u^{T_s})$ is a function of bounded variation (but not necessarily continuous), and that the slow discrete controller is initially at rest, it follows that 
\begin{equation}
U^{T_s/T_f}(j\omega)  =  \sum_{k=1}^{\infty} \frac{u_H(kT_f^+)-u_H(kT_f^-)}{2} e^{-j\omega k{T_f}}+ \frac{1}{T_f}\sum_{n=-\infty}^{\infty} H_{T_s}(j(\omega + n\frac{2\pi}{T_f})) U^{T_s}(e^{j(\omega + n\frac{2\pi}{T_f})T_s}) \label{eq:UTsTf}
\end{equation}

%\begin{equation}
%U^{T_s/T_f}(j\omega)  =  \sum_{k=1}^{\infty}\frac{u^{T_}(k+1)-u^{T_s}(k)}{2} e^{-j\omega k{T_s}}+ \frac{1}{T_f}\sum_{n=-\infty}^{\infty} H_{T_s}(j(\omega - n\frac{2\pi}{T_f})) U^{T_s}(j(\omega - n\frac{2\pi}{T_f})) \label{eq:UTsTf}
%\end{equation}
Now, since $T_s = NT_f$, the right-hand first and second terms of \eqref{eq:UTsTf} can be simplified considering that
\begin{equation}
 \sum_{k=1}^{\infty} \frac{u_H(kT_f^+)-u_H(kT_f^-)}{2} e^{-j\omega k{T_f}} =
 \sum_{k=1}^{\infty}\frac{u^{T_s}(k)-u^{T_s}(k-1)}{2} e^{-j\omega k{T_s}} = \frac{1-e^{-j\omega T_s}}{2} U^{T_s}(e^{j\omega T_s})
  \label{eq:UTs3}
\end{equation}
and
\begin{equation}
U^{T_s}(e^{j(\omega + n\frac{2\pi}{T_f})T_s}) =  U^{T_s}(e^{j(\omega T_s + n 2\pi N)}) = U^{T_s}(e^{j\omega T_s}) 
\label{eq:UTs1}
\end{equation}
and also that  
\begin{equation}
\frac{1-e^{-j\omega T_s}}{2} + \frac{1}{T_f}
\sum_{n=-\infty}^{\infty} H_{T_s}(j(\omega + n\frac{2\pi}{T_f})) = \frac{1-e^{-j\omega T_s}}{1-e^{-j\omega T_f}}
\label{eq:UTs2}
\end{equation}

Using \eqref{eq:UTs3}, \eqref{eq:UTs1}, and \eqref{eq:UTs2},  the spectrum of the signal $u^{T_s/T_f}$, given by \eqref{eq:UTsTf}, is finally 
\begin{equation}
U^{T_s/T_f}(e^{j\omega T_f})  =  \left ( \frac{1-e^{-j\omega N T_f}}{1-e^{-j\omega T_f}} \right ) U^{T_s}(e^{j\omega NT_f}) = H_{T_s/T_f}(e^{j\omega T_f}) U^{T_s}(e^{j\omega NT_f}) 
 \label{eq:UTsTf2}
\end{equation}
where $H_{T_s/T_f}(e^{j\omega T_f}) $ corresponds to the frequency response of the upsampler $Q_n^*$.
%where  $U^{T_s}(e^{j\omega NT_f})$ is simply given by
%\begin{equation}
%U^{T_s}(e^{j\omega NT_f})  =  G_L(e^{j\omega NT_f}) \left ( F_L(e^{j\omega NT_f})R^{T_s}(e^{j\omega NT_f}) - Y^{T_s}(e^{j\omega NT_f}) \right )
% \label{eq:UTs4}
%\end{equation}
From \eqref{eq:YTs}, \eqref{eq:Y}, \eqref{eq:UTf}, and \eqref{eq:UTsTf2} it is obtained %the a first relationship between $Y^{T_s}$ and $R^{T_s}$ after using \eqref{eq:UTs4}, that is
\begin{equation}
Y^{T_s}(e^{j\omega T_s}) =\frac{1}{T_s}  \sum_{n=-\infty}^{\infty} P(j(\omega + n\frac{2\pi}{T_s})) H_{T_f}(j(\omega + n\frac{2\pi}{T_s})) G_R(e^{j(\omega + n\frac{2\pi}{T_s})T_f}) H_{T_s/T_f}(e^{j(\omega + n\frac{2\pi}{T_s})T_f}) U^{T_s}(e^{j(\omega + n\frac{2\pi}{T_s})NT_f})
\label{eq:YTsRTs}
%\left ( \frac{e^{j(\omega -n\frac{2 \pi}{T_s})T_s}-1}{2} + \frac{1}{N} \frac{1-e^{-j(\omega -n\frac{2 \pi}{T_s}) T_s}}{1-e^{-j(\omega -n\frac{2 \pi}{T_s}) T_f}}    \right )
\end{equation}
%\begin{equation}
%\nonumber
%\cdot G_L(e^{j(\omega - n\frac{2\pi}{T_s}) T_s}) \left ( F_L(e^{j(\omega - n\frac{2\pi}{T_s}) T_s})R^{T_s}(j(\omega - n\frac{2\pi}{T_s})) - Y^{T_s}(j(\omega - n\frac{2\pi}{T_s})) \right )
%\label{eq:YTsRTscont}
%\end{equation}
This expression allows further simplification since
\begin{equation}
H_{T_f}(j(\omega + n\frac{2\pi}{T_s})) H_{T_s/T_f}(e^{j(\omega + n\frac{2\pi}{T_s})T_f}) = \frac{1-e^{-j(\omega + n\frac{2\pi}{T_s}))T_f}}{j(\omega + n\frac{2\pi}{T_s})} \cdot  \frac{1-e^{-j(\omega + n\frac{2\pi}{T_s}) N T_f}}{1-e^{-j(\omega + n\frac{2\pi}{T_s}) T_f}} = 
\frac{1-e^{-j\omega T_s}}{j(\omega + n\frac{2\pi}{T_s})} = 
H_{T_s}(j(\omega + n\frac{2\pi}{T_s})
 \label{eq:YTsRTs2}
\end{equation}
and thus
\begin{equation}
Y^{T_s}(e^{j\omega T_s}) =\left (\frac{1}{T_s}  \sum_{n=-\infty}^{\infty} P(j(\omega + n\frac{2\pi}{T_s}))G_R(e^{j(\omega + n\frac{2\pi}{T_s})T_f})
H_{T_s}(j(\omega + n\frac{2\pi}{T_s})   \right ) U^{T_s}(e^{j\omega T_s})
\label{eq:YTsRTs3}
\end{equation}
Now, the expression between parenthesis is exactly $P_L(e^{j\omega T_s})$, that is
%5will be represented as $[P \cdot G_R \cdot H_{T_s}]^{T_s} (j\omega)$, and the result is
\begin{equation}
%Y^{T_s}(e^{j\omega T_s}) = [P \cdot G_R \cdot H_{T_s}]^{T_s} (j\omega) \cdot  U^{T_s}(e^{j\omega T_s})
Y^{T_s}(e^{j\omega T_s}) = P_L (e^{j\omega T_s}) \cdot  U^{T_s}(e^{j\omega T_s})
\label{eq:YTsRTs4}
\end{equation}
Finally, taking into account that $U^{T_s}(e^{j\omega T_s}) = G_L(e^{j\omega T_s}) \left ( F_L(e^{j\omega T_s})R^{T_s}(e^{j\omega T_s}) - Y^{T_s}(e^{j\omega T_s})\right )$, substituting in \eqref{eq:YTsRTs4}, and reordering to obtain $Y^{T_s}(e^{j\omega T_s})$, the desired result \eqref{eq:spectrumYTs} is directly obtained from: 
\begin{equation}
%Y^{T_s}(e^{j\omega T_s}) =\frac{G_L(e^{j\omega T_s}) [P \cdot G_R \cdot H_{T_s}]^{T_s} (j\omega) }{1+ G_L(e^{j\omega T_s})  [P \cdot G_R \cdot H_{T_s}]^{T_s} (j\omega) }  F_L(e^{j\omega T_s}) \cdot  R^{T_s}(e^{j\omega T_s})
Y^{T_s}(e^{j\omega T_s}) =\frac{G_L(e^{j\omega T_s}) P_L (e^{j\omega T_s})}{1+ G_L(e^{j\omega T_s})  P_L (e^{j\omega T_s}) }  F_L(e^{j\omega T_s}) \cdot  R^{T_s}(e^{j\omega T_s})
\label{eq:YTsRTs5}
\end{equation}
In addition, from \eqref{eq:error} and \eqref{eq:YTsRTs5} the error spectrum $E^{Ts}(e^{j\omega T_s})$ is directly given by
\begin{equation}
E^{T_s}(e^{j\omega T_s}) = F_L(e^{j\omega T_s})R^{T_s}(e^{j\omega T_s}) - Y^{T_s}(e^{j\omega T_s}) = S_L(e^{j\omega T_s}) F_L(e^{j\omega T_s}) R^{T_s}(e^{j\omega T_s})
\label{eq:ETsRTs}
\end{equation}
 
Moreover, since $Y(j\omega) = P(j\omega)U(j\omega)$ then the spectrum of the continuous-time signal $y$  is given by 
\begin{equation}
Y({j\omega}) =  P(j\omega) H_{T_f}(j\omega) G_R(e^{j \omega T_f}) H_{T_s/T_f} (e^{j \omega T_f}) G_L(e^{j \omega T_s})E^{T_s}(e^{j \omega T_s})
\label{eq:YRTs}
\end{equation}
where, using \eqref{eq:UTsTf2}, \eqref{eq:YTsRTs2} (for $n = 0$), and \eqref{eq:ETsRTs}, the desired result \eqref{eq:spectrumY} is directly obtained.

$\Box$

\vspace{0.5cm}

{\bf Remark 5} : To compute the closed-loop response of the DR control system of Fig. \ref{fig:controlsetup} to a harmonic reference input with frequency $\omega_0$, that is $r(t) = e^{j\omega_0 t}$, $t \in (-\infty,\infty)$  and 
\begin{equation}
R^{T_s}(e^{j\omega T_s}) = \frac{2\pi}{T_s} \sum_{k =-\infty}^{k =\infty}  \delta(\omega - \omega_0 - k\frac{2 \pi}{T_s})  \label{eq:RTscos}
\end{equation}
Eq. \eqref{eq:spectrumY} can be directly used. The result is a multiharmonic response as expected, given by 
\begin{equation}
Y(j\omega) = \frac{2\pi}{T_s} F_L(e^{j\omega_0 T_s}) \sum_{k =-\infty}^{k =\infty}  T(j\omega_0 + k\frac{2 \pi}{T_s})  \delta(\omega - \omega_0 - k\frac{2 \pi}{T_s}) \label{eq:RTsexp}
\end{equation}
from which it is directly obtained the time response
\begin{equation}
y(t) =  F_L(e^{j\omega_0 T_s}) \sum_{k =-\infty}^{k =\infty}  T(j\omega_0 + k\frac{2 \pi}{T_s})  e^{j(\omega_0 + k\frac{2 \pi}{T_s})t }\label{eq:yRTsexp}
\end{equation}
Moreover, if the reference is $r(t) = \cos(\omega_0 t)$, considering the symmetry property $T(-j\omega) = T^\ast(j\omega)$, it easily follows that the time response is
\begin{equation}
y(t) =  F_L(e^{j\omega_0 T_s}) \left ( |T(\omega_0)| \cos (\omega_0 t + \angle{T(j \omega_0}) 
+ \sum_{k =1}^{k =\infty}  |T_k(j \omega_0)|  \cos (\omega_k t + \angle{T_k(j \omega_0 }) + |T_k(-j \omega_0)|  \cos (-\omega_k t + \angle{T_k(-j \omega_0 }) 
%\cos ((\omega_0 + k\frac{2 \pi}{T_s})t + \angle{T(j (\omega_0 + k\frac{2 \pi}{T_s}))}) 
%y(t) =  F_L(e^{j\omega_0 T_s}) \left ( \sum_{k =0}^{k =\infty}  |T(j (\omega_0 + k\frac{2 \pi}{T_s}))|  \cdot
%\cos ((\omega_0 + k\frac{2 \pi}{T_s})t + \angle{T(j (\omega_0 + k\frac{2 \pi}{T_s}))}) +
%\sum_{k =1}^{k =\infty}  |T(j (- \omega_0 + k\frac{2 \pi}{T_s}))|  \cdot
%\cos ((-\omega_0 + k\frac{2 \pi}{T_s})t + \angle{T(j (-\omega_0 + k\frac{2 \pi}{T_s}))})
\right )
 \label{eq:yRTscos}
\end{equation}
consisting of the fundamental frequency $\omega_0$ and a infinite number of harmonics at frequencies $\pm \omega_k = \pm \omega_0 + k\frac{2 \pi}{T_s}$, $k = 1, 2, 3, \cdots$. Note that the exact time response can be computed by reading the Bode plot of the complementary sensitivity function $T(j\omega)$
at the frequencies given by the fundamental frequency and the harmonics frequencies.

%\begin{equation}
%Y(j\omega) =  F_L(e^{j\omega_0 T_s}) \sum_{k =-\infty}^{k =\infty}  \left (\ T(j\omega_0 + k\frac{2 \pi}{T_s})  \delta(\omega - \omega_0 - k\frac{2 %\pi}{T_s}) + T(-j\omega_0 + k\frac{2 \pi}{T_s})   \delta(\omega + \omega_0 - k\frac{2 \pi}{T_s})  \right ) \label{eq:RTscos}
%\end{equation}

\begin{figure}[htbp]
\begin{center}
{\includegraphics[width=0.65\textwidth]{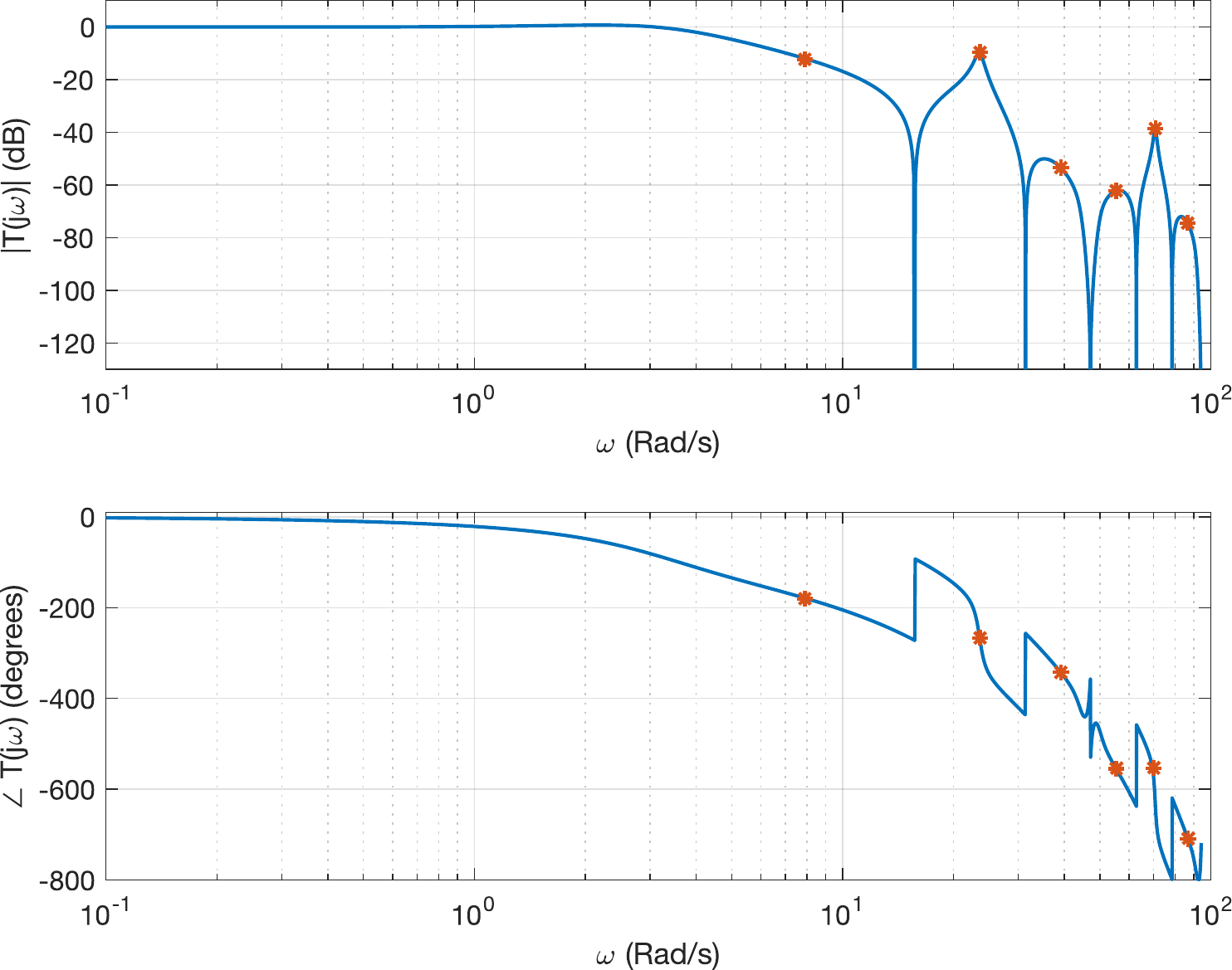}}
\caption{Bode plot of the complementary sensitivity function $T(j\omega)$, as given by \eqref{eq:T}, of Section 2.1 Example. This Bode plot contains all the needed information to compute the continuous-time response to a reference input; and, in particular the multifrequency response to a sinusoidal input. The asterisks shows the values needed to compute the response  to a sinusoidal input of frequency $\frac{\pi}{T_s} \approx 7.85$ Rad/s (in theory, an infinite number of frequencies are needed, but in practice it is enough with the first two frequencies to obtain a good estimate- see Example 6).
} 
\label{fig:bodeT}
\end{center}
\end{figure}

\vspace{0.5cm}
{\bf Example 6} : Consider the Example of Section 3.1. Figure \ref{fig:bodeT} shows the Bode plot of $T(j\omega)$, that has been computed using \eqref{eq:T}. 
Here, the time closed-loop response to a reference $r(t) = \cos (\frac{\pi}{0.4} t)$ can be obtained for a given prefilter, using \eqref{eq:yRTscos}, by computing the magnitude and angle of $T(j\omega)$ for frequencies $\frac{\pi}{0.4}$, $\frac{\pi}{0.4}  + k\frac{2 \pi}{0.4} = \frac{(2k+1) \pi}{0.4}$, and $-\frac{\pi}{0.4}  + k\frac{2 \pi}{0.4} = \frac{(2k-1) \pi}{0.4}$, for $k = 1, 2, \cdots$. Thus, the frequencies appearing at the output $y$ are:
\begin{equation}
\frac{\pi}{0.4}, \frac{3\pi}{0.4}, \frac{5\pi}{0.4}, \cdots, \frac{\pi}{0.4}, \frac{3\pi}{0.4}, \frac{5\pi}{0.4}, \cdots
 \label{eq:yRTscosEx}
\end{equation}

 Note that only for the first two frequencies $\frac{\pi}{0.4}$, $\frac{3\pi}{0.4}$, the magnitude Bode plot has significant values (for the rest of frequencies the magnitude is under $-40 \text{dB}$). These frequencies are the input frequency $\frac{\pi}{T_s}$ and the frequency $\frac{3\pi}{T_s}$ in which the ripple is produced. For the case $F_L(e^{j\omega_0 T_s}) = 1$, applying \eqref{eq:yRTscos} the result is well approximated by
\begin{equation}
y(t) \approx   2 |T(j \frac{\pi}{0.4})| \cos (\frac{\pi}{0.4} t + \angle{T(j \frac{\pi}{0.4})} 
+ 2  |T(j \frac{3 \pi}{0.4})|  \cos (\frac{3 \pi}{0.4} t + \angle{T(j \frac{(3) \pi}{0.4})} )
 \label{eq:yRTscosEx2}
\end{equation}

In Figure \ref{fig:bodeT}, the values $T(j\frac{\pi}{0.4}) = 0.2545e^{-j178.5^\circ}$, $T(j\frac{3\pi}{0.4})= 0.3166e^{-j265.3^\circ}$, $T(j\frac{5\pi}{0.4})= 0.0021e^{-j344.3^\circ}$, $\cdots$, are explicitly marked. A plot of the closed-loop output $y$ as given by \eqref{eq:yRTscosEx2} is given in Fig. \ref{fig:ysim}, where in addition it is shown the time response simulation of the DR control system (note that %a sinusoidal signal $r(t) = \cos(\frac{\pi}{T_s} t)$, $t\geq0$, and thus 
$r^{T_s}(n) = e^{j\pi} n$, for $n \geq 0$). The response given by  \eqref{eq:yRTscosEx2} is a very good approximation of the steady-state simulated response (a exact value would be obtained by considering the infinite number of harmonics).
\begin{figure}[t]
\begin{center}
{\includegraphics[width=0.85\textwidth]{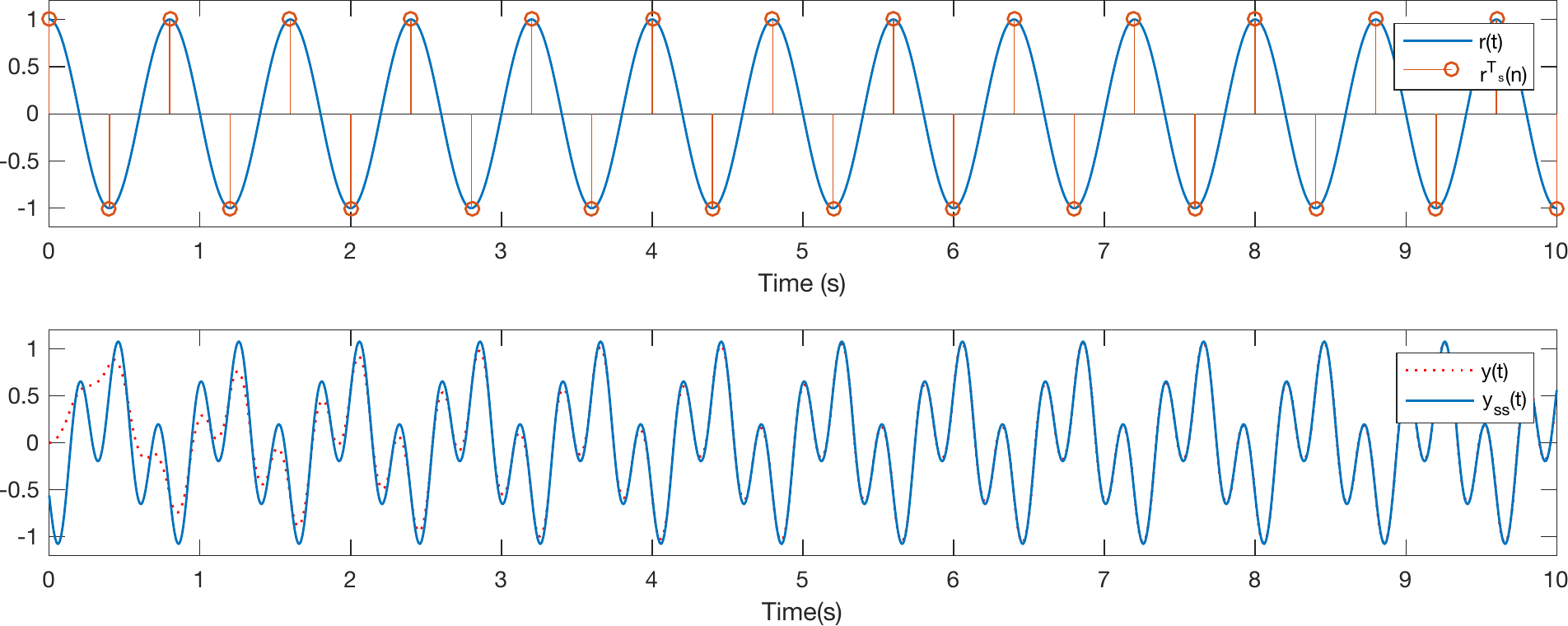}}
\caption{(up) A sinusoidal reference input with frequency $\frac{\pi}{T_s} \approx 7.85$ Rad/s, and its sampling with sampling period $T_s = 0.4$ s; (down) time simulation of the DR control system (dotted), and time response computed using the Bode plot of $T(j\omega)$ (see \eqref{eq:yRTscosEx2}). 
}
\label{fig:ysim}
\end{center}
\end{figure}

%Using directly \eqref{eq:spectrumY}, it is possible to analyze the unit step response spectrum.
For a unit step reference the spectra  are $R(j\omega)= \frac{1}{j\omega} + \pi\delta(\omega)$ and
$R^{T_s}(e^{j \omega T_s}) = \frac{1}{1 - e^{-j\omega T_s}} + \frac{\pi}{T_s}\sum_{k=-\infty}^{\infty} \delta(\omega - k\frac{2\pi}{T_s})$. Now, consider a first order prefilter with $F(s) = \frac{1}{0.1s+1}$  and $F_L({z_s}) = \frac{1-e^{-10T_s}}{z_s-e^{-10T_s}} $. The spectrum of the step response is now (note that  $F_L({e^{j\omega T_s}}) = 1$ for $\omega = k\frac{2\pi}{T_s}$ and any integer $k$)
\begin{equation}
Y(\omega) = T(j\omega)F_L(e^{j \omega T_s}).\frac{1}{1 - e^{-j\omega T_s}} + \frac{\pi}{T_s}\sum_{k=-\infty}^{\infty} T(j k\frac{2\pi}{T_s}) \delta(\omega - k\frac{2\pi}{T_s})
 \label{eq:Yw}
\end{equation}
%For this example, $T(0) = 1$ and $T(j k\frac{2\pi}{T_s}) = 0$ for $k \neq 0$. Thus,
%\begin{equation}
%Y(\omega) = \delta(\omega) + T(j\omega).\frac{1}{1 - e^{j\omega T_s}} 
% \label{eq:Yw2}
%\end{equation}
which has a significant component $|Y(\frac{3 \pi}{0.4})| \approx 0.1527$ at the ripple frequency as expected. It turns out that a simple way to avoiding ripples and to obtain a good step tracking over the continuous-time domain is to limit the value of $|T(j\omega)|$ at some design frequencies.  More precisely, 
the continuous-time tracking specification will be related with making small $|E(j\omega)/R(j\omega)|$ over the working frequencies interval, where from (4) and  (17) it is obtained
\begin{equation}
|E(j\omega)/R(j\omega)| = |F(j\omega)| \cdot |1 - T(j\omega)\frac{F_L(e^{j\omega T_s})}{F(j\omega)}\frac{R^{T_s}(e^{j\omega T_s}) }{R(j\omega)} |  
\label{eq:ER}
\end{equation}

%and considering a first order prefilter with transfer functions $F(s) = \frac{1}{0.1s+1}$  and $F_L({z_s}) = \frac{(1-e^{-10T_s})z_s}{(z_s-1)(z_s-e^{-10T_s})} $

%Note that for this example, at the ripple frequency it is obtained $|E(j\frac{3 \pi}{0.4}))/R(j\frac{3 \pi}{0.4}))|  \approx 4.7272$. 
This will be the approach to be developed in Section \ref{Sect:design}, jointly with several other stability and performance design specifications. Note that $E(j\omega)/R(j\omega)$ will be referred to as continuous sensitivity function or simply sensitivity function, and will be denoted by $S(j\omega)$.  Note that it is not a frequency response in the usual sense, since it depends on the ratio of the continuous-time reference and its sampling.
%In practice, to avoid the use of impulses as in \eqref{eq:Yw}, the step signal will be substituted by a slightly different signal $R(j\omega) = \frac{a}{j\omega(j\omega+a)}$ %and $R^{T_s}(e^{j\omega T_s}) = \frac{e^{j\omega T_s}(1-e^{-aT_s})}{(e^{j\omega T_s}-1)(e^{j\omega T_s}-e^{-aT_s})}$
%for a value of $a>0$ high enough in relation with the inverse of the dominant constant time of the plant. 
Fig. \ref{fig:Scexample} shows the magnitude Bode plot of \eqref{eq:ER} exhibiting a peak of the sensitivity function of almost $11 \text{dB}$ at the ripple frequency. Clearly, for avoiding the ripple, a specification appropriately limiting the sensitivity function magnitude has to be posed in the control design problem.

 %In Fig. \ref{fig:Scexample}, the reference signal $R(j\omega) = \frac{10}{j\omega(j\omega+10)}$ is used, 

\begin{figure}[t]
\begin{center}
{\includegraphics[width=0.65\textwidth]{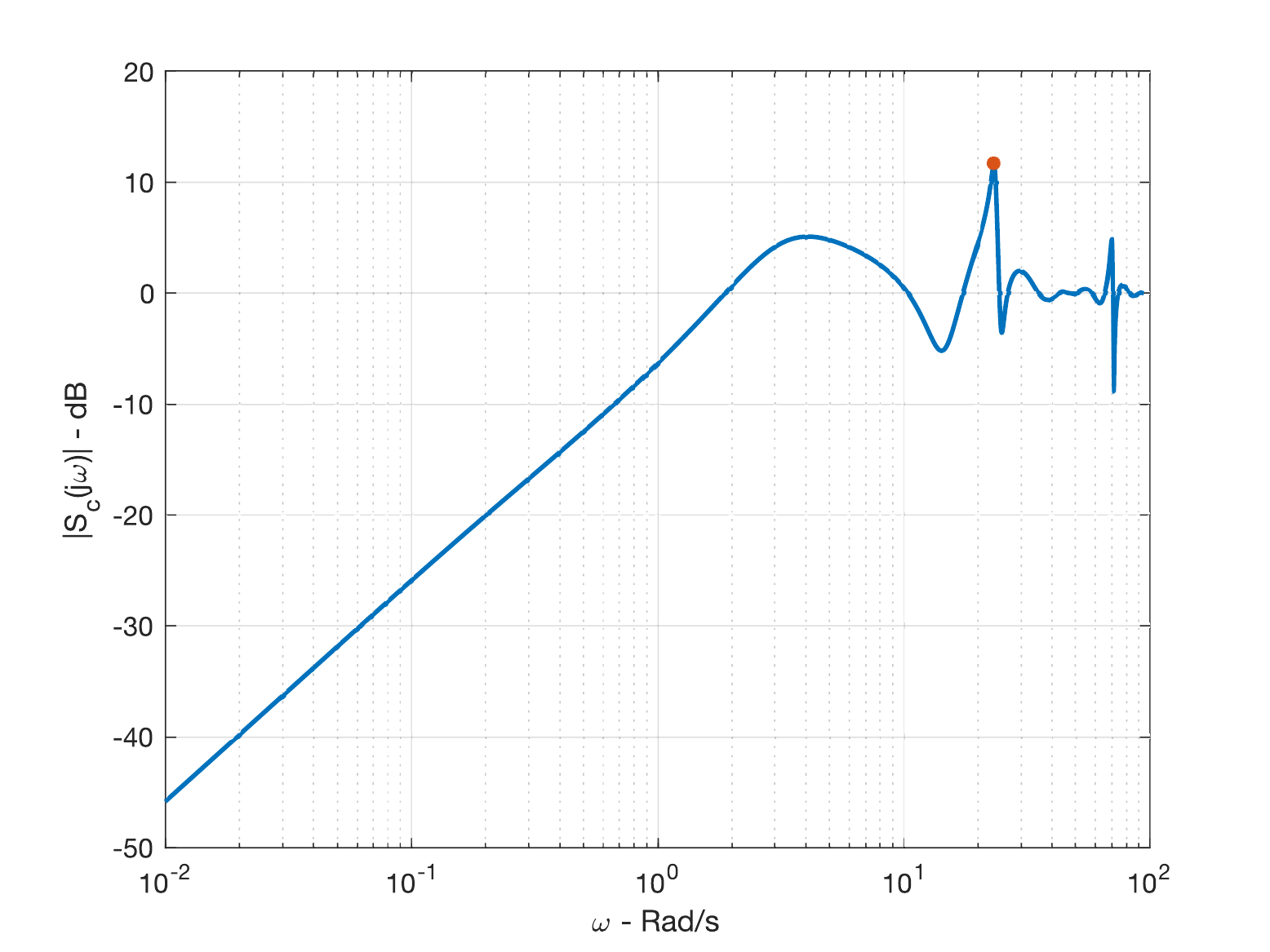}}
\caption{Magnitude Bode plot of the continuous sensitivity function for the DR control system of Section 3.1-Example, see also Example 6. The ripple is clearly exhibited  as a peak of aproximately $11 \text{dB}$ at the frequency $3\pi/T_s$. This sensitivity function is not standard in the sense that it is different for each reference signal, is this case it is related to a step reference. 
}
\label{fig:Scexample}
\end{center}
\end{figure}

%%%%%%%%%%%%%%%%%%%%%%%%%%%%%%%
\section{Multirate controller design based on QFT}
\label{Sect:design}
The starting point is a uncertain plant that can be modelled as a set $\mathcal{P}$ of transfer functions. This set may represent physical models with both parametric and non-parametric uncertainty, a set of frequency responses obtained from identifications experiments, etc. It is only required that the plant $\mathcal{P}$ be represented by a set of templates $\mathcal{P}_\omega$ that collects all the frequency responses at a  frequency $\omega \geq 0$. More specifically, $\mathcal{P}_\omega = \{P(j\omega): P(s) \in \mathcal{P} \}$.  Usually, templates are represented in the Nichols Plane (NC), and it will be assumed that they are simply connected regions of NC and that corresponds to plants with the same number of unstable poles. These restrictions are not overly restrictive and will considerably simplify the design problem, since it will be enough to work with the boundary of the templates. 
 
Now, related with Fig. 1, the DR control problem consists of designing the controllers $G_R$ and $G_L$ for an uncertain system $\mathcal{P}$, satisfying design specifications such as stability and tracking for every plant in the set $\mathcal{P}$. More specifically, in this work the design strategy is to design $G_L$ once $G_R$ has been previously design (tipically for a nominal plant). The open loop gain function is $L(e^{j\omega T_s}) = G_L(e^{j\omega T_s})P_L(e^{j\omega T_s})$, and a nominal value $L_0(e^{j\omega T_s}) = G_L(e^{j\omega T_s})P_{L0}(e^{j\omega T_s})$ is obtained for some nominal plant transfer function  $P_0 \in \mathcal{P}$. %that is $H_{L0} = [P_0 \cdot  G_R \cdot H_{T_s}]^{T_s}$, being the function to be designed.
Also, for $G_R$ and a given $P \in {\mathcal P}$ define the discrete uncertainty $\Delta_L(e^{j\omega T_s})$ as 
 \begin{equation}
\Delta_L(e^{j\omega T_s}) = \frac{P_L(e^{j\omega T_s})}{P_{L0}(e^{j\omega T_s})}
 \label{eq:QL}
\end{equation}
where $P_L = S_NP_RG_RQ_N^*$ and $P_R = S_TPH_T$ (see Section 3.2). In addition, the discrete uncertainty set is defined as $\mathcal{Q}_L = \{ \Delta_L(e^{j\omega T_s}) : P \in \mathcal{P}  \}$.
 Note that its nominal value is ${\Delta}_{L0} = 1$. Moreover, the uncertainty $\Delta(j\omega)$ is defined as
  \begin{equation}
\Delta({j\omega}) = \frac{P({j\omega})G_R(e^{j \omega T_f})}{P_{L0}(e^{j\omega T_s})}
 \label{eq:Q}
\end{equation}
%where $H(j\omega) = P(j\omega) \cdot G_R(e^{j \omega T_f}) \cdot H_{T_s}(j\omega) $,
and the uncertainty set as $\mathcal{Q}= \{ \Delta({j\omega}) :  P \in \mathcal{P}  \}$. 

The QFT design will be based on the loop gain-phase shaping of the nominal loop gain $L_0(e^{j\omega T_s})$ for the dual control system to satisfy robust design specifications. In the following, robust stability and tracking specifications are considered.
 
\subsection{Robust stability}
A direct application of Prop. 1 result in that the DR control system is robustly stable, that it is stable for every $P \in \mathcal{P}$, if it is stable for the nominal plant $P_0$, and in addition for any $\omega \in [0,\pi/T_s]$ and $\Delta_L \in \mathcal{Q}_L$ it is satisfied that
\begin{equation}
1 + L_0(e^{j\omega T_s}) \Delta_L(e^{j\omega T_s})  \neq 0.
 \label{eq:rstability}
\end{equation}
This follows from the fact that all the plants in $\mathcal{P}$ have the same number of unstable poles and thus all the open loop gain functions must cross the ray $\bf{R}_0$ the same net number of times. A more restrictive robust stability condition, including stability margins is that for some positive  real number $\mu <1$ 
\begin{equation}
\left |
1  + L_0(e^{j\omega T_s}) \Delta_L(e^{j\omega T_s})
\right| \geq \mu
 \label{eq:rstability2}
\end{equation}
for any $\omega \in [0,\pi/T_s]$  and any $\Delta_L \in \mathcal{Q}_L$. Note that this is equivalent for the discrete sensitivity function to satisfy $ |S(e^{j\omega T_s})| \leq 1/\mu$. 

For a given frequency $\omega \in [0,\pi/T_s]$ and any $\Delta_L \in \mathcal{Q}_L$,  (45) defines a forbidden region for $L_0(e^{j\omega T_s})$ at the frequency $\omega$ around the critical point $(-180^\circ, 0 \text{ dB})$ in the Nichols plane, whose boundary will be referred to as stability bound. Note that, 
in particular, stability bounds guarantees worst-case phase and gain margins as given by PM = $180^\circ + 2\cos^{-1}(\mu/2)$ and GM = $1/(1-\mu)$, respectively. $\mu$ will be referred to as the (worst-case) stability margin.

\vspace{0.5 cm}
{\bf Example 7}: %Robust stability of Section 3.1-Example is analyzed here, where the system with transfer function \eqref{eq:Pex} is the nominal plant. 
Consider the uncertain plant $\cal{P}$ given by 
\begin{equation}
\mathcal{P} = \{ \frac{a}{(s+0.5)(s+a)}: a \in [0.5,2.5] \}
\label{eq:Pexunc}
\end{equation}
where the nominal plant $P_0$ corresponds to $a = 1.5$. The question is if the DR controller given by \eqref{eq:GL}-\eqref{eq:GR}, that has been show to stabilize the DR control system for the nomimal case (see Fig. 4), is also able to guaranty stability for any plant in the uncertain plant set $\mathcal{P}$. A stability margin $\mu = 0.5$ is chosen, corresponding to PM = $30^\circ$ and GM = $2$. 

The analysis will be performed in three (equivalent) ways, for emphasizing the use of stability bounds specially for readers not familiarized with QFT: 

\begin{itemize}
\item (Discrete sensitivity) It is directly computed  $|S(e^{j\omega T_s})|$ for $a \in [0.5,2.5]$. It is not difficult to see that in fact the specification $|S(e^{j\omega T_s})| \leq 2 $ is not satisfied for high frequencies close to the Nyquist frequency $\frac{\pi}{T_s}$ and  $a > 2.07$ (Fig. \ref{fig:stabilityA}). Thus the DR control system is not stable with a stability margin $\mu = 0.5$. Note that it would be stable for $a \in [0.5,2]$.
\begin{figure}[h]
\begin{center}
{\includegraphics[width=0.6\textwidth]{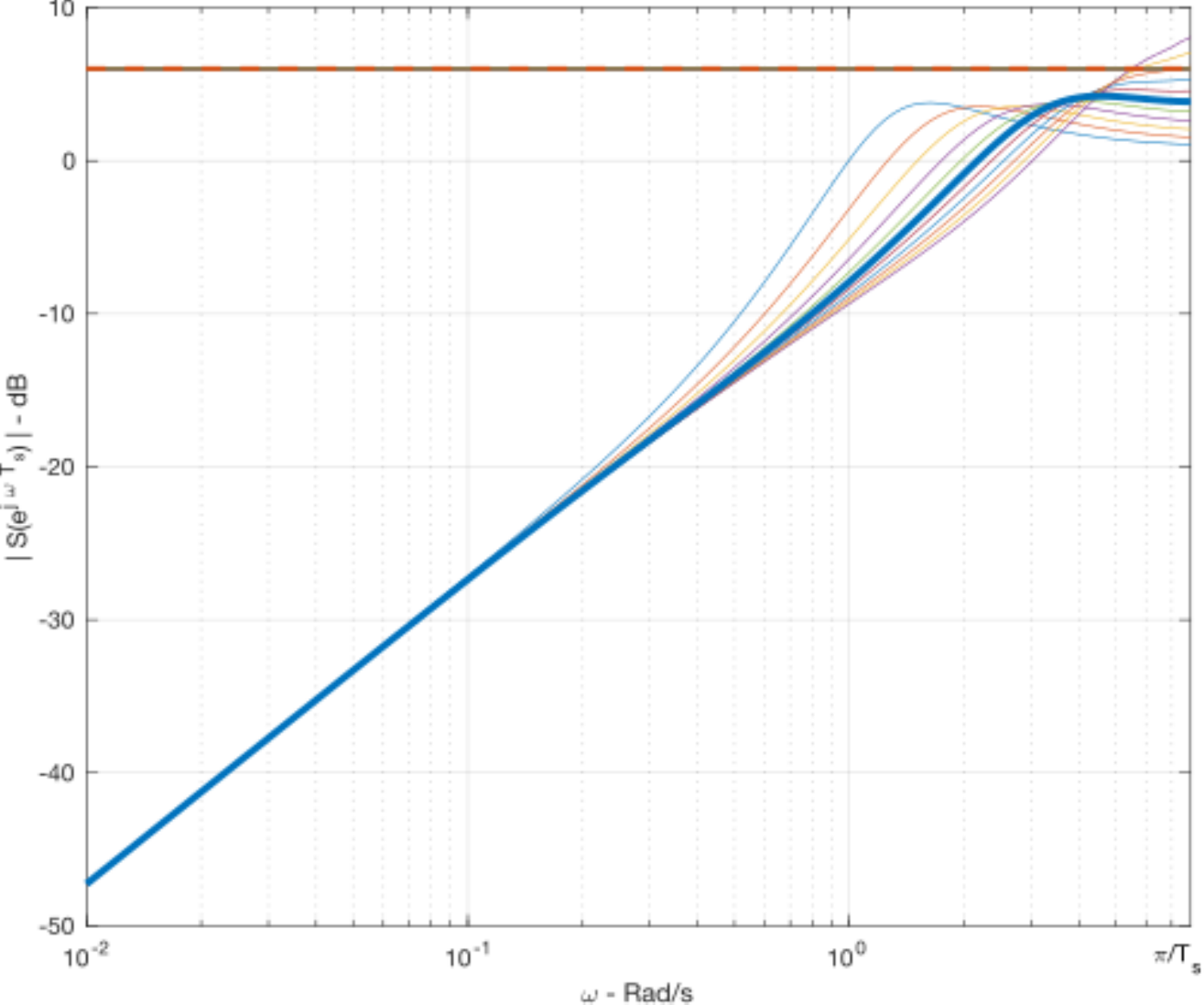}}
\caption{ Discrete sensitivity magnitude against frequency, for the DR control system of Example 7. The system does not satisfy the worst-case stability margin $\mu = 0.5$ ($1/\mu = 2 \approx 6$ dB) for $a > 2.07$.
}
\label{fig:stabilityA}
\end{center}
\end{figure}
\item (Open-loop gain functions) Here stability is based on the computation of $L(e^{j\omega T_s}) = G_L(e^{j\omega T_s})P_{L0}(e^{j\omega T_s})\Delta_{L}(e^{j\omega T_s})$, for every $\Delta_L \in \mathcal{Q}_L$. Figure \ref{fig:stabilityB}-left shows the Nichols plots for some values of the parameter $a$. Note that in this case the stability specification $|1 + L(e^{j\omega T_s}) | \geq \mu$ results in a forbidden region for any $L(e^{j\omega T_s})$ in the Nichols plane (its bound is shown in Fig. \ref{fig:stabilityB}-left). It is clear that some $L(e^{j\omega T_s})$ enter in that forbidden region for some values of the parameter $a$ and thus the DR control system is not stable with the specified stability margin $\mu$.

%\begin{figure}[htbp]
%\begin{center}
%{\includegraphics[width=1.0\textwidth]{Figuras/robuststabilityB.pdf}}
%\caption{
%}
%\label{fig:stabilityB}+
%\end{center}
%\end{figure}

\begin{figure}[h]
\centering
\begin{subfigure}{0.55\textwidth}
  \includegraphics[width=\linewidth]{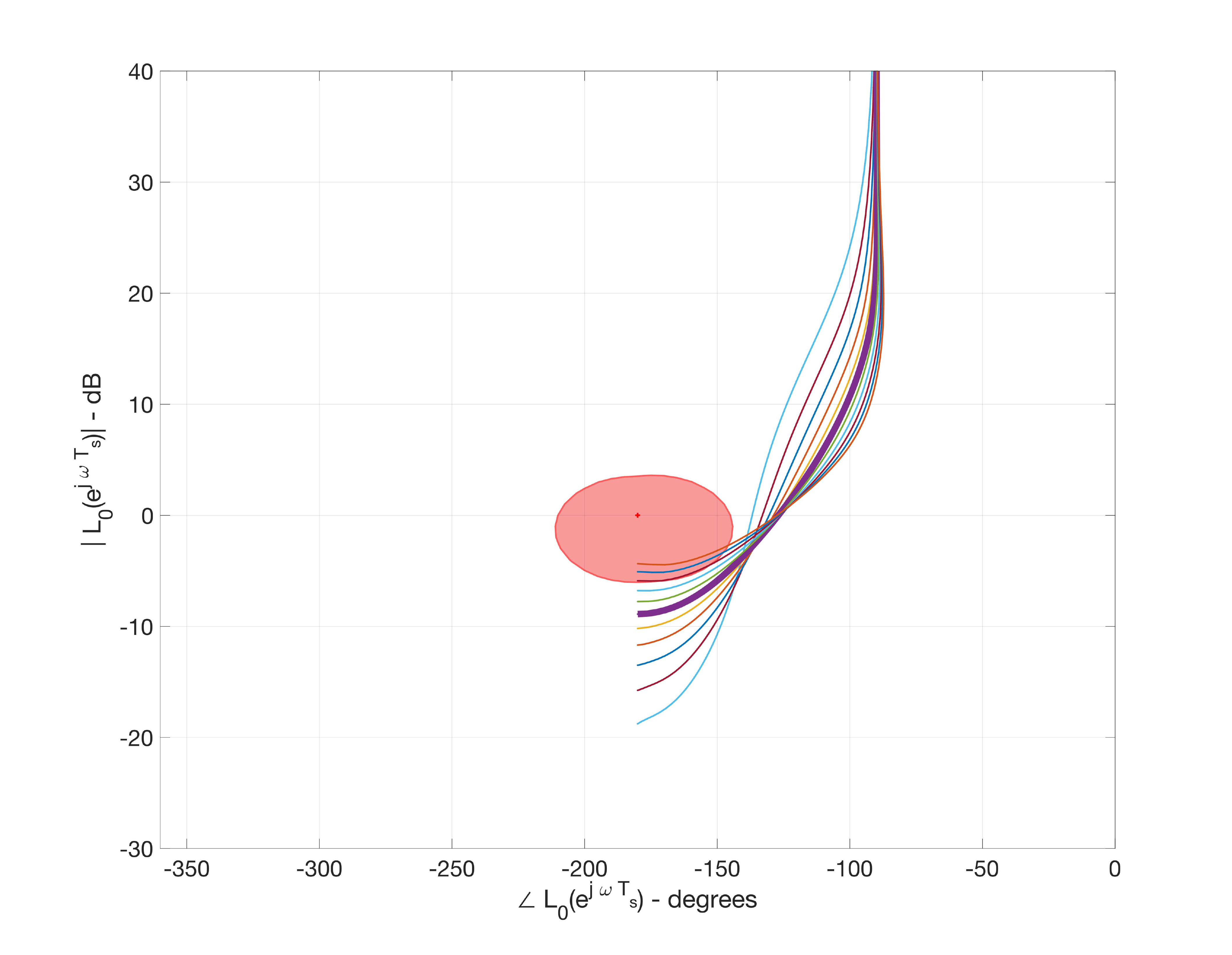}
  %\caption{A subfigure}
  \label{fig:sub1}
\end{subfigure}%
\begin{subfigure}{0.45\textwidth}
  \centering
  \includegraphics[width=\linewidth]{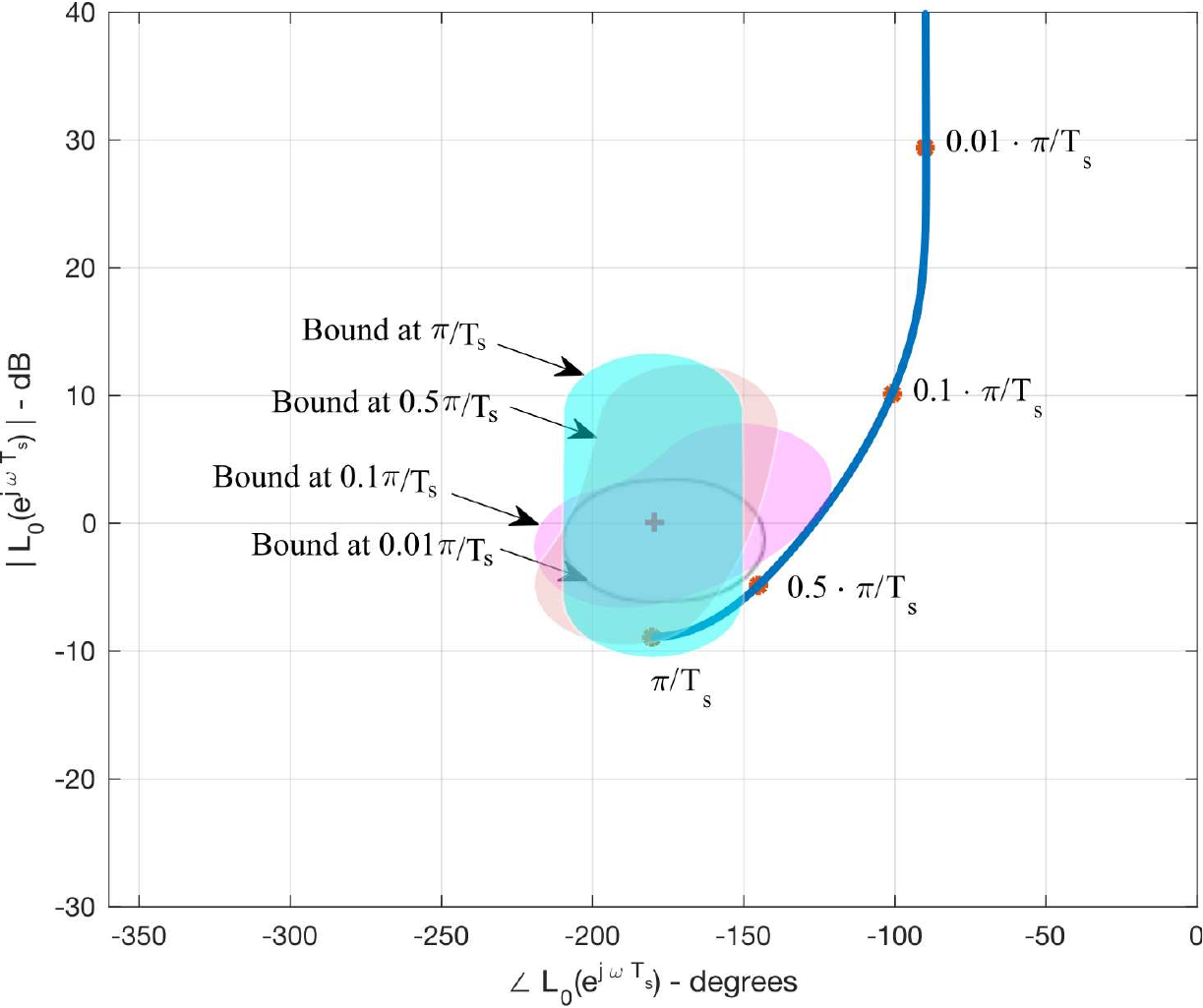}
  %\caption{A subfigure}
  \label{fig:sub2}
\end{subfigure}
\caption{(left) Nichols plots of the open-loop gain functions $L(e^{j\omega T_s})$, for some sample values of the parameter $a \in [0.5,2.5]$ (nominal open loop gain corresponds to $a = 1.5$ -thick line-), and forbidden region in NP for robust stability; (right) Forbidden regions in NP for the nominal open loop gain function, limited by boundaries at several frequencies $\omega \in \{ 0.01,0.1,0.5,1 \}\cdot \pi/T_s $. In both cases, the plots segment from $\omega = 0^-$ to $\omega = 0^+$ (see Fig. 5) has not been explicitly shown.}
\label{fig:stabilityB}
\end{figure}

\item (Nominal open-loop gain function) The above stability analysis may be appropriate for analysis but are not well suited for design, since in general it is not obvious how a modification of the controllers (in our case the slow controller $G_L$) would shape the sensitivity functions or the open-loop gain functions to satisfy the stability specification. A more convenient way both for analysis and design is proposed by using a QFT approach. Basically, the stability specification 
\eqref{eq:rstability2} is translated to a set of forbidden regions of the nominal open-loop gain function $L_0(e^{j\omega T_s})$ ideally for every frequency $\omega \in [0,\pi/T_s]$ (in practice, it is enough with a finite number of working frequencies, and some iteration may be needed if the design is not validated). Fig. \ref{fig:stabilityB}-right shows the forbidden regions bounds at several working frequencies. It results that the nominal open-loop function enters the forbidden region for $\omega = \pi/T_s$ and thus the DR control system is unstable. To stabilize the DR control system the nominal open-loop function should be conveniently shaped to be out of the forbidden regions at any frequency.

\end{itemize}

\subsection{Robust tracking}
Reference tracking specifications are considered both in discrete-time and in continuous-time. %Using Prop. 3, two types of tracking specifications are considered in discrete-time and continuous-time, respectively.

%\begin{itemize} 
%\item

\vspace{0.25cm}
\noindent {\bf Discrete-time tracking}\\
Tracking is specified at the slow sampling period $T_s$. The transfer function from $r^{T_s}$ to $e^{T_s}$ is $S_L(e^{j\omega T_s})F_L(e^{j\omega T_s})$, and the discrete-time tracking specification is $| {E^{T_s}(e^{j \omega T_s})}/{R^{T_s}(e^{j \omega T_s})}| \leq \delta_1(\omega)$ for any reference signal $R^{T_s}(e^{j \omega T_s})$, for any $\omega \in [0,\pi/T_s]$ and $P \in \mathcal{P}$. Here $\delta_1:[0,\pi/T_s]\rightarrow \mathbb{R}_{\geq 0}$ is a given function that defines the tracking specification. It easily follows that this is equivalent to 
\begin{equation} 
\left |
1  + L_0(e^{j\omega T_s}) \Delta_L(e^{j\omega T_s})
\right | \geq \frac{|F_L(e^{j\omega T_s})|}{\delta_1(\omega)}
 \label{eq:rdtracking}
\end{equation}
for any $\omega \in [0,\pi/T_s]$ and $\Delta_L \in \mathcal{Q}_L$. Note that for a given prefilter $F_L$ and a tracking specification $\delta_1$, \eqref{eq:rdtracking} takes the same form that \eqref{eq:rstability2}, that is it defines forbidden regions in the NP for the nominal open-loop gain function  $L_0(e^{j\omega T_s})$, for any $\omega \in [0,\pi/T_s]$. 
%\item 

\vspace{0.25cm}
\noindent
{\bf Continuous-time tracking}. \\
Tracking may be also specified in the continuous-time domain. Here the only limitation is that a tracking specification must be posed for some given reference. This limitation is directly related with the time-varying nature of the DR control system, and it can be alleviated by using as much tracking specifications as needed and using the worst-case. The tracking specification is  $| {E({j \omega})}/{R({j \omega})}| \leq \delta_2(\omega)$ for a given reference $R$, any $\omega > 0$, and any $P \in \mathcal{P}$. Note that the specification will result in restrictions over the nominal open-loop function $L_0(e^{j\omega T_s}) $ for frequencies below and beyond the Nyquist frequency $\pi/T_s$, and thus continuous-time tracking specifications for $\omega > \pi/T_s$ will be folded over the interval  $[0,\pi/T_s]$.

As a consequence, and to the authors knowledge this is a previously unexplored case in QFT, for the frequencies in which $L_0(e^{j\omega T_s})$ can be designed, that is for $\omega \in [0,\pi/T_s]$, in practice there will be a finite number of restrictions or boundaries to be satisfied resulting from the folding of specifications for frequencies beyond  $\pi/T_s$ ({note that usually it is enough with continuous-time tracking specifications for frequencies not much larger that the crossover frequency}). The following result gives a procedure for obtaining continuous-time tracking bounds; as usual, the worst-case boundary will be used for shaping the nominal open loop gain function.

\vspace{0.5cm}
{\bf Proposition 8}: Consider the DR control system of Fig.1, and assume that it is stable and that Asumption 1 holds. For a given frequency  $\omega > 0$ and a reference $R$, the continuous-time tracking specification $| {E({j \omega})}/{R({j \omega})}| \leq \delta_2(\omega)$ for any $P \in \mathcal{P}$, is equivalent to the following specification: if $\omega \in [k\frac{2\pi}{T_s}, (2k+1)\frac{\pi}{T_s}]$ for some $k = 0, 1, 2, \cdots$ then 
\begin{equation}
\left |
\frac{1 + L_0(e^{j \omega^\dagger T_s}) A(j\omega)}
{1  + L_0(e^{j\omega^\dagger T_s}) \Delta_L(e^{j\omega^\dagger T_s})}
\right |
\leq \frac{\delta_2(\omega)}{|F(j \omega)|}
\label{eq:prop7}
\end{equation}
is satisfied by $L_0(e^{j\omega^\dagger T_s})$ at a frequency $\omega^\dagger = \omega - k\frac{2\pi}{T_s} \in [0,\pi/T_s]$ and any $\Delta \in \mathcal{Q}$ and $\Delta_L \in \mathcal{Q}_L$; alternatively, if $\omega \in [(2k+1)\frac{\pi}{T_s}, (k+1)\frac{2\pi}{T_s}]$ for some $k = 0, 1, 2, \cdots$, then  
\begin{equation}
\left |
\frac{1 + L_0^\ast(e^{j \omega^\dagger T_s}) A(j\omega)}
{1  + L_0^\ast(e^{j\omega^\dagger T_s}) \Delta_L^\ast(e^{j\omega^\dagger T_s})}
\right |
\leq \frac{\delta_2(\omega)}{|F(j \omega)|}
\label{eq:prop72}
\end{equation}
is satisfied by $L_0(e^{j\omega^\dagger T_s})$ at a frequency $\omega^\dagger = -\omega + (k+1)\frac{2\pi}{T_s} \in [0,\pi/T_s]$ and any $\Delta \in \mathcal{Q}$ and $\Delta_L \in \mathcal{Q}_L$. In both cases, 
\begin{equation}
A(j \omega) =  \Delta_L(e^{j \omega T_s}) - \frac{F_L(e^{j \omega T_s})R^{T_s}(e^{j \omega T_s})}{F(j\omega) R(j\omega)} \Delta(j\omega)
\label{eq:Aprop7}
\end{equation}

\vspace{0.5cm}
{\bf Proof}: Using \eqref{eq:error}, \eqref{eq:SL}, \eqref{eq:T}, and \eqref{eq:spectrumY}, %the continuous-time tracking specification $| {E({j \omega})}/{R({j \omega})}| \leq \delta_2(\omega)$ directly results in  
${E({j \omega})}/{R({j \omega})}$ is given by 
\begin{equation}
\frac{E(j \omega)}{R(j \omega)} =  F(j\omega)  - \frac{Y(j\omega)}{R(j \omega)}  = F(j\omega) - \frac{P(j \omega) G_R(e^{j\omega T_f}) H_{T_s}(j \omega)G_L(e^{j\omega T_s})}{1+P_L(e^{j\omega T_s})G_L(e^{j\omega T_s})}F_L(e^{j\omega T_s}) \frac{R_L(e^{j\omega T_s})}{R(j\omega)}
\label{eq:prop7a}
\end{equation}
%where by definition (see Section 4, page 9) $H(j\omega) = P(j \omega) G_R(e^{j\omega T_f s}) H_{T_s}(j \omega)$. 
Moreover, from \eqref{eq:QL} and \eqref{eq:Q}, \eqref{eq:prop7a} is equal to
\begin{equation}
\frac{E(j \omega)}{R(j \omega)} =  F(j\omega) - \frac{\Delta(j\omega)L_0(e^{j\omega T_s})}{1+\Delta_L(e^{j\omega T_s})L_0(e^{j\omega T_s})}F_L(e^{j\omega T_s}) \frac{R_L(e^{j\omega T_s})}{R(j\omega)}
\label{eq:prop7b}
\end{equation}

Now, to obtain \eqref{eq:prop7}, ${E(j \omega)}/{R(j \omega)} $ must be expressed in the form
%$| {E({j \omega})}/{R({j \omega})}| \leq \delta_2(\omega)$ directly results in  
\begin{equation}
\frac{E(j \omega)}{R(j \omega)} =
\frac{1 + L_0(e^{j \omega T_s}) A(j\omega)}
{1  + L_0(e^{j\omega T_s}) \Delta_L(e^{j\omega T_s})} F(j\omega)
\label{eq:prop7c}
\end{equation}
and thus equalizing the right-hands of \eqref{eq:prop7b} and \eqref{eq:prop7c} it easily follows that (the frequency arguments are removed by simplicity)
\begin{equation}
(1 + L_0 A) F = F (1+\Delta_L L_0) - H G_L F_L \frac{R_L}{R} = (1+\Delta_L L_0 - \Delta L_0 \frac{F_L R_L}{F R}) F
\label{eq:prop7d}
\end{equation} 
 and \eqref{eq:Aprop7} directly follows. Note that is has been proved that the continuous-time tracking specification is equivalent to the nominal open-loop function $ L_0(e^{j \omega T_s})$ to satisfy the inequality
 \begin{equation}
\left |
\frac{1 + L_0(e^{j \omega T_s}) A(j\omega)}
{1  + L_0(e^{j\omega T_s}) \Delta_L(e^{j\omega T_s})}
\right |
\leq \frac{\delta_2(\omega)}{|F(j \omega)|}
\label{eq:prop7e}
\end{equation}
 for any $\omega \geq 0$, any $\Delta_L \in \mathcal{Q}_L$, and any $\Delta \in \mathcal{Q}$. 
 
To end the proof, periodicity and simmetry properties of $ L_0(e^{j \omega T_s})$ are recalled to obtain the folding frequency $\omega^\dagger \in [0,\pi/T_s]$ at which the inequality poses a restriction over the nominal open-loop function. If  $\omega \in [k\frac{2\pi}{T_s}, (2k+1)\frac{\pi}{T_s}]$ for some $k = 0, 1, 2, \cdots$ then at the frequency $\omega^\dagger = \omega - k\frac{2\pi}{T_s} \in [0,\frac{\pi}{
T_s}]$, periodicity of nominal open-loop gain function results in that $L_0(e^{j \omega T_s}) = L(e^{j (\omega^\dagger+ k\frac{2\pi}{T_s} )T_s}) = L_0(e^{j \omega^\dagger T_s}) $. Alternatively, if $\omega \in [(2k+1)\frac{\pi}{T_s}, (k+1)\frac{2\pi}{T_s}]$ for some $k = 0, 1, 2, \cdots$, then 
at the frequency $\omega^\dagger = -\omega + (k+1)\frac{2\pi}{T_s} \in [0,\frac{\pi}{T_s}]$, using simmetry and periodicity arguments it directly follows that
$L_0(e^{j \omega T_s}) = L_0(e^{j (-\omega^\dagger+ (k+1)\frac{2\pi}{T_s} )T_s}) = L_0(e^{-j \omega^\dagger T_s}) = L_0^\ast (e^{j \omega^\dagger T_s}) $. The same property holds for $\Delta_L(e^{j\omega T_s})$. Considering the inequality \eqref{eq:prop7e} in both cases directly gives \eqref{eq:prop7} and \eqref{eq:prop72}, repectively.
$\Box$

\vspace{0.5cm}
{\bf Example 9:} The Example of Section 3.1 is now analyzed by using discrete-time and continuous-time tracking specifications using the above QFT specifications. By simplicity, firstly it is considered the case of no-uncertainty, that is the plant is given by (5). And the slow and fast controllers are given by (6) and (7) respectively. Also, the prefilter is $F(s) = \frac{1}{0.1s+1}$ and its discretization is $F_L({z_s}) = \frac{1-e^{-10T_s}}{z_s-e^{-10T_s}} $. 

Firstly, consider a discrete-time tracking specification like (47), that combined with a stability specification like (45), gives a restriction over the discrete-time sensitivity function like $|S_L(e^{j \omega T_s})| \leq \min\{\delta_1(\omega)/|F_L(e^{j\omega T_s})|,1/\mu\}$, where $\delta_1(\omega) = \omega/2$ and $\mu = 0.5$ has been chosen. Moreover, the design frequencies $\{0.01, 0.03, 0.1,0.3,0.5,0.7,1\}\cdot \pi/T_s$ are chosen (note that for the discrete-time sensitivity function the Nyquist frequency $\pi/T_s$ is the highest frequency). In Fig. \ref{fig:dtracking}-left both QFT bounds and the open-loop gain function $L_0$ are plotted. It is clear that the tracking (and stability) specification is satisfied at the working frequencies. However, the design must be validated for the rest of frequencies, this is shown in the \ref{fig:dtracking}-rigth where the design is validated for the discrete-time tracking specification (as it is expected from the results of Section 3.1).

\begin{figure}[htbp]
\begin{center}
{\includegraphics[width=1.05\textwidth]{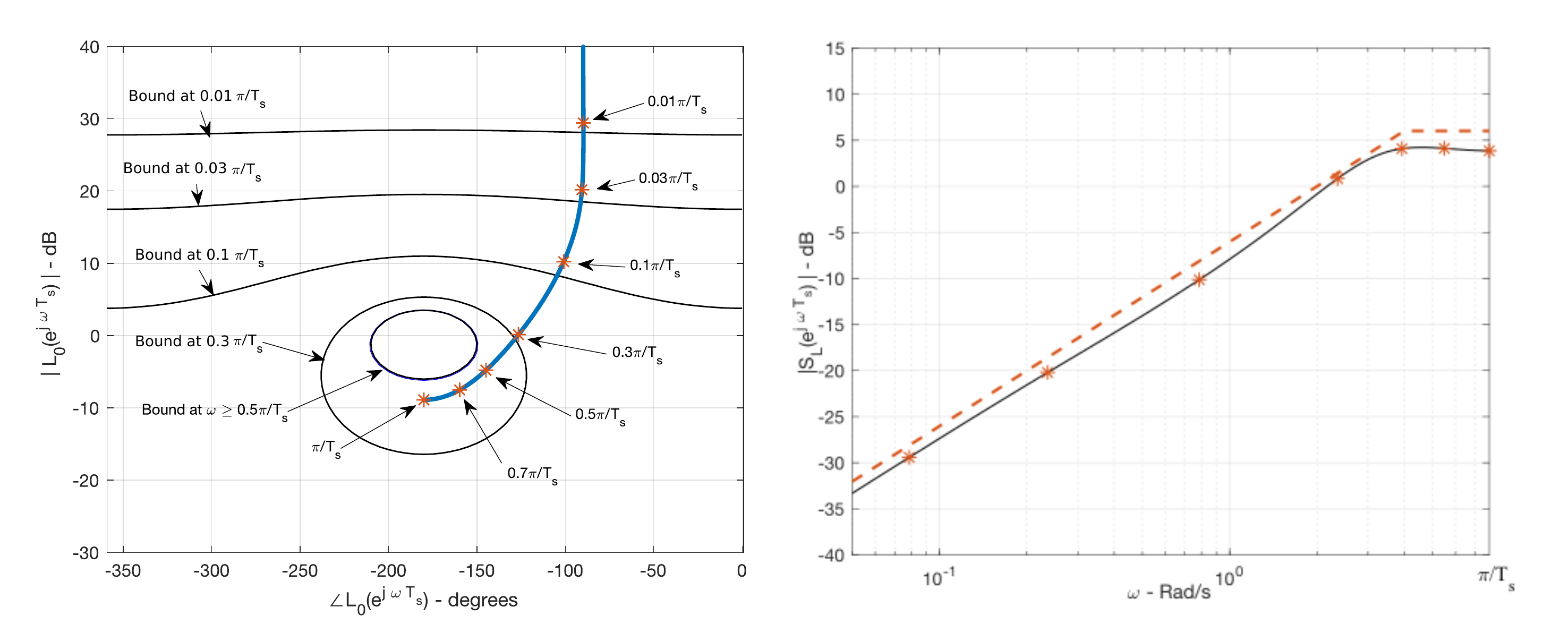}}
\caption{ (left) Discrete-time tracking boundaries at the design frequencies set $\{0.01, 0.03, 0.1,0.3,0.5,0.7,1\}\cdot \pi/T_s$, and nominal open loop gain function $L(e^{j \omega T_s})$ -asterisks denote its value at the design frecuencies-. (right) Discrete-time tracking specification as upper bound over $|S_L(e^{j \omega T_s})|$ (dotted line), and magnitude Bode plot (solid line) of the resulting $S_L(e^{j \omega T_s})$ for the DR control system -asterisks also denote its values at the design frequencies-.} 
\label{fig:dtracking}
\end{center}
\end{figure}

Now, continuous-time tracking specifications are considered, including frequencies below and beyond the Nyquist frequency $\pi/T_s$. %After the results of Section 3.1, it is expected that tracking specifications at frequencies $2.75 \pi/T_s$ and $3 \pi/T_s$ (the ripple frequency) are not satisfied. 
The continuous-time step tracking specification is $| {E({j \omega})}/{R({j \omega})}| \leq \min\{\delta_2(\omega)/|F(j\omega)|,\mu\}$. where $\delta_2(\omega)= \omega/2$ and $\mu = 0.5$. %, and    % This aproximation is used to avoid numerical problems in the computation of the term $A(j\omega)$ as given by \eqref{eq:Aprop7}, where 
%the reference is $R(j\omega) = \frac{10}{j\omega(j \omega+10)}$, where in addition $R^{T_s}(e^{j\omega T_s}) = \frac{e^{j\omega T_s}(1-e^{-10T_s})}{(e^{j\omega T_s}-1)(e^{j\omega T_s}-e^{-10T_s})}$. 
Two sets of frequencies are separately considered in the following. 

First, frequencies below the Nyquist frequency; in this case, the same frequencies that the previously used for discrete-time step tracking are used, that is $\{0.01, 0.03, 0.1,0.3,0.5,0.7,1\}\cdot \pi/T_s$. The resulting boundaries (according to Prop. 8) are shown in Fig. \ref{fig:boundsSbellow}-left; note that the nominal open loop gain satisfies the restrictions posed by the boundaries, except for the frequency $\omega_4 = 0.3\pi/T_s$ where it slightly crosses the boundary. Fig. \ref{fig:boundsSbellow}-right shows a Bode plot of the continuous sensitivity magnitude over the frequency interval $[0,\pi/T_s]$, validating the design except for the interval $[0.2,0.5]\cdot \pi/T_s$ where it slightly crosses the specification bound. In practice, this design is reasonably good and it may be concluded that the DR design, that correctly performs according to discrete-time tracking specfications (see Fig. \ref{fig:dtracking}), also will satisfactorily track steps as far as frequencies below the slow Nyquist frequency is concerned. 

Second, frequencies beyond the Nyquist frequency. The set $\{2.5, 2.75, 3, 3.8, 5, 8.9\}\cdot \pi/T_s$ (note that $3\pi/T_s$ is the ripple frequency. see Fig. \ref{fig:Scexample}) has been chosen. By using Prop. 8, boundaries are obtained at the folded frequencies $\{0.5, 0.75,1, 0.2, 1,0.9\} \cdot \pi/T_s$, see Fig. \ref{fig:boundsSbeyond}-left. Note that in particular there are boundary crossings at $\omega_9^\dagger = 0.75\pi/T_s$ and $\omega_{10}^\dagger = \pi/T_s$, these are the responsible for the ripple in the step response. Fig. \ref{fig:boundsSbeyond}-right is a magnitude Bode plot of the sensitivity function over the interval $[0,100]$ Rad/s. As a conclusion, the DR design is not validated for frequencies beyond the (slow) Nyquist frequency.
  
\vspace{0.25cm}  
\noindent
{\bf Remark 10:}  It is worthwhile to emphasize that when analyzing or designing DR controllers for satisfying continuous-time tracking specifications, like in Example 9, the shaping of the (nominal) open-loop gain at every working frequency in $[0,\pi/T_s]$ is the design element, however a particular value a some frequency in that interval is responsible for the shaping of the continuous sensitivity function not only at that same frequency, but also in (infinitely) many frequencies beyond $\pi/T_s$. In Example 9, this is reflected for example by the fact that $L(e^{j\pi/T_s})$ (the high-frequency value of the open-loop gain) is constrained by continuous tracking specifications at several frequencies, significatively at $\pi/T_s$, $3\pi/Ts$ (the ripple frequency), and $5\pi/Ts$ (also less importantly at higher frequencies $7\pi/Ts$, $9\pi/Ts$, $\cdots$). These constraints are exactly boundaries $\#7$, $\#10$, and $\#12$ (Figs. \ref{fig:boundsSbellow} and \ref{fig:boundsSbeyond}). As a direct consequence, for ripple avoiding the open-loop gain should be redesigned at $\pi/T_s$ to satisfy the worst-case boundary, which in this case reduces to boundary $\#10$ (Fig. \ref{fig:boundsSbeyond}).
\begin{figure}[h]
\centering
\begin{subfigure}{.5\textwidth}
  \centering
  \includegraphics[width=\linewidth]{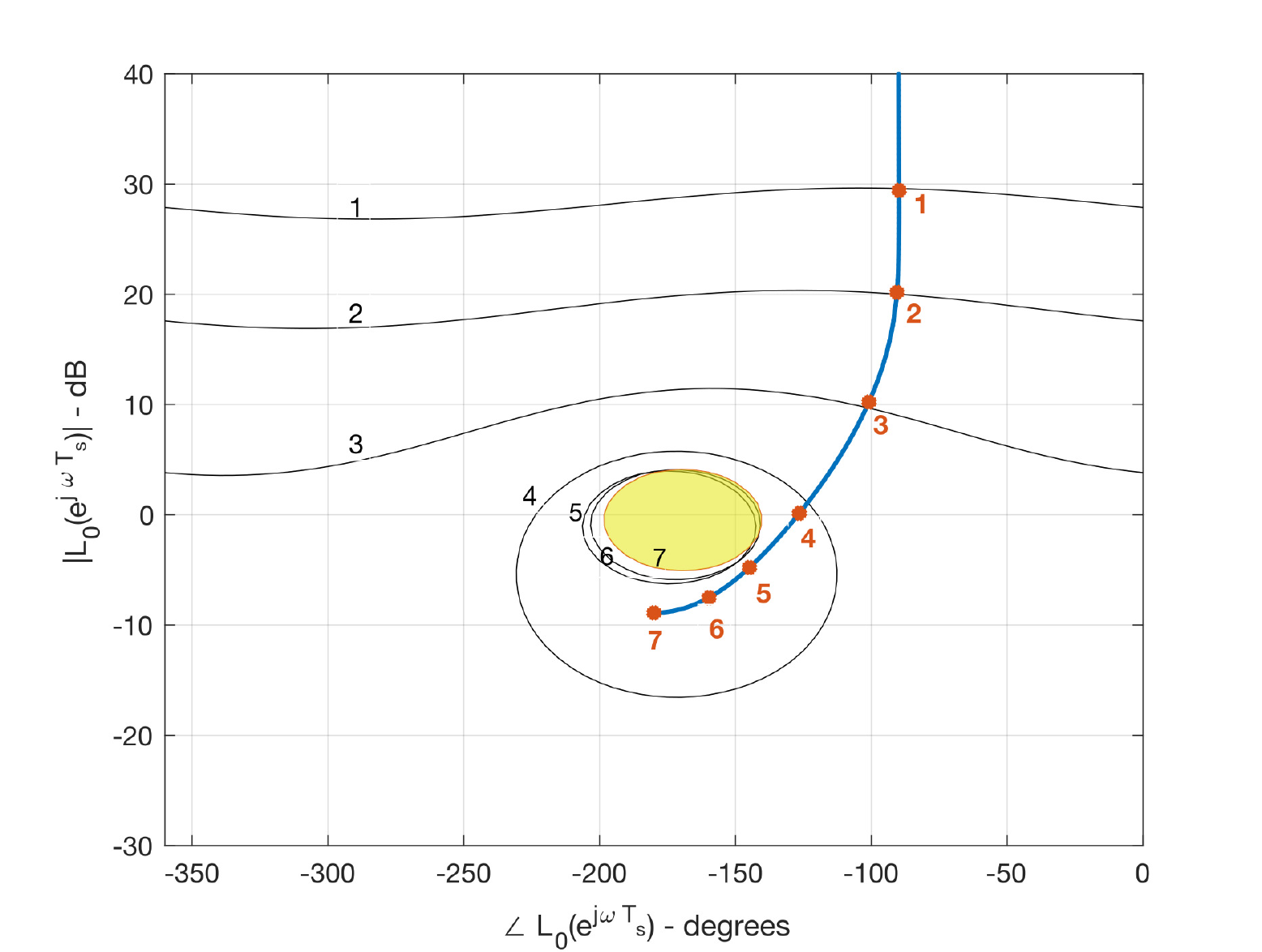}
  %\caption{A subfigure}
  \label{fig:sub1}
\end{subfigure}%
\begin{subfigure}{0.5\textwidth}
  \centering
  \includegraphics[width=\linewidth]{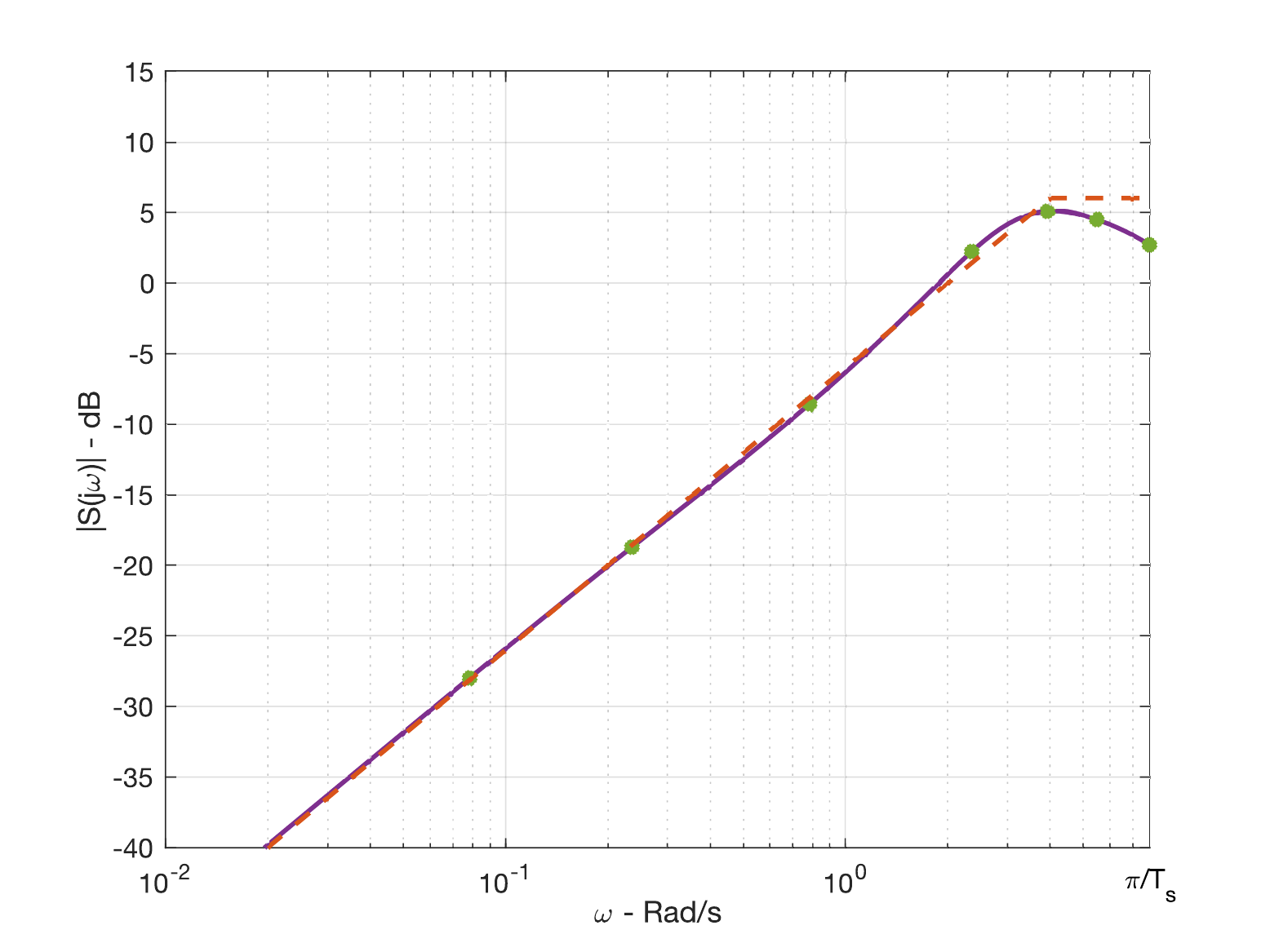}
  %\caption{A subfigure}
  \label{fig:sub2}
\end{subfigure}
\caption{ Continuous-time tracking for frequencies below the Nyquist frequency $\pi/T_s$. (Left) Boundaries at the working frequencies $\omega_{1,2,3,4,5,6,7} = (0.01,0.02,0.1,0.3,0.5,0.7,1)\cdot \pi/T_s$, indexed by the labels $1, 2, \cdots,7$, and nominal open loop gain $L_0(e^{j\omega T_s})$ (thick line) with asterisks denoting its values at the corresponding working frequencies. Open boundaries ($\#1, $\#2, $\#3$, with solid line) define a forbidden region below, and close boundaries ($\#4$..$\#7$) define define a forbidden region inside. (Right) Magnitude Bode plot of the sensitivity function $S(j\omega)$ (solid line), and specification bound (dotted line). 
}
\label{fig:boundsSbellow}
\end{figure}

\begin{figure}[ht]
\begin{center}
{\includegraphics[width=\textwidth]{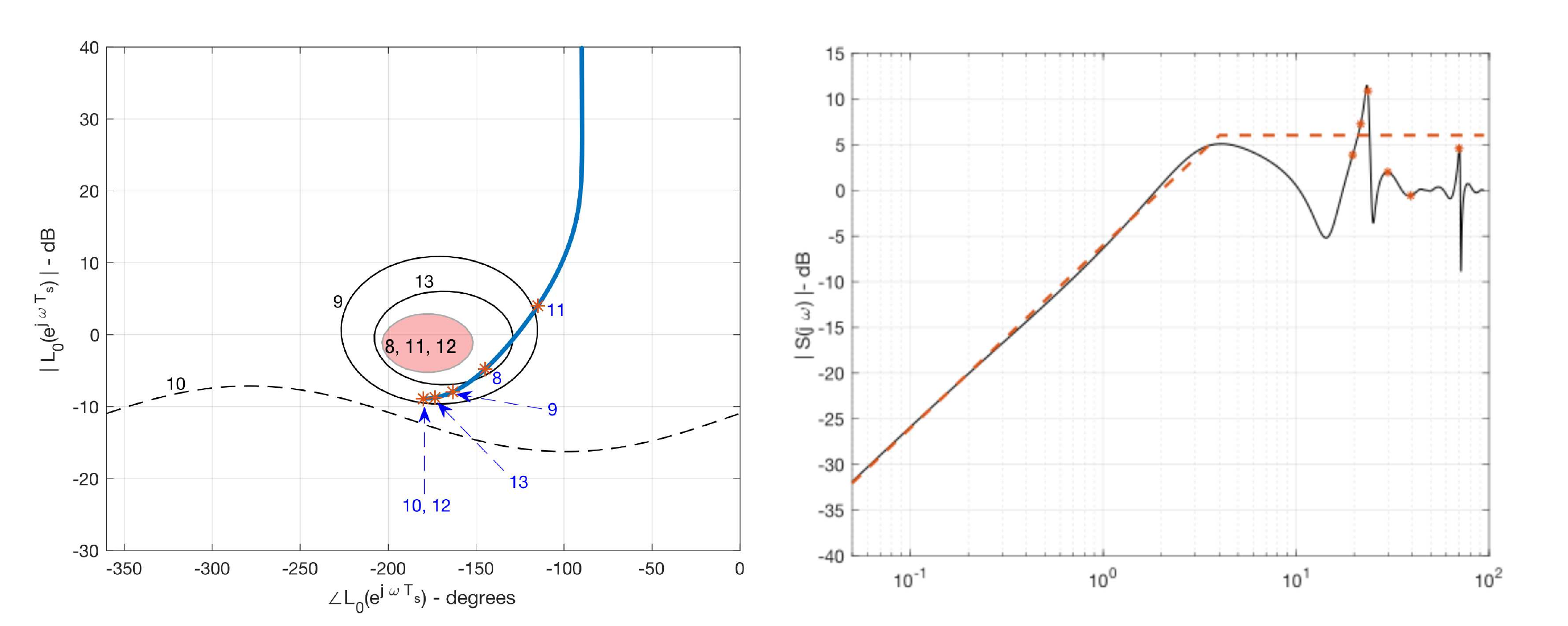}}
\caption{ Continuous-time tracking for frequencies beyond $\pi/T_s$. (Left) Specifications at the working frequencies $\omega_{8,9,10,11,12,13} = (2.5,2.75,3,3.8,5,8.9)\cdot \pi/T_s$, result in boundaries at the folded frequencies $\omega^\dagger_{8,9,10,11,12,13}  = \{0.5, 0.75,1, 0.2, 1,0.9\} \cdot \pi/T_s$. The open boundary $\#10$ (dotted line) define the forbidden region above. (Right) Magnitude Bode plot of the sensitivity function $S(j\omega)$ (solid line), and specification bound (dotted line). 
 }
\label{fig:boundsSbeyond}
\end{center}
\end{figure}

\vspace{0.25cm}  
\noindent
{\bf Example 11:} In this example, the DR control system of Section 3.1-Example is redesigned to avoid the ripple in the step response. 
 According to Example 9 (see also Remark 10), the design action will consists of redesigning the slow controller by reshaping the open-loop gain, to satisfy the constraint posed by boundary $\#10$, obviously without significatively altering it at the rest of frequencies. Looking at Fig. \ref {fig:boundsSbeyond}, the design problem is about to shape the open-loop gain close to the Nyquist frequency to be below the boundary $\#10$. A simple solution is to add a notch filter to the slow controller.  The following filter has been used, with design parameters $K$, $\alpha_1$, and $\alpha_2$: \begin{equation}
N(z) = K\frac{(z-\alpha_1)(z+ \alpha_2)}{(z-0.5)(z+0.5)}
\end{equation} 
After some trial and error, a good solution has been found (the previous design has not been significatively changed at low frequencies), resulting in $\alpha_1 = 0.52$ , $\alpha_1 = 0.76$, and $K = 0.75$. Note that the dc-gain of the notch filter is $0.85$, which means that the open-loop gain has been detuned at low frequencies to allow it to satisfy restrictions posed by the boundaries at frequencies beyond the Nyquist frequency, including the ripple frequency. Fig. \ref{fig:designOK}-left shows the Nichols plot of the open-loop gain including the notch filter, that is $L_0(e^{j\omega T_s)} = N(e^{j\omega T_s})G_L(e^{j\omega T_s})P_{L0}(e^{j\omega T_s})$. Note that although at low frequencies the open-loop gain has been slightly detuned, there is no much significative differences with the design of Fig. \ref{fig:boundsSbellow}; however, at frequencies close to the Nyquist frequency $\pi/T_s$ the open-loop gain satisfies the restrictions posed by boundaries, in particular the boundary $\#10$ which is the dominant boundary corresponding to the ripple frequency. The validation of the design is performed by checking the value of the continuous sensitivity $S(j\omega)$ for frequencies up to $\omega = 100$ Rad/s. In contrast to Fig. \ref{fig:boundsSbeyond}-right, it is shown in Fig. \ref{fig:designOK}-right how $S(j\omega)$ clearly satisfies tracking specification for frequencies beyond $\pi/T_s$. As it is above discussed, the detuning of the open-loop gain at low frequencies is also clearly seen when comparing both Figures, but it is not considered relevant in practice. Of course, a better design without the need of detuning could be performed but at the cost of using a slow controller with a much higher order. Finally, a time simulation of the initial DR control system and its redesign to avoid the ripple is shown in Fig. \ref{fig:timesim}.  
\begin{figure}[h]
\centering
\begin{subfigure}{.5\textwidth}
  \centering
  \includegraphics[width=\linewidth]{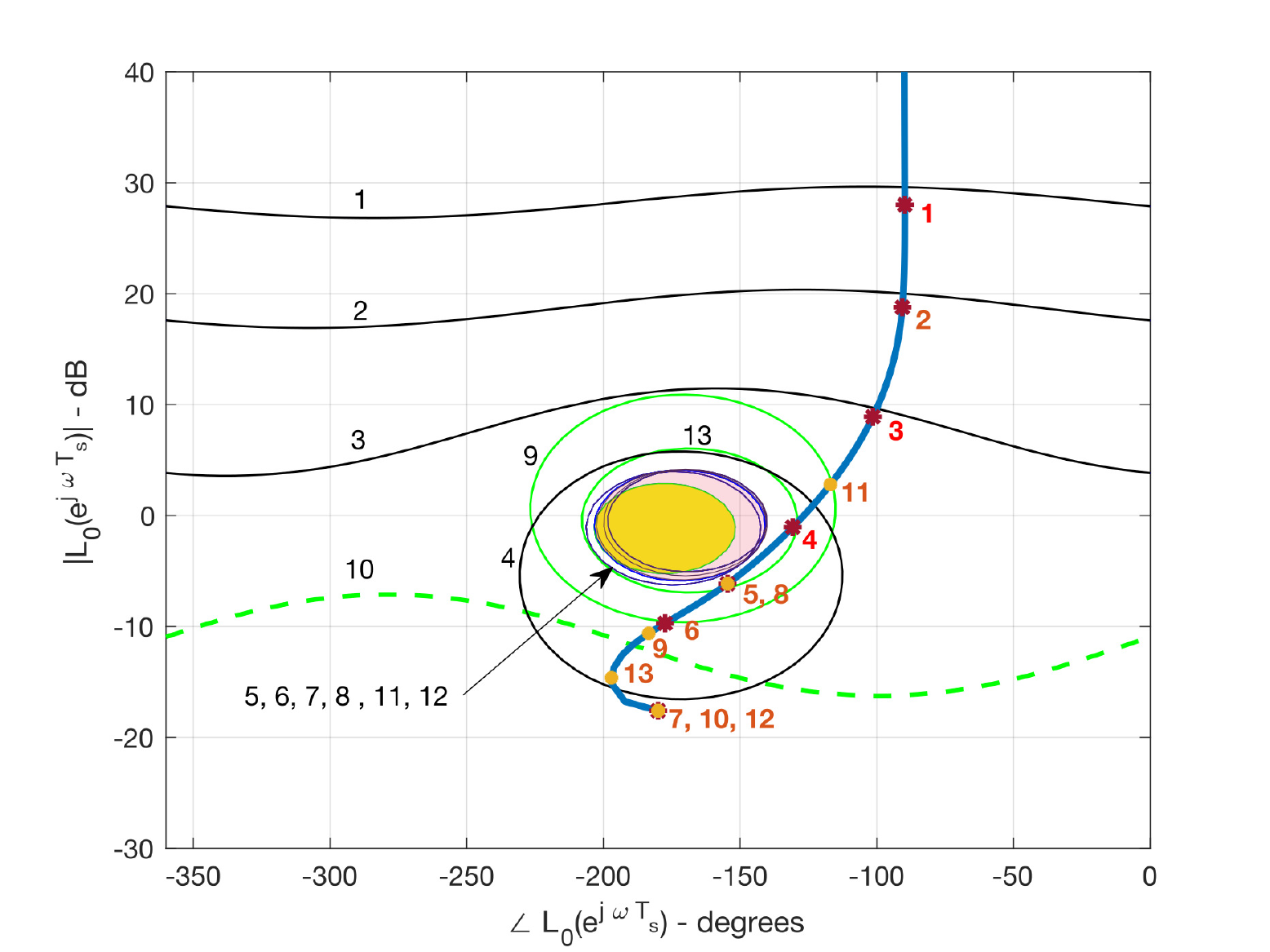}
  %\caption{A subfigure}
  \label{fig:sub1}
\end{subfigure}%
\begin{subfigure}{0.5\textwidth}
  \centering
  \includegraphics[width=\linewidth]{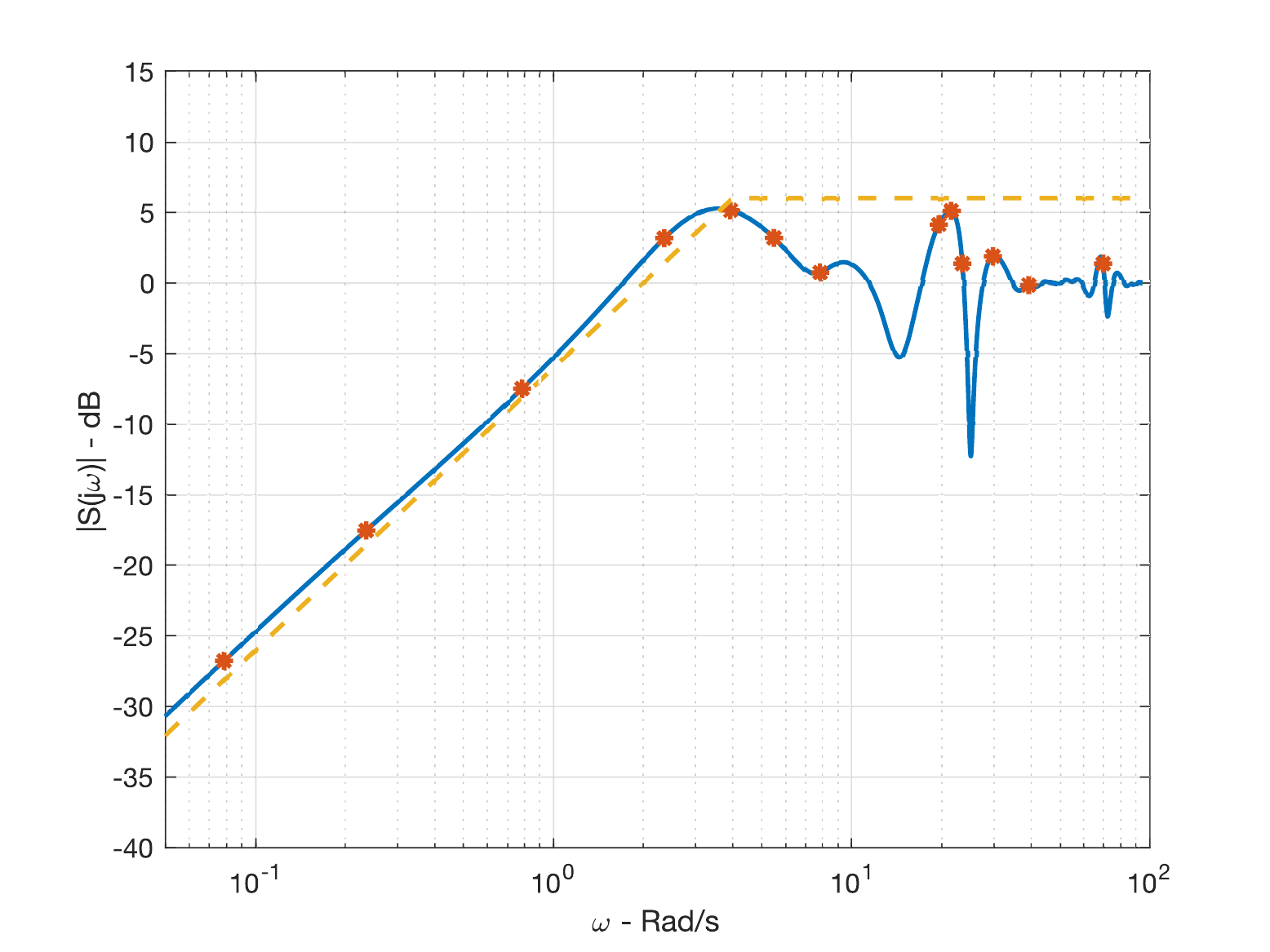}
  %\caption{A subfigure}
  \label{fig:sub2}
\end{subfigure}
\caption{ Continuous-time tracking for frequencies below and beyond the Nyquist frequency $\pi/T_s$. Redesign by using a notch filter. (Left) Boundaries at the working frequencies $\omega_{1,2,\cdots,13} = (0.01,0.02,0.1,0.3,0.5,0.7,1, 2.5,2.75,3,3.8,5,8.9)\cdot \pi/T_s$, indexed by the labels $1, 2, \cdots,13$, and nominal open loop gain $L_0(e^{j\omega T_s})$ over $[0,\pi/T_s]$ (thick line) with asterisks denoting its values at the corresponding working frequencies (note that working frequencies beyond $\pi/T_s$ are folded).  (Right) Magnitude Bode plot of the sensitivity function $S(j\omega)$ (solid line), and specification bound (dotted line). 
}
\label{fig:designOK}
\end{figure}

\begin{figure}[h]
\begin{center}
{\includegraphics[width=0.5\textwidth]{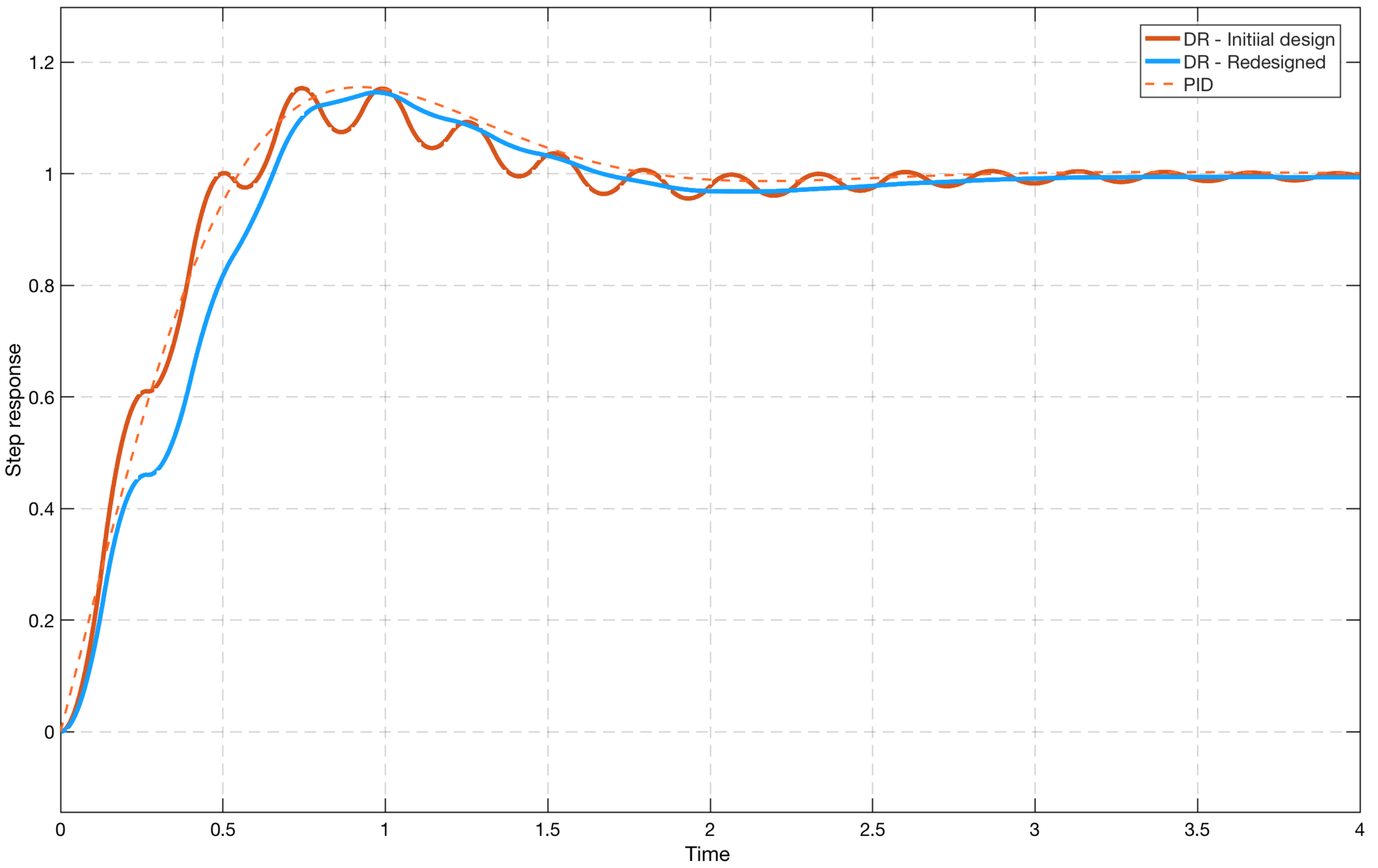}}
\caption{ Step response of the initial DR control system, the redesigned DR control system, and the corresponding to the PID controller. 
 }
\label{fig:timesim}
\end{center}
\end{figure}
%\newpage
%\end{itemize}  
\section{Application}
In this section, the QFT design procedure developed above is going to be applied to an unstable system with parametric uncertainty: a reaction wheel balancing. In contrast to Examples 8-10, where besides robust stability the focus was in the performance at frequencies beyond the (slow) Nyquist frequency, in this application case the design challenge will be at low frequencies, below the Nyquist frequency, with the added difficulty posed by the fact that the open-loop system is unstable.

The inverted pendulum is a classical control problem, frequently used as a test-bed to evaluate different control strategies. Among several versions of the inverted pendulum, in the reaction wheel inverted pendulum (RWIP) the motor is located at the top of the pendulum instead than at its base. A flywheel connected to the motor axis generates the torque that keeps the pendulum in its unstable equilibrium position, cancelling the unavoidable disturbances and following a desired reference. Figure \ref{fig:wheel} shows a CAD model of the RWIP.
\begin{figure}[ht]
\begin{center}
{\includegraphics[width=0.6\textwidth]{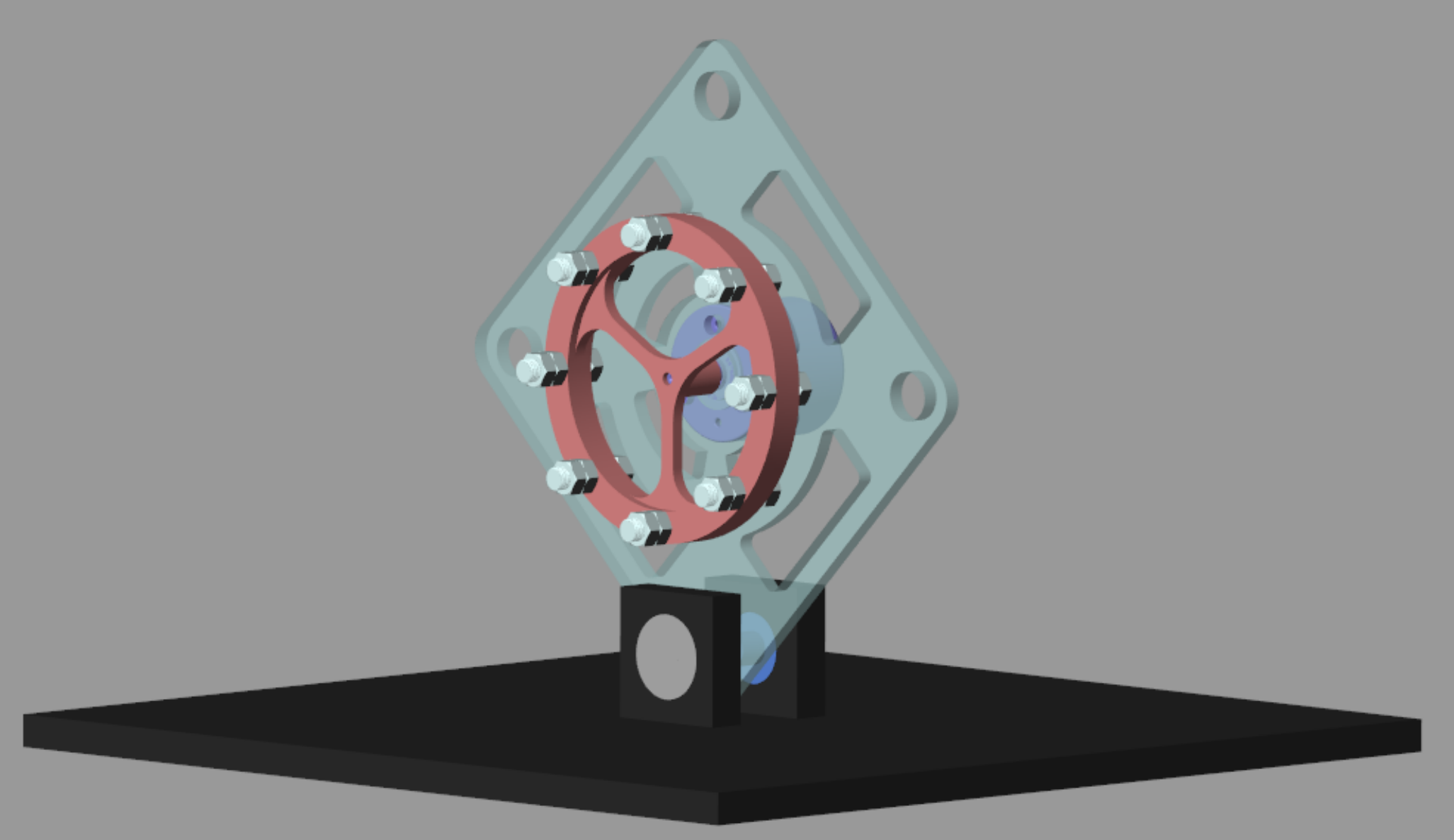}}
\caption{Reaction Wheel Set-up.}
\label{fig:wheel}
\end{center}
\end{figure}
An angular position sensor is located at the joint between the pendulum and the base. This sensor is used to measure the angular position of the pendulum, $\theta(t)$. 

%The first goal of the control structure is to keep this signal as close to zero as possible, canceling disturbances. A secondary goal
A DR control design problem will defined by using stability and tracking specifications, the goal is to move the pendulum following a specified reference, typically a sinusoid. The case of no prefilter, that is $F(s) = 1$ will be considered.
To reach these goals the motor applies a certain torque to the flywheel. The torque causes the angular velocity of the wheel, $w(t)$. The acceleration of the flywheel generate a torsion torque that rotates the pendulum around the joint. 
In order to design the appropriated controller, a mathematical model of the RWIP is needed. The non-linear equation that describes the relationship between the angular velocity of the flywheel and the angular position of the pendulum is as follows: 
\begin{equation}
J_f \dot w (t) + \widehat{ML} g sin (\theta(t))-B\dot \theta(t)=J_T \ddot\theta (t)
\label{eq1:wheel}
\end{equation}
being $J_T$ the moment of inertia of the RWIP that can be calculated using the Steiner theorem as follows:
\begin{equation}
J_T=m_p {l_p}^2+m_f {l_f}^2+J_p+J_f
\end{equation}
where constants $m_p$ and $m_f$ are the mass of the pendulum (including the motor attached to it) and the flywheel. Constants $l_p$ and $l_f$ are the distance between the rotation point of the pendulum and the center of gravity of pendulum and flywheel (note that in the proposed structured shown in Figure \ref{fig:wheel} these distances are equal). Constants $J_p$ and $J_f$ are the moments of inertia of the pendulum and flywheel, which depends on its density and geometry. Table \ref{table:wheel} shows the values of these parameters, measured in the CAD model of the proposed RWIP.
\begin{table}[t]
\begin{center}
\begin{tabular}{| l | c | c |}
\hline
Constant & Value & Units \\ \hline
$J_p$ & 413 & $kg \hspace{1 mm} mm^2$ \\ \hline
 $m_p$ & 0.233 & $kg$ \\  \hline
 $l_p$ & 84.85 & $mm$ \\  \hline
 $J_f (nominal)$ & 290 & $kg \hspace{1 mm} mm^2$ \\  \hline 
 $m_f$ & 0.147 & $kg$ \\  \hline
 $l_f$ & 84.85 & $mm$ \\  \hline
 $B$ & 0.1 & $N \hspace{1 mm} m/s$ \\  \hline 
 $g$ & 9.81 & $m/s^2$ \\  \hline 
\end{tabular}
\caption{Physical constant values for the RWIP} %considered.}
\label{table:wheel}
\end{center}
\end{table}
$\widehat{ML}$ is the product of the masses and distances of the different parts of the RWIP:
\begin{equation}
\widehat{ML}=m_p l_p+m_f l_f
\end{equation}
Constant $B$ is the viscous friction of the motor joint that must be experimentally determined in the motor that moves the flywheel. Finally, $g$ is the gravitational constant.  
Nonlinear dynamics from \eqref{eq1:wheel} can be easily linearized for small movements around the working point $\theta(t)=0$, i.e. the upwards unstable equilibrium position. The result is the plant transfer function $P(s)$ given by 
%\begin{equation}
%J_f \dot w (t) +\widehat{ML} g \theta(t)-B\dot \theta(t)=J_T \ddot\theta (t)
%\end{equation}
%The linear differential equation can be converted into a transfer function, by using Laplace transform:
\begin{equation}
P(s) = \frac{\theta(s)}{w(s)}=\frac{J_f s}{J_T s^2 + B s -ML g}
\end{equation}
Moreover, different weights in the flywheel are obtained by allowing a varying number of screws (Fig. 16). The nominal value $J_f= 290 \hspace{1mm} kg\hspace{1  mm} mm^2$ may be reduced until a third of it resulting in an uncertain parameter $J_f \in \left [ 1/3,1 \right] \hspace{1mm} 290 \hspace{1mm} kg\hspace{1  mm} mm^2$ ({the correct values of $J_f$ are obtained by means of NX Siemens software}). In the following, all the time simulations have been performed using {\it Simscape Multibody} of Simulink considering all physical constants of our set-up.

The design specifications are robust stability and robust continuous-time tracking: the DR control system must be stable and satisfy some stability margin $\mu$, and also tracks a sinusoidal reference of amplitude 10 degrees and frequency 0.1 Hz; and for any $J_f \in \left [ 1/3,1 \right] \hspace{1mm} 290 \hspace{1mm} kg\hspace{1  mm} mm^2$. More specifically, a stability margin $\mu = 1/\sqrt{2}$ (corresponding to worst-case margins PM = $\approx 40^\circ$ and GM $ \approx 10$ dB) has been chosen. Also, the continuous-time tracking specification is based on a second order model with $\xi = 0.5$ and $\omega_n = 5$, it is given by $\left | \frac{E(j \omega)}{R(j \omega)} \right | \leq  \delta_2 (\omega)$, where
\begin{equation}
\delta_2 (\omega) = \left\lbrace
\begin{array}{ll}
1-\frac{5^2}{(j\omega)^2 + 5(j\omega)+5^2} & \textup{if } \omega \leq 5\hspace{1mm}Rad/s  \\
\sqrt{2} &  \textup{if }\omega > 5\hspace{1mm}Rad/s  
\end{array}
\right.
\end{equation}
and the reference is the sinusoidal signal $R(j\omega) = A\frac{b}{(j\omega+a)^2 + b^2}$ with $A = 10\pi/180$ and $b = 2\pi/10$. Moreover, there are design restrictions regarding the controller digital implementation: the angle measurement should be performed at most each $T_s=8 \hspace{1mm} ms$, while the control action may be updating with $T_f=4 \hspace{1mm} ms$. 
\vspace{0.2cm}

\subsection{PID-based DR controller}
The design procedure starts with a continuous-time PID that has been tuned to satisfy the design specifications for the nominal plant (with $J_f = 290  \hspace{1mm} kg\hspace{1  mm} mm^2$). The result is $ G_r(s) = K_r(1+T_ds+1/(sT_i))$, with $K_r = 42.2$, $T_d = 0.031$, and $T_i=3$.
Now, this PID is used for a first DR controller design, consisting of a slow (integral) and fast (derivative) parts discretization. In this case, the proportional constant was included in the integral part (although it does not matter to include it in the derivative part). The result is

%When the digital implementation was planed, the proposed performance required a sampling period smaller than $T=7 \hspace{1mm} ms$. A dual-rate with angle measurement each $T_s=8 \hspace{1mm} ms$ and control updating $T_f=4 \hspace{1mm} ms$, that is $N=2$ was considered by slow (integral) and fast (derivative) parts discretization following know formula. In this case the proportional constant was included in the integral part (although it does not matter to include it in the derivative part)

\begin{equation}
 G_L(z_s)=K_r\frac{z_s-(1-(t/T_i))}{z_s-1}=42.2\frac{z_s-0.9973}{z_s-1}
\label{eq:PIDL} 
\end{equation}

\begin{equation}
 G_R(z_f)=\frac{(1+(T_d/T_f))z_f - (T_d/T_f)}{z_f}=\frac{8.817z_f-0.8866}{z_f}
\label{eq:PIDR} 
\end{equation}

A time simulation of the DR control system with the above controllers is shown in Fig. \ref{fig:disenom}, for the nominal case. The result is that the sensitivity function does not satisfy the tracking specification, note that in particular $|E(j0.63)/R(j0.63)| \approx - 10$ dB, which is far from the design specification of aproximately $-20$ dB obtained from (61) (see also Fig. \ref{fig:disenom}-right) . For other values of the parameter $J_f$ the performance is even worst. 
%With this dual-rate controller, the set-up control is achieved, but there is no consideration of reference tracking and perturbation rejection. These problems are shown in Figure \ref{fig:disenom}, when a sinusoidal reference of $0.1 \hspace{1 mm} Hz$ and amplitude of $10^{\circ}$ is considered . 
\begin{figure}[t]
\centering
\begin{subfigure}{.45\textwidth}
  \centering
  \includegraphics[width=\linewidth]{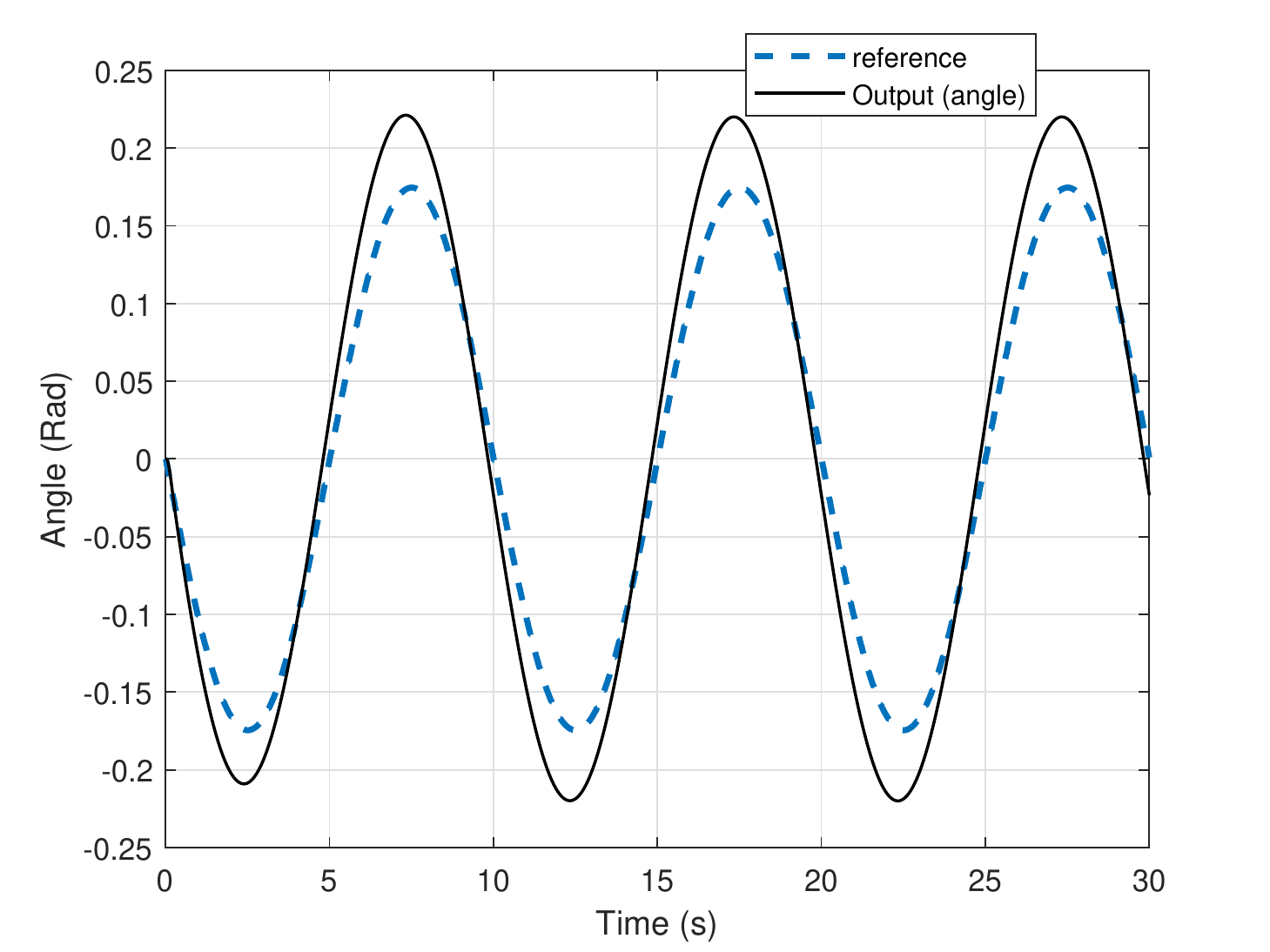}
  %\caption{Tracking reference with initial dual-rate controller: Output}
  %\label{fig:dise_nom}
\end{subfigure}%
\begin{subfigure}{0.475\textwidth}
  \centering
  \includegraphics[width=\linewidth]{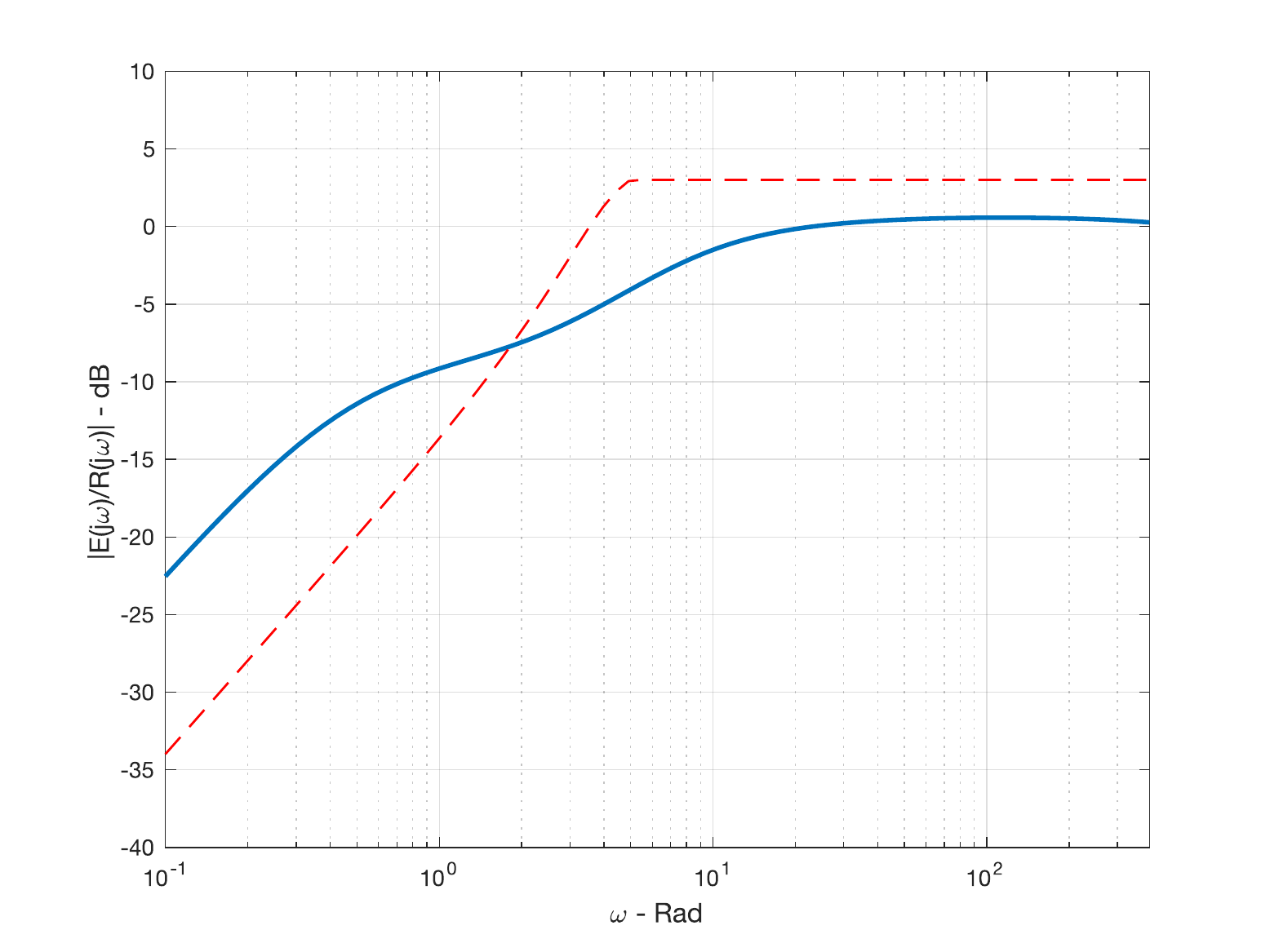}
  %\caption{Tracking reference with initial dual-rate controller: control action}
  %\label{fig:dise_nomu}
\end{subfigure}
\caption{Tracking of a sinusoidal reference with the PID-based DR controller \eqref{eq:PIDL}-\eqref{eq:PIDR}: left) RWIP angle (solid) and reference (dotted), and right) Sensitivity magnitude $|E(j\omega)/R(j\omega)|$ for the sinusoidal reference (solid), and tracking specification bound $\delta_2(\omega)$ (dotted).}
\label{fig:disenom}
\end{figure}
%\begin{figure}
%\begin{center}
%{\includegraphics[width=\textwidth]{Figuras/dr_nominal.png}}
%\caption{Tracking reference with initial dual-rate controller: Output}
%\label{fig:dise_nom}
%\end{center}
%\end{figure}
%\begin{figure}
%\begin{center}
%{\includegraphics[width=\textwidth]{Figuras/dr_nominaluu.png}}
%\caption{Tracking reference with initial dual-rate controller: control action}
%\label{fig:dise_nomuu}
%\end{center}
%\end{figure}
%% JT=mp*lp^2+mv*lv^2+Jp+Jv ;
%ML=mp*lp+mv*lv ; 
%
%B=0.1 ; Jm=0.001 ; Kt=10 ;
%Otras pruebas B=6e-3 ; Jm=0.001 ; Kt=3.46 ; 
%Gp1=Jv*s/(JT*s^2+B*s-ML*g) ;

\subsection{QFT design of the dual-rate controller}

The next design step consists of designing a QFT DR controller for the RWIP following the design procedure developed in Section 4. The fast controller (63) is used, jointly with the uncertain plant model and the design specifications (robust stability and robust tracking of a sinusoidal reference), to design a new slow controller, that will be referred to as $G_{L,QFT}$. %, based of the design procedure developed in Section 4, and with the goal of satisfying the robust stability and tracking specification defined above. 

%, the QFT procedure was considered assuming a uncertainty in the moment of inertia of the flywheel (due to different weights) $J_f=\left [ 1/3,1 \right] \hspace{1mm} 290 \hspace{1mm} kg\hspace{1  mm} mm^2$,
%designing a slow controller for the reference tracking (the sinusoidal said before) assuming the specification
%
%$$
%\left | \frac{E(w)}{R(w)} \right | \leq  \delta_2 (\omega)   
%$$
%being
%
%\begin{equation*}
%\delta_2 (\omega) = \left\lbrace
%\begin{array}{ll}
%1-\frac{5^2}{s^2 + 5 s+5^2} & \textup{if } \omega \leq 5\hspace{1mm}Rad/s  \\
%\sqrt{2} &  \textup{if }\omega > 5\hspace{1mm}Rad/s  
%\end{array}
%\right.
%\end{equation*}
%\\
% leq \frac{5^2}{s^2 + 5 s+5^2}

\vspace{0.25cm}
\noindent
{\bf Robust stability}
Note that the plant, given by (60), always have one unstable pole (the uncertain parameter $J_f$ only affects its gain), thus the procedure developed in Section 4.1 (based on Prop.1) can be directly used. DR control system stability is guarantied if: the nominal DR control system is stable, and the 
nominal open-loop gain is out of the forbidden regions defined by the stability bounds. 

Firstly, for stability of the nominal case Prop. 1 has to be used. Assumptions 1 and 2 are easily checked (details are omitted by brevity). This first stability condition is satisfied if the Nichols plot of the nominal open-loop gain makes a net number of crossings of ${\bf R}_0$ equal to 1 (the number of open-loop unstable poles). Since there is an open-loop integrator (given by (62)), then there is a half crossing $-1/2$ (see Remark 2). %that proceeds from the identation of $z_s = 1$ in the Nyquist path. 
Thus, for a stable design with an open-loop integrator, the Nichols plot of the open-loop gain must perform a crossing  $+1$ of the ray ${\bf R}_0$, in this way the net number of crossings is $-1/2 + 1 = 1/2$, and its double is the number of unstable poles. For example, the PID-based dual controller guaranties a nominal stable design (see Fig. 18-left, its Nichols plot of the nominal open-loop gain satisfies the crossing condition). Note that the net number of crossings must be equal to $1/2$ for any nominal stable design. 

Secondly, stability bounds are computed according to (45) for the stability margin $\mu = 1/\sqrt{2}$ as specified. In this case, since there is only uncertainty in the gain plant all the stability bounds are identical for any frequency. The forbidden region corresponds to the shadow region in Fig. 18-left. Note that, although  
the PID-based DR controller makes the nominal control system stable, its Nichols plot enters the forbidden region and thus the DR is not robustly stable with the specified margin. Thus, the slow controller must be redesigned to avoid the forbidden region at every frequency in order to satisfy design specifications.

\begin{figure}[t]
\centering
\begin{subfigure}{.45\textwidth}
  \centering
  \includegraphics[width=\linewidth]{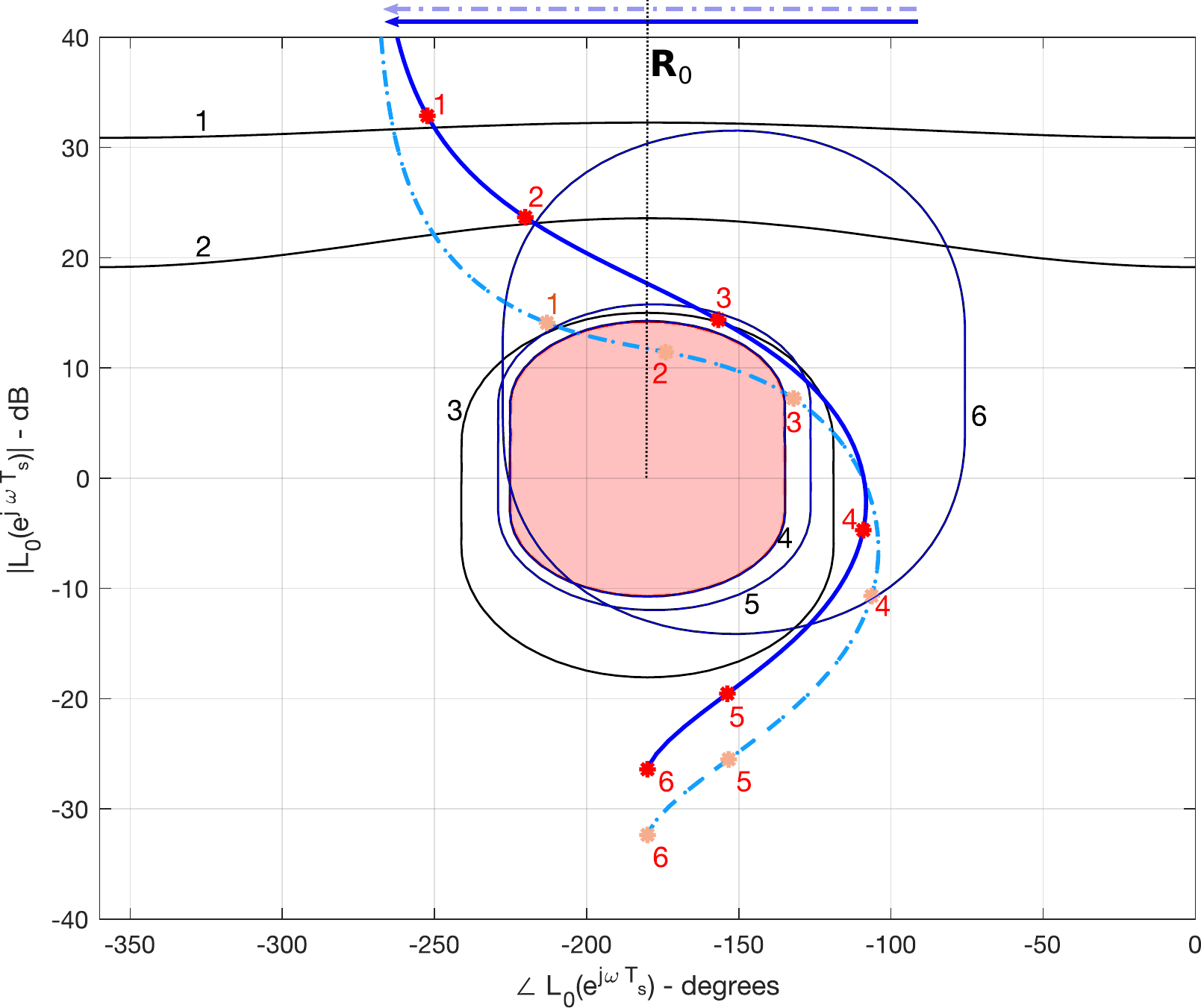}
  %\caption{Tracking reference with initial dual-rate controller: Output}
  %\label{fig:dise_nom}
\end{subfigure}%
\begin{subfigure}{0.525\textwidth}
  \centering
  \includegraphics[width=\linewidth]{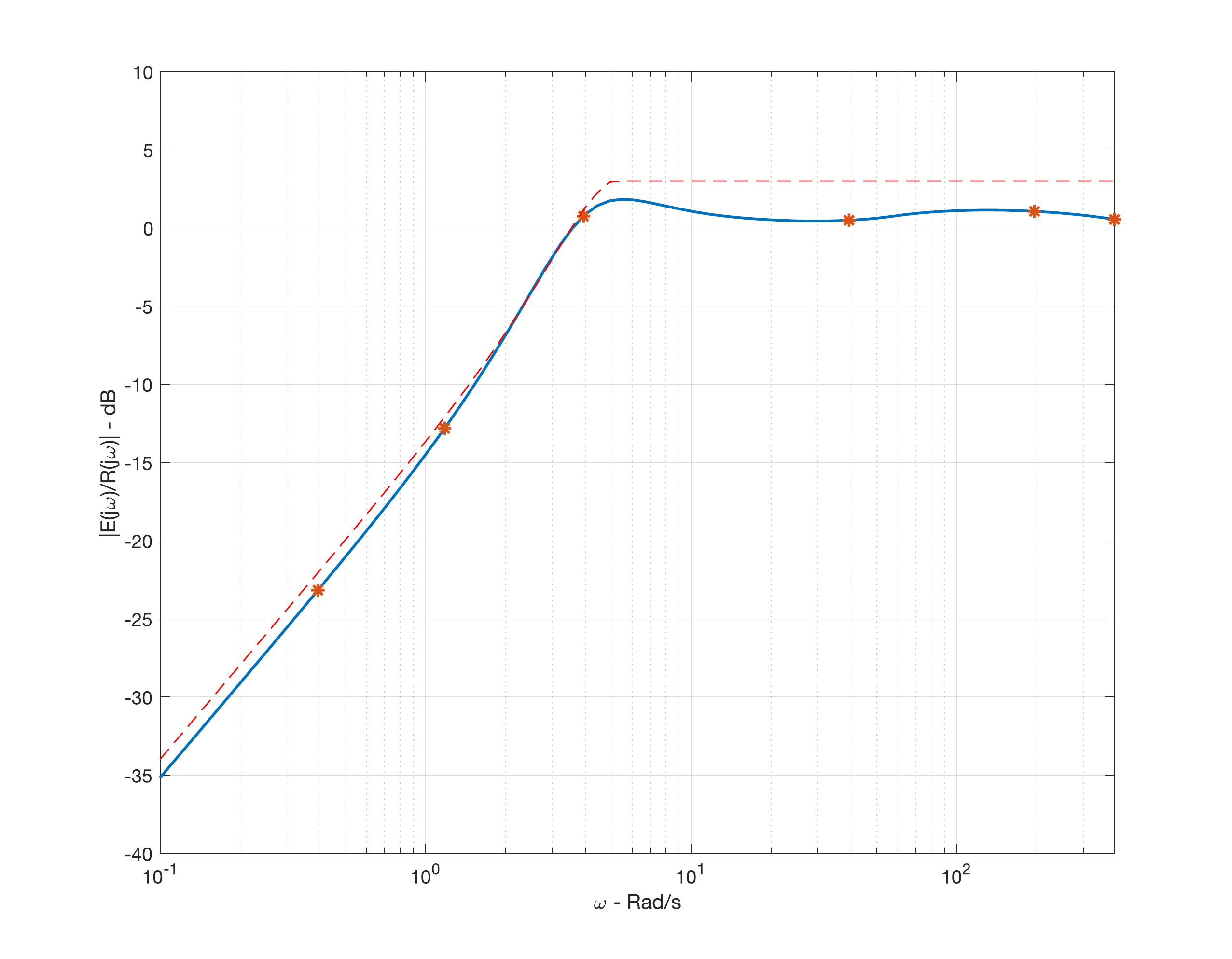}
  %\caption{Tracking reference with initial dual-rate controller: control action}
  %\label{fig:dise_nomu}
\end{subfigure}
\caption{QFT design of the dual rate controller: (left) stability and continuous-time tracking boundaries, and loop-shaping of the PID-based DR controller, and the modified QFT controller; (right) validation of robust continuous-time tracking specifications for the QFT design.  }
\label{fig:disenom2}
\end{figure}

\vspace{0.25cm}
\noindent
{\bf Robust tracking} Here, the design procedure starts with the computation the tracking bounds that define the forbidden regions in the Nichols plane. The chosen design frequencies are $\omega \in \{0.001,0.003,0.01,0.1,0.5,1\}\pi/T_s$. In this problem, frequencies beyond $\pi/T_s$ result in forbidden regions that are not significative, since they are less demanding than for example the corresponding to stability bounds (details are omitted by brevity), and thus the design will be focused on frequencies below $\pi/T_s$.  
The computed stability bounds are shown in Fig. 18-left. Note that the PID-based DR controller does not satisfy the tracking specifications for the design frequencies $0.001\pi/T_s$,  $0.003 \pi/T_s$, and $0.01\pi/T_s$.

\vspace{0.5cm}
Using the slow PID-based controller (62) as starting point, it needs to be redesign to satisfy both robust stability and tracking specifications. Clearly, its gain should be increased (this is equivalent to move upward the Nichols plot in Fig. 18-left) to satisfy low frequencies bounds; however, something else is needed since otherwise high-frequency bounds may be crossed and thus tracking specifications would not be satisfied at those frequencies. A solution has been obtained both modifying the controller gain and its zero. The result is 
%In Figure \ref{fig:design} the QFT slow controller design is explained, leading to:
\begin{equation}
 G_{L,QFT}(z_s) = 84.4\frac{z_s-0.9823}{z_s-1} 
\end{equation}
Note that the Nichols plot of the open-loop gain corresponding to (64) satisfied the crossing condition, avoids the forbidden stability region, and also does not enters the forbidden regions defined by the tracking bounds. Regarding tracking, this must be validated for frequencies different to the design frequencies. Fig. 18-right shows how the sensitivity function satisfies the tracking bound, and thus the design is validated. 

%\begin{figure}[htbp]
%\begin{center}
%{\includegraphics[width=0.75\textwidth]{Figuras/pendulo.pdf}}
%\caption{QFT design procedure}
%\label{fig:design}
%\end{center}
%\end{figure}
%and the Sensitivity validation 
%%taking account 
%%$$ S=1-\frac{5^2}{s^2 + 5 s+5^2}
%%$$
%is showed in Figure \ref{fig:designS}.
%\begin{figure}[htbp]
%\begin{center}
%{\includegraphics[width=0.45\textwidth]{Figuras/penduloS.pdf}}
%\caption{$G_L$ QFT design procedure}
%\label{fig:designS}
%\end{center}
%\end{figure}
Finally,  time simulation plots with this new DR controller (63)-(64) are shown in Fig. \ref{fig:todos}. The cases for $J_f=290\hspace{1mm} kg\hspace{1  mm} mm^2$ (nominal value), $J_f= 0.4\times 290\hspace{1mm} kg\hspace{1  mm} mm^2 = 122\hspace{1mm} kg\hspace{1  mm} mm^2$, and $J_f= 0.3\times 290\hspace{1mm} kg\hspace{1  mm} mm^2 = 88\hspace{1mm} kg\hspace{1  mm} mm^2$ (case without screws), have been considered. Note that the design performs correctly in spite of the uncertainty.  As it may be expected, the design results in a more demanding control action of the motor for decreasing values of $J_f$.
%that is nominal $J_f\times 0.4$ and $J_f=88\hspace{1mm} kg\hspace{1  mm} mm^2$ ($J_f\times 0.3$) -case without screws- are shown in Figure \ref{fig:todos}. 

%The last case with a slight violation of the uncertainty interval is clearly worst and the control action is unavailable for our motor.
%A lower $J_f$ implies a more difficult control. 

%\begin{figure}[h]
%\centering
%\begin{subfigure}{.45\textwidth}
%  \centering
%  \includegraphics[width=\linewidth]{Figuras/dr_slowqft.eps}
%  %\caption{Tracking reference with QFT designed dual-rate controller: Output}
%  \label{fig:dise_qfta}
%\end{subfigure}
%\begin{subfigure}{0.45\textwidth}
%  \centering
%  \includegraphics[width=\linewidth]{Figuras/dr_slowqftuu.eps}
%  %\caption{Tracking reference with QFT designed dual-rate controller: control action}
%  \label{fig:dise_qftau}
%\end{subfigure}
%\caption{Tracking reference with QFT designed dual-rate controller $J_f=290\hspace{1mm} kg\hspace{1  mm} mm^2$: a) Output (Rad) and b) control action %(RPM)}
%\label{fig:diseqft}
%\end{figure}

\begin{figure}[t]
\centering
\begin{subfigure}{.45\textwidth}
  \centering
  \includegraphics[width=\linewidth]{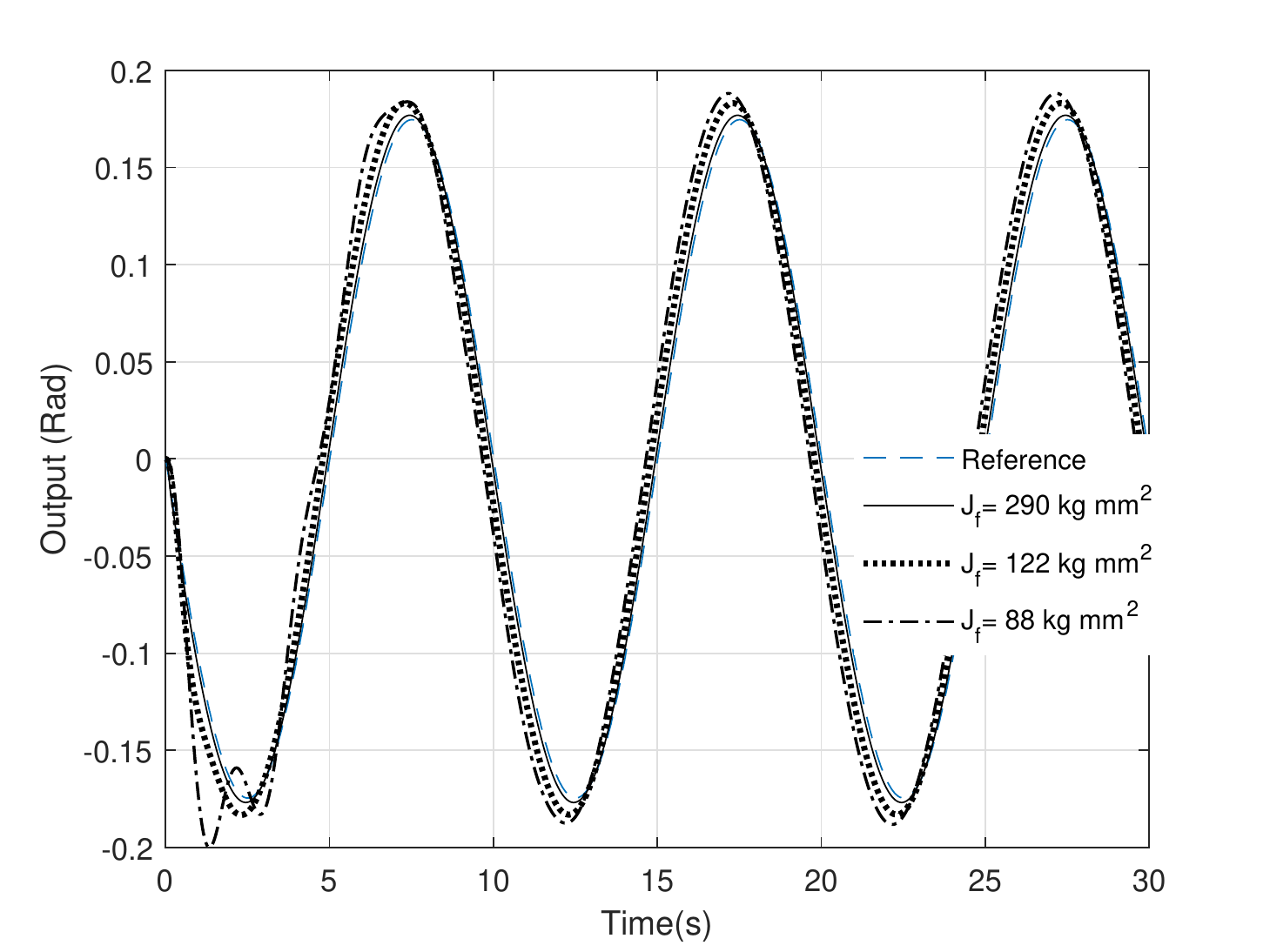}
  %\caption{Tracking reference with QFT designed dual-rate controller: Output}
  \label{fig:dise_qftb}
\end{subfigure}
\begin{subfigure}{0.45\textwidth}
  \centering
  \includegraphics[width=\linewidth]{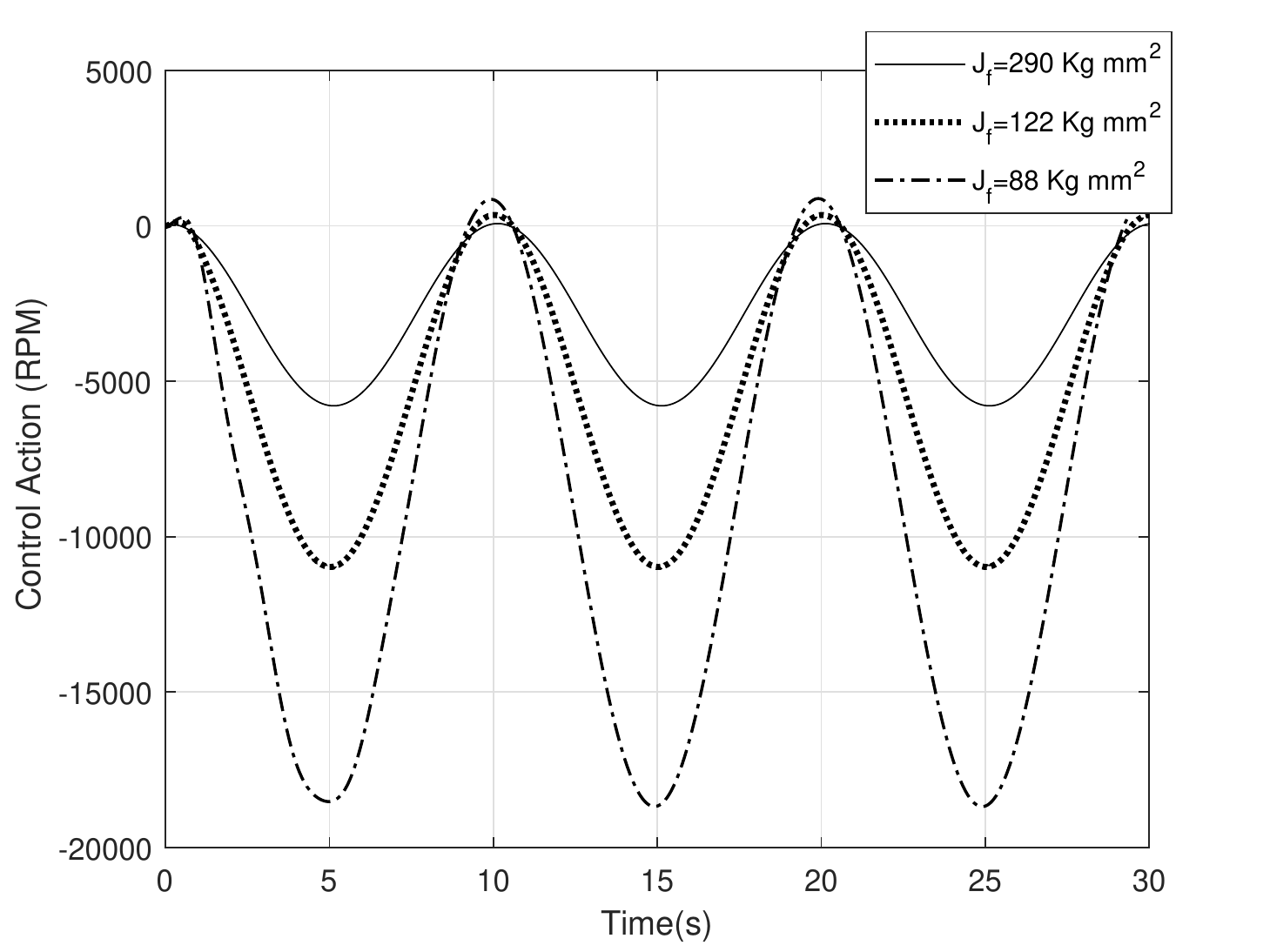}
  %\caption{Tracking reference with QFT designed dual-rate controller: control action}
  \label{fig:dise_qftbu}
\end{subfigure}
\caption{Tracking reference with QFT designed DR controller. Comparison among cases $J_f=290\hspace{1mm} kg\hspace{1  mm} mm^2$, $J_f=122\hspace{1mm} kg\hspace{1  mm} mm^2$, $J_f=88\hspace{1mm} kg\hspace{1  mm} mm^2$: a) Output (Rad) and b) control action (RPM)}
\label{fig:todos}
\end{figure}

%\begin{figure}[h]
%\centering
%\begin{subfigure}{.45\textwidth}
%  \centering
%  \includegraphics[width=\linewidth]{Figuras/qft_Jx03034.eps}
%  %\caption{Tracking reference with QFT designed dual-rate controller: Output}
%  \label{fig:dise_qft}
%\end{subfigure}
%\begin{subfigure}{0.45\textwidth}
%  \centering
%  \includegraphics[width=\linewidth]{Figuras/qft_Jx03034uu.eps}
%  %\caption{Tracking reference with QFT designed dual-rate controller: control action}
%  \label{fig:dise_qftu}
%\end{subfigure}
%\caption{Tracking reference with QFT designed dual-rate controller $J_f=88\hspace{1mm} kg\hspace{1  mm} mm^2$: a) Output and b) control action}
%\label{fig:qftJ03}
%\end{figure}
%
%\begin{figure}
%\begin{center}
%{\includegraphics[width=\textwidth]{Figuras/dr_slowqft.png}}
%\caption{Tracking reference with QFT designed dual-rate controller: Output}
%\label{fig:dise_qft}
%\end{center}
%\end{figure}
%\begin{figure}
%\begin{center}
%{\includegraphics[width=\textwidth]{Figuras/dr_slowqftuu.png}}
%\caption{Tracking reference with QFT designed dual-rate controller: control action}
%\label{fig:dise_qftu}
%\end{center}
%\end{figure}
%As it can be seen, the specs are fulfilled.

%%%%%%%%%%%%%%%%%%%%%%%%%%%%%%%yaniv
%\vspace{5cm}
\section{Conclusions}
In spite of the large number of contributions on dual-rate control systems, there has been a lack of efficient techniques for their analysis and design in the frequency domain. In this work, a QFT approach is proposed to cope with this problem. Besides allowing the formulation a Nyquist-like stability result, also including worst-case stability margins, robust tracking specifications are considered both in the discrete-time domain and in the continuous-time domain. As a result, a new QFT-based technique has been developed for the design of robust DR control systems, using as a design element the slow discrete-time controller.  Several detailed examples, and finally a case study (a reaction wheel inverted pendulum), have been developed including cases with/without uncertainty, and with continuous-time tracking specifications below/beyond the Nyquist frequency. To the authors knowledge, this work is the first interdisciplinary work on the areas of QFT and multirate control, that surprisingly have been isolated over the years. Throughout this work, several illustrative examples have been developed with a tutorial style, with the goal of building a bridge between both areas.

%\nocite{*}% Show all bib entries - both cited and uncited; comment this line to view only cited bib entries;
\bibliographystyle{elsarticle-harv}
\bibliography{multirate,multirateQFT}%

\begin{thebibliography}{63}
\expandafter\ifx\csname natexlab\endcsname\relax\def\natexlab#1{#1}\fi
\expandafter\ifx\csname url\endcsname\relax
  \def\url#1{\texttt{#1}}\fi
\expandafter\ifx\csname urlprefix\endcsname\relax\def\urlprefix{URL }\fi

\bibitem[{Alcaina et~al.(2019)Alcaina, Cuenca, Salt, Zheng, and
  Tomizuka}]{alcaina2019energy}
Alcaina, J., Cuenca, {\'A}., Salt, J., Zheng, M., Tomizuka, M., 2019.
  Energy-efficient control for an unmanned ground vehicle in a wireless sensor
  network. Journal of Sensors 2019.

\bibitem[{Araki and Hagiwara(1986)}]{araki1986pole}
Araki, M., Hagiwara, T., 1986. Pole assignment by multirate sampled-data output
  feedback. International Journal of Control 44~(6), 1661--1673.

\bibitem[{Araki and Ito(1993)}]{araki1993frequency}
Araki, M., Ito, Y., 1993. Frequency-response of sampled-data systems i:
  open-loop consideration. IFAC Proceedings Volumes 26~(2), 259--262.

\bibitem[{Araki and Yamamoto(1986)}]{araki1986multivariable}
Araki, M., Yamamoto, K., 1986. Multivariable multirate sampled-data systems:
  state-space description, transfer characteristics, and nyquist criterion.
  Automatic Control, IEEE Transactions on 31~(2), 145--154.

\bibitem[{Bamieh et~al.(1991)Bamieh, Pearson, Francis, and
  Tannenbaum}]{bamieh1991ltl}
Bamieh, B., Pearson, J., Francis, B., Tannenbaum, A., 1991. {A lifting
  technique for linear periodic systems with applications to sampled-data
  control}. Systems \& Control Letters 17~(2), 79--88.

\bibitem[{Ba{\~n}os(2007)}]{Banos07}
Ba{\~n}os, A., 2007. Nonlinear quantitative feedbak theory. International
  Journal of Robust and Nonlinear Control 17, 181--202.

\bibitem[{Ba{\~n}os and Horowitz(2000)}]{BanosHorowitz00}
Ba{\~n}os, A., Horowitz, I.~M., 2000. Qft design of multi-loop nonlinear
  control systems. International Journal of Robust and Nonlinear Control
  10~(15), 1263--1277.

\bibitem[{Boje(2003)}]{Boje03}
Boje, E., 2003. Pre-filter design for tracking error specifications in qft.
  International Journal of Robust and Nonlinear Control 13, 637--642.

\bibitem[{Borghesani et~al.(2000)Borghesani, Chait, and
  Yaniv}]{BorghesaniChaitYaniv00}
Borghesani, C., Chait, Y., Yaniv, O., 2000. QFT frequency domain control design
  toolbox. Terasoft.

\bibitem[{Braslavsky(1995)}]{BraslavskyTesis}
Braslavsky, J.~H., 1995. Frequency-domain analysis of sampled-data control
  systems. The University of Newcastle.

\bibitem[{Cervera and Ba{\~n}os(2008)}]{CerveraBanos08}
Cervera, J., Ba{\~n}os, A., 2008. Automatic loop shaping in qft using crone
  structures. Journal of Vibration and Control 14~(9), 1513--1529.

\bibitem[{Chait et~al.(1999)Chait, Chen, and V}]{ChaitChenHollot99}
Chait, Y., Chen, Q., V, H.~C., 1999. Automatic loop-shaping of qft controllers
  via linear programming. Journal of Dynamic Systems, Measurement, and Control
  121~(3), 351--357.

\bibitem[{Chen and Francis(1991)}]{ChenFrancis91}
Chen, T., Francis, B., 1991. Input-ouput stability of sampled-data systems.
  IEEE Transactions on Automatic Control 36~(1), 50--58.

\bibitem[{Chen and Qiu(1994)}]{chen1994h}
Chen, T., Qiu, L., 1994. $h^\infty$ design of general multirate sampled-data
  control systems. Automatica 30~(7), 1139--1152.

\bibitem[{Chen and Ballance(1998)}]{ChenBallance98}
Chen, W.~H., Ballance, D.~J., 1998. Stability analysis on the nichols chart and
  its application in qft. Tech. Rep. CSC-98013, Center for Systems and Control,
  Univ. Glasgow.

\bibitem[{Cimino and Pagilla(2010)}]{cimino2010design}
Cimino, M., Pagilla, P., 2010. Design of linear time-invariant controllers for
  multirate systems. Automatica 46~(8), 1315--1319.

\bibitem[{Coffey and Williams(1966)}]{coffey1966stability}
Coffey, T.~C., Williams, I.~J., 1966. Stability analysis of multiloop,
  multirate sampled systems. AIAA Journal 4~(12), 2178--2190.

\bibitem[{Cohen et~al.(1994)Cohen, Chait, Yaniv, and
  Borghesani}]{cohen1994stability}
Cohen, N., Chait, Y., Yaniv, O., Borghesani, C., 1994. Stability analysis using
  nichols charts. International Journal of Robust and Nonlinear Control 4~(1),
  21--46.

\bibitem[{Cuenca et~al.(2011)Cuenca, Salt, Sala, and
  Piz{\'a}}]{cuenca2011delay}
Cuenca, {\'A}., Salt, J., Sala, A., Piz{\'a}, R., 2011. A delay-dependent
  dual-rate pid controller over an ethernet network. Industrial Informatics,
  IEEE Transactions on 7~(1), 18--29.

\bibitem[{Cuenca et~al.(2019)Cuenca, Zhan, Salt, Alcaina, Tang, and
  Tomizuka}]{cuenca2019remote}
Cuenca, {\'A}., Zhan, W., Salt, J., Alcaina, J., Tang, C., Tomizuka, M., 2019.
  A remote control strategy for an autonomous vehicle with slow sensor using
  kalman filtering and dual-rate control. Sensors 19~(13), 2983.

\bibitem[{Dasgupta(1999)}]{Dasgupta99}
Dasgupta, S., 1999. An approach to multirate control. In: Proc. 38th Conference
  on Decision and Control. pp. 3446--3451.

\bibitem[{Eitelberg(2000)}]{Eitelberg00}
Eitelberg, E., 2000. Quantitative feedback design for tracking error tolerance.
  Automatica 36, 319--326.

\bibitem[{Elso et~al.(2013)Elso, Gil-Martinez, and
  Garc{\'\i}a-Sanz}]{ElsoGilGarcia13}
Elso, J., Gil-Martinez, M., Garc{\'\i}a-Sanz, M., 2013. A quantitative feedback
  solution to the multivariable traking error problem. International Journal of
  Robust and Nonlinear Control 24~(16), 2331--2346.

\bibitem[{Er and ANDERSON(1991)}]{er1991practical}
Er, M.-J., ANDERSON, B.~D., 1991. Practical issues in multirate output
  controllers. International Journal of control 53~(5), 1005--1020.

\bibitem[{Francis and Georgiou(1988)}]{francis1988stability}
Francis, B., Georgiou, T., 1988. Stability theory for linear time-invariant
  plants with periodic digital controllers. Automatic Control, IEEE
  Transactions on 33~(9), 820--832.

\bibitem[{Garc{\'\i}a-Sanz(2008)}]{GarciaSanz08}
Garc{\'\i}a-Sanz, M., 2008. The qft control toolbox (qftct) for matlab.

\bibitem[{Garc{\'\i}a-Sanz(2017)}]{GarciaSanzBook}
Garc{\'\i}a-Sanz, M., 2017. Robust Control Engineering: practical QFT
  solutions. CRC Press.

\bibitem[{Garc{\'\i}a-Sanz and Guillen(2000)}]{GarciaSanzGuillen00}
Garc{\'\i}a-Sanz, M., Guillen, J.~C., 2000. Automatic loop-shapping of qft
  robust controllers via genetic algorithms. In: 3rd IFAC Symposium on Robuts
  Control Design.

\bibitem[{Gutman(????)}]{Gutman96}
Gutman, P.~O., ???? Qsyn: the Toolbox for robust control systems design for use
  with Matlab.

\bibitem[{Horowitz and Ba{\~n}os(2001)}]{HorowitzBanos01}
Horowitz, I., Ba{\~n}os, A., 2001. Advances in nonlinear systems. No. 264 in
  LNCIS. Springer-London, Ch. Fundamentals of nonlinear quantitative feedback
  theory, pp. 63--134.

\bibitem[{Horowitz(1963)}]{Horowitz63}
Horowitz, I.~M., 1963. Synthesis of feedback systems. Academic Press.

\bibitem[{Horowitz(1975)}]{Horowitz75}
Horowitz, I.~M., 1975. A synthesis theory for linear time-varying feedback
  systems with plant uncertainty. IEEE transactions on Automatic Control
  20~(4), 454--464.

\bibitem[{Horowitz(1976)}]{Horowitz76}
Horowitz, I.~M., 1976. Synthesis of feedback systems with nonlinear
  time-varying uncertain plants to satisfy quantitative performance
  specifications. Proceedings of the IEEE 64, 123--130.

\bibitem[{Horowitz(1993)}]{Horowitz93}
Horowitz, I.~M., 1993. Quantitative feedback design theory-QFT. QFT
  Publications.

\bibitem[{Horowitz and Liao(1986)}]{HorowitzLiao86}
Horowitz, I.~M., Liao, Y.~K., 1986. Quantitative feedback design for
  sampled-data systems. International Journal of Control 44, 665--675.

\bibitem[{Horowitz and Sidi(1972)}]{HorowitzSidi72}
Horowitz, I.~M., Sidi, M., 1972. Synthesis of feedback systems with large plant
  ignorance for prescribed time-domain tolerances. International Journal of
  Control 16~(2), 287--309.

\bibitem[{Hutchinson(1994)}]{hutchinson1994multi}
Hutchinson, S., 1994. Multi-rate analysis and design of visual feedback digital
  servo-control system. Urbana 51, 61801.

\bibitem[{Jury(1977)}]{jury1977sampled}
Jury, E.~I., 1977. Sampled-data control systems. Krieger Publishing Co., Inc.

\bibitem[{Kalman and Bertram(1959)}]{kalman1959general}
Kalman, R.~E., Bertram, J., 1959. General synthesis procedure for computer
  control of single-loop and multiloop linear systems (an optimal sampling
  system). Transactions of the American Institute of Electrical Engineers, Part
  II: Applications and Industry 77~(6), 602--609.

\bibitem[{Keller and Anderson(1990)}]{keller1990new}
Keller, J.~P., Anderson, B.~D., 1990. A new approach to the discretization of
  continuous-time controllers. In: 1990 American Control Conference. IEEE, pp.
  1127--1132.

\bibitem[{Khargonekar et~al.(1985)Khargonekar, Poolla, and
  Tannenbaum}]{khargonekar1985robust}
Khargonekar, P., Poolla, K., Tannenbaum, A., 1985. Robust control of linear
  time-invariant plants using periodic compensation. Automatic Control, IEEE
  Transactions on 30~(11), 1088--1096.

\bibitem[{Kranc(1957)}]{kranc1957input}
Kranc, G., 1957. Input-output analysis of multirate feedback systems. Automatic
  Control, IRE Transactions on 3~(1), 21--28.

\bibitem[{Lall and Dullerud(2001)}]{lall2001lsr}
Lall, S., Dullerud, G., 2001. {An LMI solution to the robust synthesis problem
  for multi-rate sampled-data systems}. Automatica 37~(12), 1909--1922.

\bibitem[{Lee and Morari(1992)}]{lee1992robust}
Lee, J.~H., Morari, M., 1992. Robust inferential control of multi-rate
  sampled-data systems. Chemical Engineering Science 47~(4), 865--885.

\bibitem[{Li et~al.(2003)Li, Shah, Chen, and Qi}]{li2003application}
Li, D., Shah, S.~L., Chen, T., Qi, K.~Z., 2003. Application of dual-rate
  modeling to {CCR} octane quality inferential control. IEEE Transactions on
  Control Systems Technology 11~(1), 43--51.

\bibitem[{Lozano-Perez(2012)}]{lozano2012autonomous}
Lozano-Perez, T., 2012. Autonomous robot vehicles. Springer Science \& Business
  Media.

\bibitem[{Morant and Albertos(1986)}]{morant1986model}
Morant, F., Albertos, P., 1986. Model reference control of a cement mill. In:
  Digital Computer Applications to Process Control. Elsevier, pp. 297--301.

\bibitem[{Nataraj and Kubal(2006)}]{NatarajKubal06}
Nataraj, P. S.~V., Kubal, N., 2006. Automatic loop shaping in qft using hybrid
  optimization and constraint propagation techniques. International Journal of
  Robust and Nonlinear Control 17, 251--264.

\bibitem[{Rubin and Gutman(2019)}]{RubinGutman19}
Rubin, D., Gutman, P., 2019. Open Qsyn.

\bibitem[{S{\aa}gfors et~al.(1998)S{\aa}gfors, Toivonen, and
  Lennartson}]{saagfors1998h}
S{\aa}gfors, M.~F., Toivonen, H.~T., Lennartson, B., 1998. $h^\infty$ control
  of multirate sampled-data systems: A state-space approach. Automatica 34~(4),
  415--428.

\bibitem[{Salt and Albertos(2005)}]{salt2005mbm}
Salt, J., Albertos, P., 2005. {Model-Based Multirate Controllers Design}.
  Control Systems Technology, IEEE Transactions on 13~(6), 988--997.

\bibitem[{Salt and Alcaina(2019)}]{salt2019dual}
Salt, J., Alcaina, J., 2019. Dual-rate sampled-data systems. some interesting
  consequences from its frequency response analysis. International Journal of
  General Systems 48~(5), 554--574.

\bibitem[{Salt et~al.(2014)Salt, Cuenca, Palau, and
  Dormido}]{salt2014multiratesen}
Salt, J., Cuenca, {\'A}., Palau, F., Dormido, S., 2014. A multirate control
  strategy to the slow sensors problem: An interactive simulation tool for
  controller assisted design. Sensors 14~(3), 4086--4110.

\bibitem[{Salt and Sala(2014{\natexlab{a}})}]{SaltSala14}
Salt, J., Sala, A., 2014{\natexlab{a}}. A new algorithm for dual-rate systems
  frequency response computation en discrete control systems. Applied
  Mathematical Modelling 38, 5692--5704.

\bibitem[{Salt and Sala(2014{\natexlab{b}})}]{salt2014new}
Salt, J., Sala, A., 2014{\natexlab{b}}. A new algorithm for dual-rate systems
  frequency response computation in discrete control systems. Applied
  Mathematical Modelling 38~(23), 5692--5704.

\bibitem[{Sim et~al.(2002)Sim, Hong, and Lim}]{sim2002multirate}
Sim, T., Hong, G., Lim, K., 2002. Multirate predictor control scheme for visual
  servo control. IEE Proceedings-Control Theory and Applications 149~(2),
  117--124.

\bibitem[{Sklansky and Ragazzini(1955)}]{sklansky1955analysis}
Sklansky, J., Ragazzini, J., 1955. Analysis of errors in sampled-data feedback
  systems. Transactions of the American Institute of Electrical Engineers, Part
  II: Applications and Industry 74~(2), 65--71.

\bibitem[{Thompson(1986)}]{thompson1986gain}
Thompson, P.~M., 1986. Gain and phase margins of multi-rate sampled-data
  feedback systems. International Journal of control 44~(3), 833--846.

\bibitem[{Whitbeck and Didaleusky(1980)}]{whitbeck1980multirate}
Whitbeck, R.~F., Didaleusky, D., 1980. Multirate digital control systems in
  simulation applications. Report AFWAL-TR-80-3101, Vols. I, II and III, 1980.

\bibitem[{Yaniv and Chait(1993{\natexlab{a}})}]{YanivChait93}
Yaniv, O., Chait, Y., 1993{\natexlab{a}}. Direct control design in sampled-data
  uncertain systems. Automatica 29~(2), 365--372.

\bibitem[{Yaniv and Chait(1993{\natexlab{b}})}]{yaniv1993direct}
Yaniv, O., Chait, Y., 1993{\natexlab{b}}. Direct control design in sampled-data
  uncertain systems. Automatica 29~(2), 365--372.

\bibitem[{Yaniv and Horowitz(1986)}]{YanivHorowitz86}
Yaniv, O., Horowitz, I.~M., 1986. A quantitative design method for mimo linear
  feedback systems having uncertain plants. International Journal of Control
  43~(2), 401--421.

\bibitem[{Zhang et~al.(2017)Zhang, Shi, Wang, and Yu}]{zhang2017analysis}
Zhang, D., Shi, P., Wang, Q.-G., Yu, L., 2017. Analysis and synthesis of
  networked control systems: A survey of recent advances and challenges. ISA
  transactions 66, 376--392.

\end{thebibliography}

%\clearpage

%\section*{Author Biography}

%\begin{biography}{\includegraphics[width=66pt,height=86pt,draft]{empty}}{\textbf{Author Name.} This is sample author biography text this is sample author biogultraphy text this is sample author biography text this is sample author biography text this is sample author biography text this is sample author biography text this is sample author biography text this is sample author biography text this is sample author biography text this is sample author biography text this is sample author biography text this is sample author biography text this is sample author biography text this is sample author biography text this is sample author biography text this is sample author biography text this is sample author biography text this is sample author biography text this is sample author biography text this is sample author biography text this is sample author biography text.}
%\end{biography}

\end{document}